\documentclass[11pt]{article}

\usepackage{jheppub}
\usepackage{amsmath,amssymb,latexsym,bm,amsfonts}

\usepackage{cleveref}

\usepackage{graphicx}

\usepackage{slashed}

\usepackage{xcolor}
\usepackage{color}
\usepackage{float}

\usepackage{braket}
\usepackage{diagbox}
\usepackage{array,multirow}

\usepackage{verbatim}
\usepackage{multirow}
\usepackage{appendix}

\numberwithin{equation}{section}

\newcommand{\be}{\begin{equation}}
\newcommand{\ee}{\end{equation}}
\newcommand{\bear}{\begin{array}}
\newcommand{\eear}{\end{array}}

\title{Flux vacua in type IIB compactifications on orbifolds: their finiteness and minimal string coupling}

\author[a,b]{Ignatios Antoniadis,}
\author[a]{Anthony Guillen,}
\author[c,d]{Osmin Lacombe}

\affiliation[a]{Laboratoire de Physique Th\'eorique et Hautes Energies - LPTHE\\
	Sorbonne Universit\'e, CNRS, 4 Place Jussieu, 75005 Paris, France}
\affiliation[b]{High Energy Physics Research Unit, Faculty of Science, Chulalongkorn University, Bangkok 1030,
Thailand}
\affiliation[c]{Dipartimento di Fisica e Astronomia, Universit\`a di Bologna, via Irnerio 46, 40126 Bologna, Italy}
\affiliation[d]{INFN, Sezione di Bologna, viale Berti Pichat 6/2, 40127 Bologna, Italy}

\emailAdd{antoniad@lpthe.jussieu.fr}
\emailAdd{aguillen@lpthe.jussieu.fr}
\emailAdd{deriusosmin.lacombe@unibo.it}

\date{} 

\abstract{We perform a detailed study of (supersymmetric) moduli stabilisation in type IIB toroidal orientifolds with fluxes. We provide strong evidence towards exhaustion of the finite number of inequivalent vacua for a given total 3-form flux charge $N_{\rm flux}$. We also find that the minimal value of the string coupling $g_s$ is given in terms of $N_{\rm flux}$ and present strong evidence for the asymptotic relation $g_{s,\,\rm{min}}\sim c/N_{\rm flux}^\alpha$, with $\alpha=1,2$, valid for not too small $N_{\rm flux}$, where $c$ is an order 1 coefficient and $\alpha$ depends on the orbifold. Imposing tadpole cancellation, $N_{\rm flux}$ is bounded from the number of orientifold O3-planes. Combined with the flux quantisation, this forbids orientifold vacua without anti-brane charge. On the other hand, the presence of negative D3-brane charge induced by magnetised D7-branes breaks supersymmetry and relaxes the bound, allowing significantly smaller values for $g_s$.}

\begin{document}

\maketitle

\newpage

\section{Introduction}

String theory provides a consistent quantum framework for unifying gauge and gravitational interactions and describing particle physics and cosmology involving phenomena at very different scales. An important prerequisite towards this goal is stabilising the string moduli and thus fixing the compactification parameters down to four dimensions in a controllable way.  A well known systematic mechanism for the geometric (closed string) moduli stabilisation  is based on turning on 3-form fluxes for the field-strengths of the NS-NS (Neveu-Schwarz) and R-R (Ramond) 2-form gauge potentials, in the framework of type IIB string compactifications on Calabi-Yau threefolds, which preserve ${\cal N}=2$ supersymmetry in four dimensions~\cite{Dasgupta:1999ss,Giddings:2001yu,Frey:2002hf, Kachru:2002he}. The fluxes can be chosen in a way to break supersymmetry down to ${\cal N}=1$ and lead to a superpotential depending on the complex structure moduli and the axio-dilaton modulus~\cite{Gukov:1999ya}. The resulting scalar potential can be minimised in a supersymmetric way, fixing all complex structure deformations of the compactification manifold, as well as the string coupling, in terms of the discrete flux quanta obeying the Dirac quantisation. Indeed, the number of complex structure moduli is equal to the number of holomorphic $(2,1)$ cycles, given by the Hodge number $h^{2,1}$, which when supplemented with the axio-dilaton modulus and the flux around the unique $(3,0)$ cycle, leads to $h^{2,1}+1$ complex equations for the same number of complex moduli variables. An a-posteriori consistency condition for the validity of the above mechanism is to obtain a small value for the string coupling $g_s$ justifying the neglect of string quantum corrections.

The multitude of possible background fluxes stabilising the complex structure and axio-dilaton leads to a large landscape of vacua. These vacua have undergone great scrutiny in the past decades. The study of their statistics was initiated in the seminal works of \cite{Ashok:2003gk,Denef:2004ze}, and was then gradually complemented by searches using different algorithmic methods or looking for more specific phenomenological properties of flux vacua \cite{Conlon:2004ds,Hebecker:2006bn,Cole:2019enn,Cole:2021nnt,Cicoli:2022chj}. In parallel, several works searched for all possible solutions on explicit examples \cite{Martinez-Pedrera:2012teo,Dubey:2023dvu,Plauschinn:2023hjw,Ishiguro:2020tmo,Ishiguro:2023wwf}. Exhausting flux vacua in explicit examples is a way to showcase explicitly both the finiteness of the flux landscape \cite{Grimm:2020cda,Bakker:2021uqw,Grimm:2021vpn,Grimm:2023lrf}, to compute exact vacua statistics, and to test the strong constraints coming from the tadpole cancellation \cite{Betzler:2019kon,Bena:2020xrh,Bena:2021wyr}. It is one of the motivation of this work.

After complex structure moduli and axio-dilaton stabilisation, one is left with a constant superpotential $W_0$ that leads to a supersymmetric anti-de Sitter (AdS) vacuum in the particular case of no K\"ahler class moduli, counted by the Hodge number $h^{1,1}$. However, in the general case of $h^{1,1}\ne 0$, supersymmetry is broken along the K\"ahler class moduli directions but the scalar potential vanishes at lowest order due to the no-scale structure of the corresponding effective supergravity~\cite{Cremmer:1983bf, Ellis:1983ei}. An extra ingredient is therefore needed for a controlled stabilisation of the K\"ahler moduli. A non perturbative superpotential, induced for instance by gaugino condensation in a strongly coupled gauge sector~\cite{Derendinger:1985kk} or by D-brane instantons~\cite{Kachru:2003aw}, leads again, generically, to an AdS supersymmetric minimum. This minimum may be uplifted to small positive value by adding for instance anti-D3-branes breaking supersymmetry to a non-linear version~\cite{Kachru:2003aw}, or $\alpha'$ corrections and D-terms~\cite{Burgess:2003ic,Balasubramanian:2005zx,Conlon:2005ki,Antoniadis:2006eu}, but their full consistency was challenged by swampland conjectures \cite{Ooguri:2018wrx} and other constraints, see e.g. \cite{Palti:2019pca,vanBeest:2021lhn} for reviews.

An alternative perturbative method was proposed recently, based on quantum corrections to the K\"ahler potential that grow logarithmically with the size of the 2-dimensional space transverse to D7-branes~\cite{Antoniadis:2018hqy}. This is due to the propagation of massless closed strings corresponding to local tadpoles whose existence is not forbidden by global tadpole cancellation. Explicit computations can be done in the case of geometric untwisted moduli in orientifolds of orbifold compactifications~\cite{Antoniadis:2019rkh}. Such models contain at most three K\"ahler class moduli, all from the untwisted orbifold sector, as well as at most three mutually orthogonal stacks of D7-branes magnetised along their four-dimensional internal world-volume. Using as parameters the values of $g_s$ and $W_0$ determined by the first step of moduli stabilisation with 3-form fluxes, as well as the Fayet-Iliopoulos (FI) D-terms induced by the magnetic fluxes, it was shown that the resulting scalar potential can develop a shallow de Sitter (dS) minimum and produce a novel model of inflation starting around the inflection point~\cite{Antoniadis:2020stf}.

Towards the ambitious goal of providing an explicit, calculable and physically interesting model of complete moduli stabilisation, in this work we restrict to the first step. We perform an exhaustive investigation of complex structure and axio-dilaton moduli stabilisation in ${\cal N}=1$ orientifolds of toroidal orbifold compactifications of type IIB string theory in the presence of 3-form closed string fluxes~\cite{Cascales:2003zp,Blumenhagen:2003vr,Font:2004cy,Lust:2005dy,Lust:2006zh,Lust:2006zg}, as well as of 2-form open string internal magnetic fields along the world-volume of D7-branes. Such compactifications have been the playground for many examples of partial moduli stabilisation and Standard Model embedding, but no systematic study of their flux vacua was performed in the past. Recent works \cite{Ishiguro:2020tmo,Ishiguro:2023wwf} have made great progress in the systematic study of the landscape of orbifolds flux vacua, initiated earlier in \cite{Font:2004cy}. They addressed several orbifolds, and restricted to solutions with vanishing superpotential. We go beyond these works and confirmed their results whenever our works intersect. 

Orbifold compactifications involve two types of closed string moduli: 
\begin{itemize}
\item[-] toroidal deformations of the six-dimensional internal metric and R-R antisymmetric tensor which are invariant under the orbifold action. They arise from the untwisted orbifold sector; 
\item[-] deformations blowing up the orbifold singularities into smooth Calabi-Yau manifolds. They arise from the twisted orbifold sector. These twisted deformations are associated to a discrete symmetry which is unbroken at the orbifold point \cite{Cascales:2003zp,DeWolfe:2004ns,Grimm:2024fip}, corresponding to vanishing vacuum expectation value (VEV) of the twisted deformations.
\end{itemize}

When implemented by a corresponding transformation of the fluxes around the twisted cycles, the discrete symmetry of the twisted sector remains an invariance of the effective supergravity. It follows, as we show explicitly in a particular example, that all twisted deformations can be stabilised at the orbifold point by choosing vanishing fluxes around the twisted cycles. 
For such fluxes we are therefore left with the stabilisation of the untwisted complex structure moduli, which are at most three. We analyse their stabilisation in great detail. 

Our study is based mostly on analytic and partly on numerical computations, focusing mainly on two aspects: (i) the multiplicity of inequivalent vacua by modding out $S$ and $U$ duality transformations and (ii) the minimum value of the string coupling which controls the magnitude of quantum corrections and thus the validity of the stabilisation mechanism. 

Our two main results are:
\begin{enumerate}
\item we provide strong evidence that we exhausted the finite number of inequivalent vacua for a given total 3-form flux number $N_{\rm flux}$ by constructing them explicitly; the existence of a finite number of inequivalent vacua is predicted by finiteness theorems \cite{Grimm:2020cda,Bakker:2021uqw,Grimm:2021vpn,Grimm:2023lrf}, but the value of this number is not. 
\item the minimum value of the string coupling depends only on $N_{\rm flux}$ and satisfies the asymptotic relation 
$$g_{s,\,\rm{min}}\sim {c\over N_{\rm flux}^\alpha}\quad;\quad\alpha=1,2$$
\end{enumerate}
with $c$ an order one model dependent parameter and $\alpha$ depends on the number of complex structure moduli of the orbifold. The 3-form fluxes are subject to the D3-brane tadpole condition imposing the vanishing of the total charge and since $N_{\rm flux}$ can be shown to be positive, it is bounded by the number of orientifold O3-planes to which is added the induced D3-brane charge. In presence of magnetised D7-branes the latter can have both signs, or be vanishing. The number of D7-branes is also subject to the D7-brane tadpole condition, which is non-trivial if the orbifold has $\mathbb {Z}_2$ elements implying the existence of O7-planes.

The outline of our paper is the following. In Section~\ref{sec:3}, we present a short review of the toroidal orbifolds, including the possible presence of discrete torsion, and describe their complex structures (subsection~\ref{orbiifoldconstruction} and~\ref{sec:6}). It turns out that there are zero, one or three complex structure moduli. We also introduce the possible 3-form fluxes and the induced superpotential, as well as the effective ${\cal N}=1$ supergravity action and describe the complex structure moduli stabilisation mechanism (subsections~\ref{sec:4} and~\ref{sec:modstab}). We then discuss the D3-brane charge tadpole condition (subsection~\ref{sec:tadpoleintegersmultiple}). Section~\ref{sec:gsNflux} contains our detailed analysis  of moduli stabilisation, finiteness of inequivalent string vacua and computation of the minimal value of the string coupling as a function of the total D3-brane flux $N_{\rm flux}$, for the cases of zero (subsection~\ref{nocs}) and one complex structure modulus (subsection~\ref{sec:1}). In Section~\ref{T6Z2xZ2}, we study in great detail the moduli stabilisation in the only case of three untwisted complex structure moduli, which is the orbifold $\mathbb {Z}_2\times\mathbb {Z}_2$. This orbifold has also three untwisted K\"ahler moduli as well as 48 twisted moduli which can be either complex structure or K\"ahler, depending on whether there is or not discrete torsion, the two cases being exchanged by mirror symmetry~\cite{Vafa:1994rv}. We first discuss the stabilisation of twisted moduli (in the presence of discrete torsion) at the orbifold point (subsection~\ref{subsec:twistedmoduliZ2Z2}) and then the stabilisation of the untwisted complex structure moduli (subsection~\ref{sec:2}). We proceed with the counting of independent string vacua and the computation of the minimal value of the string coupling (subsection~\ref{sec:5}), while we exclude the existence of solutions with flux integers hierarchy and parametric control on $g_s$ (subsection~\ref{sec:parametriccontrol}). Finally, we study the presence of magnetised D7-branes (subsection~\ref{subsec:magnetizedbranes}). Section~\ref{conclusions} contains our conclusions, while appendix~\ref{app:1} displays tables with the complex structure data of the orbifolds used in our analysis.

\section{Toroidal orbifolds with fluxes}\label{sec:3}

In this work, we consider type IIB string compactifications on an internal space $Y$ chosen to be an orientifold of a $T^6/G$ toroidal orbifold, with $G=\mathbb{Z}_N, \mathbb{Z}_N\times \mathbb{Z}_M$.

\subsection{Orbifolds construction, cohomology basis and complex structures} \label{orbiifoldconstruction}

\paragraph{Orbifold group and action} We start this section by reviewing the construction of toroidal orbifolds, largely based on \cite{Reffert:2006du}. The construction starts from a $6$-torus $T^6 = R^6/\Lambda$, where $\Lambda$ is a six dimensional lattice. In the torus, points are thus identified as $x^i \sim x^i + l^i$ where $l \in \Lambda$. Once the lattice is specified, we can choose an automorphism $\Gamma$ as the orbifold group, or point group, to quotient by. Imposing that the resulting orbifold has $\mathrm{SU}(3)$-holonomy, for $\mathcal{N}=1$ supersymmetry purposes, restricts $\Gamma$ to be a subgroup of $\mathrm{SU}(3)$. If we further restrict to abelian orbifold groups, and require that $\Gamma$ acts crystallographically on the torus lattice, we end up with a short list of groups:
\begin{alignat}{2}
    &\Gamma = \mathbb{Z}_N    \quad &&\text{with}\quad N = 3,4,6,7,8,12, \nonumber\\
    &\Gamma = \mathbb{Z}_N\times\mathbb{Z}_M \quad &&\text{with}\quad M = kN 
    \quad\text{and}\quad N, M = 2,3,4,6.
\end{alignat}

The action of the group on the torus has a simple expression in complex coordinates $(z^1, z^2, z^3)$. For the group generator element $\theta_N\equiv(n_1,n_2,n_3)$, it reads:
\begin{equation}
    \theta_N: (z^1, z^2, z^3) \rightarrow (e^{2i\pi n_1/N}z^1, e^{2i\pi n_2/N}z^2, e^{2i\pi n_3/N}z^3). \label{orbifoldActioncomplex}
\end{equation}
The groups $\mathbb{Z}_N$ are generated by one element $\theta_N$ and the groups $\mathbb{Z}_N\times\mathbb{Z}_M$ are generated by two elements $\theta_N, \theta_M$. Note that there are two inequivalent embeddings for the $\mathbb{Z}_N$ with $N=6,8,12$. For instance $\theta_{6,I} = (1,1,-2)$ or $\theta_{6,II} = (1,2,-3)$. To illustrate the previous notation $\mathbb{Z}_{6,I}$ acts as
\begin{equation}\label{eq:14}
\theta_{6,I}: (z^1, z^2, z^3) \rightarrow (e^{2i\pi/6}z^1, e^{2i\pi/6}z^2, e^{-4i\pi/6}z^3).
\end{equation}
In table \ref{tab:2}, we give the list of toroidal orbifolds considered in \cite{Reffert:2006du}, along with the corresponding torus lattices and group actions $(\theta_N, \theta_M)$. 

\begin{table}[h!]
    \centering
    \begin{tabular}{|c|c||c|c||c|c|c|c|}
    \hline
        orbifold & torus lattice $\Lambda$ & $\theta_N$ & $\theta_M$ & $h^{1,1}$ & $h^{2,1}$ & $\tilde h^{1,1}$ & $\tilde h^{2,1}$ \\
        \hline\hline
        $\mathbb{Z}_3$ & $\mathrm{SU}(3)^3$ & $(1,1,-2)$ & 
        & $9$ & $0$ & $27$ & $0$ \\
        $\mathbb{Z}_{4,a}$ & $\mathrm{SU}(4)^2$ & $(1,1,-2)$ & 
        & $5$ & $1$ & $20$ & $0$ \\
        $\mathbb{Z}_{4,b}$ 
        & $\mathrm{SU}(2)\times\mathrm{SU}(4)\times\mathrm{SO}(5)$ & $(1,1,-2)$ 
        & & $5$ & $1$ & $22$ & $2$ \\
        $\mathbb{Z}_{4,c}$ & $\mathrm{SU}(2)^2\times\mathrm{SO}(5)^2$ 
        & $(1,1,-2)$ & & $5$ & $1$ & $26$ & $6$ \\
        $\mathbb{Z}_{6,Ia}$ & $G_2\times\mathrm{SU}(3)^2\ (*)$ & $(1,1,-2)$ & 
        & $5$ & $0$ & $20$ & $1$ \\
        $\mathbb{Z}_{6,Ib}$ & $G_2^2\times\mathrm{SU}(3)$ & $(1,1,-2)$ & & $5$ 
        & $0$ & $24$ & $5$ \\
        $\mathbb{Z}_{6,IIa}$ & $\mathrm{SU}(2)\times\mathrm{SU}(6)$ & $(1,2,-3)$ 
        & & $3$ & $1$ & $22$ & $0$ \\
        $\mathbb{Z}_{6,IIb}$ & $\mathrm{SU}(3)\times\mathrm{SO}(8)$ & $(1,2,-3)$ 
        & & $3$ & $1$ & $26$ & $4$ \\
        $\mathbb{Z}_{6,IIc}$ & $\mathrm{SU}(2)^2\times\mathrm{SU}(3)^2\ (*)$ 
        & $(1,2,-3)$ & & $3$ & $1$ & $28$ & $6$ \\
        $\mathbb{Z}_{6,IId}$ & $G_2\times\mathrm{SU}(2)^2\times\mathrm{SU}(3)$ 
        & $(1,2,-3)$ & & $3$ & $1$ & $32$ & $10$ \\
        $\mathbb{Z}_7$ & $\mathrm{SU}(7)$ & $(1,2,-3)$ 
        & & $3$ & $0$ & $21$ & $0$ \\
        $\mathbb{Z}_{8,Ia}$ & $\mathrm{SU}(4)\times\mathrm{SU}(4)\ (*)$ 
        & $(1,2,-3)$ & & $3$ & $0$ & $21$ & $0$ \\
        $\mathbb{Z}_{8,Ib}$ & $\mathrm{SO}(5)\times\mathrm{SO}(9)$ & $(1,2,-3)$ 
        & & $3$ & $0$ & $24$ & $3$ \\
        $\mathbb{Z}_{8,IIa}$ & $\mathrm{SU}(2)\times\mathrm{SO}(10)$ 
        & $(1,3,-4)$ & & $3$ & $1$ & $24$ & $2$ \\
        $\mathbb{Z}_{8,IIb}$ & $\mathrm{SO}(4)\times\mathrm{SO}(9)$ & $(1,3,-4)$ 
        & & $3$ & $1$ & $28$ & $6$ \\
        $\mathbb{Z}_{12,Ia}$ & $E_6$ & $(1,4,-5)$ & & $3$ & $0$ & $22$ & $1$ \\
        $\mathbb{Z}_{12,Ib}$ & $\mathrm{SU}(3)\times F_4$ & $(1,4,-5)$ & 
        & $3$ & $0$ & $26$ & $5$ \\
        $\mathbb{Z}_{12,II}$ & $\mathrm{SO}(4)\times F_4$ & $(1,5,-6)$ & & $3$ & $1$ & $28$ & $6$ \\
        \hline
        $\mathbb{Z}_2\times \mathbb{Z}_2$ & $\mathrm{SU}(2)^6$ & $(1,0,-1)$ 
        & $(0,1,-1)$ & $3$ & $3$ & $48$ & $0$ \\
        $\mathbb{Z}_2\times \mathbb{Z}_4$ 
        & $\mathrm{SU}(2)^2\times\mathrm{SO}(5)^2$ & $(1,0,-1)$ & $(0,1,-1)$
        & $3$ & $1$ & $58$ & $0$ \\
        $\mathbb{Z}_2\times \mathbb{Z}_{6,I}$ 
        & $G_2\times\mathrm{SU}(2)^2\times\mathrm{SU}(3)$ & $(1,0,-1)$ 
        & $(0,1,-1)$ & $3$ & $1$ & $48$ & $2$ \\
        $\mathbb{Z}_2\times \mathbb{Z}_{6,II}$ & $G_2^2\times\mathrm{SU}(3)$ 
        & $(1,0,-1)$ & $(1,1,-2)$ & $3$ & $0$ & $33$ & $0$ \\
        $\mathbb{Z}_3\times \mathbb{Z}_3$ & $\mathrm{SU}(3)^3$ & $(1,0,-1)$ 
        & $(0,1,-1)$ & $3$ & $0$ & $81$ & $0$ \\
        $\mathbb{Z}_3\times \mathbb{Z}_6$ & $G_2^2\times\mathrm{SU}(3)$ 
        & $(1,0,-1)$ & $(0,1,-1)$ & $3$ & $0$ & $70$ & $1$ \\
        $\mathbb{Z}_4\times \mathbb{Z}_4$ & $\mathrm{SO}(5)^3$ & $(1,0,-1)$ 
        & $(0,1,-1)$ & $3$ & $0$ & $87$ & $0$ \\
        $\mathbb{Z}_6\times \mathbb{Z}_6$ & $G_2^3$ & $(1,0,-1)$ & $(0,1,-1)$
        & $3$ & $0$ & $81$ & $0$ \\
        \hline
    \end{tabular}
    \caption{List of simple toroidal orbifolds borrowed from \cite{Reffert:2006du}, along with the corresponding torus lattices, group actions $(\theta_N, \theta_M)$ and number of untwisted and twisted K\"ahler and complex structure moduli, respectively $(h^{1,1}, h^{2,1})$ and $(\tilde h^{1,1}, \tilde h^{2,1})$, in the absence of discrete torsion. In all of these orbifolds, the matrix $M$ representing the action of the group on the real coordinates is the transpose of the Coxeter element of the torus lattice, except for the three entries marked by  $(*)$ where the action is realised as a generalised Coxeter twist, see \cite{Reffert:2006du}.}
    \label{tab:2}
\end{table}
\paragraph{Untwisted cohomology basis} The $H^3(T^6,\mathbb{Z})$ complex cohomology basis is written as:
\begin{align}
\omega_{A_0} &= dz^1\wedge dz^2\wedge dz^3, \quad \omega_{A_1} = d\bar z^1\wedge dz^2\wedge dz^3, \quad \omega_{A_2} = dz^1\wedge d\bar z^2\wedge dz^3, \quad \omega_{A_3} = \cdots, \nonumber\\
\omega_{B_0} &= d\bar z^1\wedge d\bar z^2\wedge d\bar z^3, \quad \omega_{B_1} = d z^1\wedge d\bar z^2\wedge d\bar z^3, \quad \omega_{B_2} = d \bar z^1\wedge d z^2\wedge d\bar z^3, \quad \omega_{B_3} = \cdots, \nonumber\\
\omega_{C_1}&=dz^1\wedge d\bar z^1\wedge dz^2, \quad \omega_{C_2}=dz^1\wedge d\bar z^1\wedge dz^3, \quad \omega_{C_3}=\cdots, \nonumber\\
\omega_{D_1}&=dz^1\wedge d\bar z^1\wedge d\bar z^2, \quad \omega_{D_2}=dz^1\wedge d\bar z^1\wedge d\bar z^3, \quad \omega_{D_3}=\cdots,  \label{complexcohombasis}
\end{align}
We normalise the top $(3,0)$-form $\Omega$ to $\omega_{A_0}$, namely:
\begin{equation}
\Omega= \, dz^1\wedge dz^2 \wedge dz^3=\omega_{A_0}. \label{Omegacomplex}
\end{equation}
In the above basis, the cohomology structure of the 3-forms is clear: $\omega_{A_0}$ is a $(3,0)$-form, $\omega_{A_i},\omega_{C_i}$ are $(2,1)$-forms, $\omega_{B_i},\omega_{D_i}$ are $(1,2)$-forms and $\omega_{B_0}$ is a  $(0,3)$-form.

The untwisted orbifold cohomology is obtained from the torus one by keeping only the forms invariant under the action of the orbifold group. As the orbifold acts simply on the complex coordinates through \eqref{orbifoldActioncomplex}, the complex cohomology basis \eqref{complexcohombasis} is convenient to identify the orbifold cohomology basis. For instance, the $\omega_{C_i}$ and $\omega_{D_i}$ are projected out in all the orbifolds. The numbers of $(1, 1)$-forms $dz^i\wedge d\bar z^j$ and $(2, 1)$-forms $dz^i\wedge dz^j \wedge d\bar z^k$ left invariant under the orbifold action are counted by the Hodge numbers $h^{1,1}$ and $h^{2,1}$. On top of these forms, the orbifold contains additional twisted forms, counted by  $\tilde h^{1, 1}$ and $\tilde h^{2, 1}$. In these notations, the total Hodge numbers are thus $(h^{1,1}+\tilde{h}^{1,1}, h^{2,1}+\tilde{h}^{2,1})$. They also count the number of untwisted and twisted K\"ahler and complex structure moduli, and are indicated in  \cref{tab:2}. 

The cohomology can also be expressed in terms of a real basis. Introducing the notation $(i  j  k) \equiv dx^i \wedge dx^j \wedge dx^{k}$, we define our real basis as:
\begin{equation}
    \begin{tabular}{cc cc cc}
         $\alpha_0 = (135),$ & $\beta^0 = \phantom{-}(246),$ 
         & $\gamma_1 = (123),$ & $\delta^1 = -(456),$ 
& $\gamma_5 = (156),$ & $\delta^5 = -(234),$ \\
         $\alpha_1 = (235),$ & $\beta^1 = -(146),$ 
         & $\gamma_2 = (125),$ & $\delta^2 = -(346),$ 
         &  $\gamma_6 = (356),$ & $\delta^6 = -(124),$ \\
         $\alpha_2 = (145),$ & $\beta^2 = -(236),$ 
         & $\gamma_3 = (134),$ & $\delta^3 = -(256),$ \\
                   $\alpha_3 = (136),$ & $\beta^3 = -(245),$  & $\gamma_4 = (345),$ & $\delta^4 = -(126),$ 
                       \end{tabular} \label{realbasis}
    \end{equation}
One can check that in the convention $\int dx^1\wedge dx^2 \wedge \cdots \wedge dx^6 = -1$, the basis elements satisfy $\int \alpha_i\wedge\beta^j = \delta^j_i$ and $\int \gamma_i\wedge\delta^j = \delta^j_i$. It is not obvious to construct 3-forms invariant under the orbifold action from the real basis. The simplest way is to identify them in the complex basis as explained above, and then express them in terms of the forms of the real basis.

To go from the complex basis \cref{complexcohombasis} to the real basis \eqref{realbasis}, one needs the expressions of the complex coordinates $z^i$ in terms of the real ones $x^i$. These are determined by the complex structure of the orbifold, as is described hereafter.

\paragraph{The complex structure} The action of the orbifold group in real coordinates $x^i$ is represented by a six-dimensional matrix $M$:
\begin{equation}
x^i \rightarrow M^i_j x^j, \qquad M=Q^t \label{orbifoldActionreal}.
\end{equation} 
This matrix can be taken as the transpose of the Coxeter element $Q$ of the torus lattice $\Lambda$, $M=Q^t$. There are other possibilities, see \cite{Reffert:2006du} for reference. The orbifold actions in real and complex coordinates are compatible if the eigenvalues of $Q$ and  $\theta_N$ are equal. This selects only one or a few possible lattices for each orbifold group and embedding. 

The complex coordinates $z^i$ are written in terms of the real coordinates $x^i$ through the complex structure. The latter is determined by writing the complex coordinates as arbitrary linear combinations of the real coordinates $z^i = A^i_jx^j$, and imposing invariance under the orbifold action. For instance, for a $\mathbb{Z}_N$ orbifold we see from \cref{orbifoldActioncomplex,orbifoldActionreal} that we should require:
\begin{equation}
 \theta_N(z^i)= e^{2i\pi n_i/N}z^i= e^{2i\pi n_i/N} A^i_jx^j=  A^i_jM^j_kx^k. \label{groupaction}
 \end{equation}
The expression of the matrix elements $M^j_k$ thus gives a set of relations between the coefficients $A^i_j$. This fixes the complex structure up to a complex normalisation. After fixing the latter, the remaining free coefficients are the untwisted complex structure moduli. This procedure gives the complex structures listed in Appendix \ref{app:1}, a sample of which is given in table \ref{tab:3}. 

Let us show the details of the procedure for the $\mathbb{Z}_{4,b}$ orbifold. The group generator given in \cref{tab:2} is $\theta_N=(1,1,-2)$ and the matrix $M$ reads:
\begin{equation}
M=Q^t=\begin{pmatrix}1&-1&0&0&0&0\\2&-1&0&0&0&0\\ 0&0&0 &0&-1&0\\  0&0&1&0&-1&0\\0&0& 0&1&-1&0\\0&0&0&0&0&-1 
\end{pmatrix}
\end{equation}
The identification \eqref{groupaction} thus reads:
\begin{align}
i A^1_k x^k = (A^1_1+2A^1_2)x^1-(A^1_1+A^1_2)x^2+A^1_4x^3 + A^1_5 x^4-(A^1_3+A^1_4+A^1_5)x^5-A^1_6 x^6, \nonumber\\
i A^2_k x^k=  (A^2_1+2A^2_2)x^1-(A^2_1+A^2_2)x^2+A^2_4x^3 + A^2_5 x^4-(A^2_3+A^2_4+A^2_5)x^5-A^2_6 x^6,\\
- A^3_k x^k = (A^3_1+2A^3_2)x^1-(A^3_1+A^3_2)x^2+A^3_4x^3 + A^3_5 x^4-(A^3_3+A^3_4+A^3_5)x^5-A^3_6 x^6, \nonumber
\end{align}
which is solved for:
\begin{align}\label{examplecomplexrealbasis}
&z_1=A^1_1\Big(x^1+\frac{1}{2}(i-1)x^2\Big)+A^1_3 \Big(x^3+ix^4-x^5\Big),\nonumber\\
&z_2=A^2_1\Big(x^1+\frac{1}{2}(i-1)x^2\Big)+A^2_3 \Big(x^3+ix^4-x^5\Big),\\
&z_3=A^3_3\Big(x^3-x^4+x^5\Big)+A^3_6 x^6.\nonumber
\end{align}
We see that the space parameterised by $z_1$ and $z_2$ is generated by two vectors, with coefficients $A_1^1,A^1_3,A^2_1,A^2_3$. To have independent coordinates $z_1$ and $z_2$ with unit overall coefficient, one can make the choice $A_1^1=A^2_3=1$ and $A^1_3=A^2_1=0$. On the other hand, the $z_3$ coordinate is expressed in terms of two additional independent vectors, such that fixing the overall complex normalisation leaves one free coefficient. The latter corresponds to the complex structure modulus $\mathcal{U}$ which survives the orbifolding, from the initial nine complex structure moduli of $T^6$. The final complex coordinates thus read:
\begin{align}
&z_1=x^1+\frac{1}{2}(i-1)x^2, \nonumber\\
&z_2=x^3+ix^4-x^5,\label{finalbasisexample}\\
&z_3=x^3-x^4+x^5+\mathcal{U} x^6.\nonumber
\end{align}
 Notice that the orbifold action symmetry $z^1\leftrightarrow z^2$ was still clear in \cref{examplecomplexrealbasis} before making our choice for the remaining $A^i_k$. This is not the case anymore once they are fixed in \cref{finalbasisexample}.  On the other hand, if there remains a symmetry after fixing the arbitrary parameters, it has to be a symmetry of the orbifold action (see for instance the orbifold $T^6/\mathbb{Z}_3\times\mathbb{Z}_3$ in table \ref{tab:3}).

\newpage

\paragraph{Projective coordinates} We conclude this section by mentioning that the $(3,0)$-form $\Omega$ can be parameterised in the real basis through the complex structure moduli. It takes the form:
\begin{align}
\Omega&=dz_1\wedge dz_2\wedge dz_3 + \text{twisted forms} \nonumber\\
&= X^a \alpha_a + \mathcal{G}_a \beta^a, \label{omegaXaGa}
\end{align}
where $X^a$ are projective coordinates. They can be set to $(X^0, X^i, X^I) = (1, \mathcal{U}^i, \mathcal{D}^I)$ when evaluating the above relation, with $i = 1, \dots, h^{2, 1}$ parameterising the untwisted complex structure moduli and $I=1,\ldots,\tilde h^{2,1}$ the twisted ones. Similarly, the total cohomology basis $(\alpha_a, \beta^a)=(\alpha_i,\alpha_I,\beta^i,\beta^I)$ is constructed from the untwisted one \eqref{realbasis} supplemented by the twisted cohomology basis. See \cref{sec:6} for a more complete introduction to the twisted moduli. In \cref{omegaXaGa}, the function $\mathcal{G}$ is the prepotential introduced in \cref{eq:prepotential} and $\mathcal{G}_a$ denotes its derivative with respect to the $X^a$ coordinate \cite{Candelas:1990pi,Ceresole:1995ca,Taylor:1999ii,Blumenhagen:2003vr}.

\begin{table}%[ht!]
    \centering
    \hspace{2.5cm}coefficients of the complex structure\\
    \vspace{5pt}
    \begin{tabular}{|c|c||c|c|c|c|c|c|}
    \hline
        \multicolumn{1}{|l}{orbifold}& &$x_1$ & $x_2$ & $x_3$ & $x_4$ & $x_5$ & $x_6$\\
        \hline\hline
        & $z_1$ & $1$ & $e^{2i\pi/3}$ & $0$ & $0$ & $0$ & $0$ \\
        $\mathbb{Z}_{3}$ & $z_2$ & $0$ & $0$ & $1$ & $e^{2i\pi/3}$ & $0$ & $0$ \\
        & $z_3$ & $0$ & $0$ & $0$ & $0$ & $1$ & $e^{2i\pi/3}$\\
        \hline
        & $z_1$ & $1$ & $e^{3i\pi/4}/\sqrt{2}$ & $0$ & $0$ & $0$ & $0$ \\
        $\mathbb{Z}_{4,b}$ & $z_2$ & $0$ & $0$ & $1$ & $i$ & $-1$ & $0$ \\
        & $z_3$ & $0$ & $0$ & $1$ & $-1$ & $1$ & $\mathcal{U}$\\
        \hline
        & $z_1$ & $1$ & $\mathcal{U}^1$ & $0$ & $0$ & $0$ & $0$ \\
        $\mathbb{Z}_{2}\times\mathbb{Z}_2$ & $z_2$ & $0$ & $0$ & $1$ & $\mathcal{U}^2$ & $0$ & $0$ \\
        & $z_3$ & $0$ & $0$ & $0$ & $0$ & $1$ & $\mathcal{U}^3$\\
        \hline
        & $z_1$ & $1$ & $e^{2i\pi/3}$ & $0$ & $0$ & $0$ & $0$ \\
        $\mathbb{Z}_{3}\times\mathbb{Z}_3$ & $z_2$ & $0$ & $0$ & $1$ & $e^{2i\pi/3}$ & $0$ & $0$ \\
        & $z_3$ & $0$ & $0$ & $0$ & $0$ & $1$ & $-e^{i\pi/3}$\\
        \hline
    \end{tabular}
    \caption{Complex structures of some orbifolds of table \ref{tab:2}. $\mathcal{U}, \mathcal{U}^i$ are complex structure moduli.}
    \label{tab:3}
\end{table}

\subsection{Fluxes, superpotential and complex structure moduli stabilisation}\label{sec:4}

 We just detailed the construction of the toroidal orbifolds considered in this work. In what follows, we study in detail the stabilisation by background fluxes of their $h^{2,1}$ untwisted complex structure moduli. In the orbifolds we consider, $h^{2,1} \in \{0, 1, 3\}$ and the equations of stabilisation are algebraic, making an analytic treatment technically possible.

In addition, we will choose fluxes such that the $\tilde{h}^{2,1}$ twisted complex structure moduli are stabilised at the orbifold point, i.e. have vanishing VEVs. This is possible due to the orbifolds discrete symmetries, as described in \cref{sec:6} and shown explicitly for $T^6/\mathbb{Z}_2\times\mathbb{Z}_2$ in \cref{subsec:twistedmoduliZ2Z2}.

\paragraph{Background fluxes superpotential and charge} In the effective theory, the presence of background $3$-form fluxes $H_3$ (NS-NS) and $F_3$ (R-R) generate a superpotential for the complex structure moduli.
 This superpotential is expressed in terms of the $3$-form $G_3$ and reads \cite{Gukov:1999ya}:
\begin{equation}\label{eq:10}
    W = \int G_3\wedge \Omega, \quad\text{where}\quad G_3 \equiv  F_3+ \mathcal{S}H_3.
\end{equation}
In our conventions,  the axio-dilaton is defined as $\mathcal{S} \equiv C_0 + ie^{-\phi} = C_0 + i/g_s$.  The background fluxes also contribute to the D3 tadpole by inducing a positive charge $N_\mathrm{flux}$:
\begin{equation}\label{eq:9}
    N_\mathrm{flux} = \int_{T^6} H_3\wedge F_3 =m^H n^F - m^F n^H + p^H q^F - p^F q^H.
\end{equation}
The last equality is written in terms of the flux integers, that we introduce hereafter. The products of flux integers denote the sum over all basis elements, $e.g.$  $m^H n^F = \sum_{i=0}^3 m^H_i n^F_i$, see eqs. \eqref{realbasis} and \eqref{eq:8}.

\paragraph{Flux quanta and integers} The fluxes $F_3$ and $H_3$ should satisfy a Dirac quantisation condition, so that their expansion coefficients on a normalised 3-form cohomology basis should be integers. As introduced in \cref{realbasis}, we work with a normalised real cohomology basis generated by  $\alpha_i,\beta^i,\gamma_j,\delta^j$ . Hence, the quantised 3-form $G_3$ is expanded in terms of flux integer quanta as:
\begin{equation}\label{eq:8}
    G_3= m_i\alpha_i + n_i\beta^i + p_j\gamma_j + q_j\delta^j, \quad\text{where}\quad
    m_i = m_i^F + \mathcal{S}m_i^H  ,  \,\,  n_i =  n_i^F + \mathcal{S}n_i^H ,  \text{etc.}
\end{equation}
where $i = 0, \dots, 3$ and $j = 1, \dots 6$.  The flux quanta $m_i^{H,F}, n_i^{H,F}, p_i^{H,F},q^{H,F}$ are integers. In the rest of the paper, we call flux parameters the coefficients of $G_3$ on the real cohomology basis, hence $m_i,n_i,p_j, q_j$. They are not all independent. Indeed, $\alpha_i, \beta^i, \gamma_j$ and $\delta^j$ are elements of the real basis of the torus $T^6$. However,  as $G_3$ is expanded on the orbifold cohomology basis, it only depends on real 3-forms surviving the orbifolding. Such forms are linear combinations of the real basis elements, producing relations between the flux parameters $m_i, n_i, p_j, q_j$. The easiest way to identify the real 3-forms surviving the orbifolding is to match them to the complex ones through the complex structure, see \cref{orbiifoldconstruction}.

As explained below \cref{complexcohombasis}, it is indeed easier to identify the orbifold 3-forms in the complex basis because the orbifold action is simpler there. In the complex basis the $G_3$ form reads:
\begin{equation}\label{eq:7}
    G_3= A^i\omega_{A_i} + B^i\omega_{B_i}, \qquad {\rm with} \quad i = 0, \dots, 3,
\end{equation}
where the $A^i$ and $B^i$ coefficients are not integer and only the $\omega_{A_i}$ and $\omega_{B_i}$, surviving the orbifolding, are considered. In this basis, the superpotential  $W$ of  \cref{eq:10} simply reads:
\begin{equation}
W= B^0\int \omega_{B_0} \wedge \omega_{A_0}. \label{fluxsupcomplex}
\end{equation}
As we chose to normalise the real basis, both the integral and the $B^0$ coefficient depend on the complex structure. The dependence on the flux quanta comes from $B^0$.

When expanding the $(3,0)$-form  $\Omega$ in the projective coordinates \eqref{omegaXaGa}, we see from \cref{eq:10,eq:8} that the flux superpotential is expressed simply as: 
\begin{equation}\label{superpotfromG}
    W = n_a X^a + m^a \mathcal{G}_a.
\end{equation}
This latter expression is derived from the special geometry of the moduli space and involves the derivative $\mathcal{G}_a$ of the prepotential $\mathcal{G}$, see \cref{eq:prepotential}~\cite{deWit:1983xhu}.

To summarize the previous discussion, in order to express the superpotential in terms of flux integers, we have to compute the $B^0$ coefficients of \cref{eq:7}, that enters in \eqref{fluxsupcomplex}, for the $\omega_{A_i}$ and $\omega_{B_i}$ forms surviving the orbifold. This is done by equating the two expressions of $G_3$ in \eqref{eq:7} and \eqref{eq:8} with the $\omega_{A_i}$ and $\omega_{B_i}$ expanded on the real basis using the complex structures of Appendix \ref{app:1}. It produces at the same time the  $A^i$ and $B^i$ coefficients, in particular $B^0$, and the aforementioned constraints between the flux parameters $m_i,n_i,p_j$ and $q_j$. When solving these constraints and expressing some parameters as function of others, one should ensure that the quanta all remain integers. The easiest way to do so is to express all flux parameters in terms of the smallest parameters, that we call basis parameters. For instance, if one of the constraints gives $n_1=2m_1$, one should take $m_1$ as basis parameter rather than $n_1$, so that taking $m_1^F$ and $m_1^H$ integers ensures that $n_1^F$ and $n_1^H$ are integers as well.

%, we can either work in real or in complex basis. In real basis, one should expand the 3-form $\Omega=dz^1\wedge dz^2\wedge dz^3$ on the real basis and apply \cref{eq:10} with $G_3$ given in \cref{eq:8}, once the surviving 3-forms have been identified and the flux integers related to each other. In the complex basis, 

\paragraph{The $T^6/\mathbb{Z}_3$ example} In this orbifold, only the $\omega_{A_0}$ and $\omega_{B_0}$ survive the projection. This means that the complex structure is completely fixed by the orbifold: there is no untwisted complex structure modulus and $h^{2,1}=0$. The $G_3$ flux should thus be expanded as:
\begin{equation}
G_3=A^0 \omega_{A_0} + B^0 \omega_{B_0}. \label{G3T6Z3}
\end{equation}
The complex structure, i.e. the relation between complex and real coordinates, is given in \cref{tab:3} and allows to write  the $\omega_{A_0}$ and $\omega_{B_0}$  forms as:
\begin{align}
\omega_{A_0}=dz_1\wedge dz_2\wedge dz_3&=(dx^1+ e^{2i\pi/3}dx^2) \wedge (dx^3+ e^{2i\pi/3}dx^4) \wedge (dx^5+ e^{2i\pi/3}dx^6) \nonumber\\
&= \alpha_0 + \beta^0 + e^{2i\pi/3} (\alpha_1 + \alpha_2 + \alpha_3) - e^{-2i\pi/3} (\beta^1 + \beta^2 + \beta^3), \nonumber\\
\omega_{B_0}=d\bar z_1\wedge d\bar z_2\wedge d\bar z_3&=(dx^1+ e^{2i\pi/3}dx^2) \wedge (dx^3+ e^{2i\pi/3}dx^4) \wedge (dx^5+ e^{2i\pi/3}dx^6) \nonumber\\
&= \alpha_0 + \beta^0 + e^{-2i\pi/3} (\alpha_1 + \alpha_2 + \alpha_3) - e^{2i\pi/3} (\beta^1 + \beta^2 + \beta^3) .
\end{align}
Using these expansions in \cref{G3T6Z3} and matching with the real basis expansion \eqref{eq:8} we deduce that  $p_j = q_j = 0$, $m_0=n_0=A^0+B^0$, $m_1=m_2=m_3=A^0 e^{2i\pi/3} + B^0 e^{-2 i\pi/3}$, $n_1=n_2=n_3=m_0+m_1$. One can then invert these relations to obtain $B^0=-i/\sqrt{3} (e^{2i\pi/3}m_0-m_1)$, and obtain the flux superpotential \eqref{fluxsupcomplex} as:
\begin{equation}
W=B^0\int \omega_{B_0}\wedge\omega_{A_0}=-3e^{2i\pi/3}(m_0 +e^{2i\pi/3}m_1).
\end{equation}
In addition, replacing the flux quanta in \eqref{eq:9} yields $N_\mathrm{flux} = -3(m^H_0m^F_1 - m^F_0 m^H_1)$. Note that it is a multiple of $3$. This is a consequence of the orbifold geometry, without any restriction on the integers. See later discussion.

\paragraph{Parameterisations of the superpotential} For orbifolds with $h^{2,1} = 0$ of \cref{tab:2}, we parameterise the flux superpotential \eqref{eq:10} as
\begin{equation}\label{eq:11}
    W = K(a + \gamma b) ,\qquad\text{where}\quad a = a^F +  \mathcal{S}a^H ,  \quad b = b^F + \mathcal{S}b^H .
\end{equation}
Here  $a, b$ are the independent basis parameters. It turns out that $N_{\mathrm{flux}}$ is always multiple of an integer $k$ depending on the orbifold, but not of the flux quanta: 
\begin{equation}\label{Nfluxh21_0}
N_\mathrm{flux} = k(a^H b^F - a^F b^H), \qquad k\in \mathbb Z \,\, \text{ fixed by the orbifold geometry}.
\end{equation}
For orbifolds with $h^{2,1} = 1$, we can similarly parameterise:
\begin{equation}\label{eq:12}
    W = \mathcal{U}A + B,  \qquad\text{where} \quad A \equiv A^F +  \mathcal{S}A^H , B\equiv B^F +  \mathcal{S}B^H .
\end{equation}
The coefficients $A$ and $B$ are simple combinations of flux parameters.  With this notation the flux number reads:
\begin{equation}\label{Nfluxh21_1}
N_\mathrm{flux} = k(\mathrm{Re}(A^H \bar B^F) - \mathrm{Re}(A^F\bar B^H)), \qquad k\in \mathbb{Q}  \,\,\text{ fixed by the orbifold geometry}.
\end{equation}
In these cases $k$ is not necessarily integer anymore. However $N_\mathrm{flux}$ turns out to be again multiple of an integer $\ell$ depending on the orbifold, due to the specific combination of the flux parameters appearing in its expression. It indeed reads:
\begin{equation}
N_\mathrm{flux}=\ell n, \qquad n\in\mathbb{N} \,\,  \,\,\text{ fixed by the orbifold geometry}. \label{firstdefl}
\end{equation}
 For instance, the orbifold $T^6/\mathbb{Z}_{4,a}$ has $\ell=4$ and the exact expression of $N_\mathrm{flux}$ reads:
\begin{equation}
N_\mathrm{flux}=4\Big(m^H_0n^F_0 - m^F_0n^H_0 + m^H_1n^F_0 - m^F_1n^H_0 + m^H_0m^F_2 - m^F_0m^H_2 + 2(m^H_1m^F_2 - m^F_1m^H_2)\Big).
\end{equation}
In tables \ref{tab:4} and \ref{tab:5}, we provide the data for the orbifolds of table \ref{tab:2} with these parameterisations.

The only orbifold with $h^{2,1} = 3$ present in our list is $T^6/\mathbb{Z}_2\times\mathbb{Z}_2$. We take it as an example to discuss full stabilisation of untwisted and twisted complex structure moduli, and reserve it for \cref{T6Z2xZ2}.

\phantom{.}

\begin{table}%[ht!]
    \centering
    \begin{tabular}{|c||c|c||c|c||c|}
    \hline
        orbifold & $a$ & $b$ & $K$ & $\gamma$ & $k$ \\
        \hline\hline
        $\mathbb{Z}_3$ & $m_0$ & $m_1$ & $-3e^{2i\pi/3}$ & $e^{i\pi/3}$ & $-3$ \\
        \hline
        $\mathbb{Z}_{6,Ia}$ & $m_0$ & $m_1$ & $2\sqrt{3}e^{5i\pi/6}$ & $\sqrt{3}e^{i\pi/6}$ & $3$ \\
        \hline
        $\mathbb{Z}_{6, Ib}$ & $n_0$ & $n_1$ & $-\sqrt{3}e^{i\pi/6}$ & $e^{5i\pi/6}/\sqrt{3}$ & $3$ \\
        \hline
        $\mathbb{Z}_7$ & $m_0$ & $m_1$ & $7/2(7+i\sqrt{7})$ & $(1+i\sqrt{7})/4$ & $7$ \\
        \hline
        $\mathbb{Z}_{8,Ia}$ & $m_0$ & $m_1$ & $16i\sqrt{2}$ & $\sqrt{2}e^{i\pi/4}$ & $8$ \\
        \hline
        $\mathbb{Z}_{8,Ib}$ & $n_0$ & $n_1$ & $-4i\sqrt{2}$ & $e^{3i\pi/4}/\sqrt{2}$ & $4$ \\
        \hline
        $\mathbb{Z}_{12,Ia}$ & $m_0$ & $m_1$ & $12e^{i\pi/6}$ & $e^{i\pi/3}$ & $3$ \\
        \hline
         $\mathbb{Z}_{12,Ib}$ & $m_0$ & $m_1$ & $-3e^{5i\pi/6}$ & $e^{i\pi/3}$ & $3$ \\
        \hline
        $\mathbb{Z}_2\times\mathbb{Z}_{6,II}$ & $n_0$ & $n_1$ & $-\sqrt{3}e^{i\pi}/6$ & $e^{5i\pi/6}/\sqrt{3}$ & $3$ \\
        \hline
        $\mathbb{Z}_3\times\mathbb{Z}_3$ & $m_0$ & $m_1$ & $3e^{2i\pi/3}$ & $e^{i\pi/3}$ & $3$ \\
        \hline
        $\mathbb{Z}_3\times\mathbb{Z}_6$ & $n_0$ & $n_1$ & $-1$ & $e^{2i\pi/3}$ & $3$ \\
        \hline
        $\mathbb{Z}_4\times\mathbb{Z}_4$ & $n_0$ & $n_1$ & $-2$ & $e^{3i\pi/4}/\sqrt{2}$ & $4$ \\
        \hline
        $\mathbb{Z}_6\times\mathbb{Z}_6$ & $n_0$ & $n_1$ & $-1$ & $e^{5i\pi/6}/\sqrt{3}$ & $3$ \\
        \hline
    \end{tabular}
    \caption{Superpotential and $N_\mathrm{flux}$ data for orbifolds with $h^{2,1} = 0$, with the parameterisation \eqref{eq:11}.}
    \label{tab:4}
\end{table}

\begin{table}%[ht!]
    \centering
    \begin{tabular}{|c||c||c|c||c|c|}
    \hline
        orbifold & basis integers & $A$ & $B$ & $k$ & $\ell$ \\
        \hline\hline
        $\mathbb{Z}_{4,a}$ & $m_0,m_1,m_2,n_0$ & $4\sqrt{2}e^{3i\pi/4}m_0 - 8m_1$ & $4\sqrt{2}e^{3i\pi/4}n_0- 8m_2$ & $1/8$ & $4$ \\
        \hline
        $\mathbb{Z}_{4,b}$ & $m_1,m_3,n_0,n_3$ & $2m_1 - 2\sqrt{2}e^{i\pi/4}n_3$ & $ - 2\sqrt{2}e^{3i\pi/4}m_3-4in_0$ & $1/2$ & $2$\\
        \hline
        $\mathbb{Z}_{4,c}$ & $m_1,n_0,n_1,n_3$ & $m_1 - \sqrt{2}e^{i\pi/4}n_3$ & $- \sqrt{2}e^{i\pi/4}n_0 + n_1$ & $2$ & $2$\\
        \hline
        $\mathbb{Z}_{6,IIa}$ & $m_1,m_3,n_2,p_1$ & $-6m_1 + 12e^{2i\pi/3}p_1$ & $ - 6i\sqrt{3}m_3-18n_2$ & $1/18$ & $6$ \\
        \hline
        $\mathbb{Z}_{6,IIb}$ & $m_0,m_1,m_2,m_3$ & $A_{6,IIb}$ & $B_{6,IIb}$ & $-2/9$ & $3$ \\
        \hline
        $\mathbb{Z}_{6,IIc}$ & $m_0,m_3,p_2,q_2$ & $-6m_0 - 2i\sqrt{3}p_2$ & $6m_3 - 2i\sqrt{3}q_2$ & $1/6$ & $2$\\
        \hline
        $\mathbb{Z}_{6,IId}$ & $m_0,m_1,n_0,n_1$ & $- e^{i\pi/3}m_0-i\sqrt{3}m_1 $ & $ - \sqrt{3}e^{i\pi/6}n_0 + n_1$ & $2$ & $1$\\
        \hline
        $\mathbb{Z}_{8,IIa}$ & $m_0,m_1,n_0,n_2$ & $4im_0 - 4\sqrt{2}m_1$ & $-8in_0 +4(2i+\sqrt{2})n_2$ & $-1/8$ & $4$\\
        \hline
        $\mathbb{Z}_{8,IIb}$ & $m_1,m_2,n_0,n_2$ & $2(2i-\sqrt{2})m_1 - 4im_2$ & $4in_0 - 2(2i+\sqrt{2})n_2$ & $1/4$ & $2$\\
        \hline
        $\mathbb{Z}_{12,II}$ & $m_0,m_3,p_4,q_2$ & $-\sqrt{3}m_0 + \sqrt{6}e^{i\pi/4}p_4$ & $\sqrt{3}m_3 + \sqrt{6}e^{i\pi/4}q_2$ & $2/3$ & $2$ \\
        \hline
        $\mathbb{Z}_{2}\times\mathbb{Z}_4$ & $m_2,n_0,n_1,n_2$ & $im_2 - \sqrt{2}e^{i\pi/4}n_1$ & $\sqrt{2}e^{3i\pi/4}n_0 - in_2$ & $-2$ & $2$ \\
        \hline
        $\mathbb{Z}_{2}\times\mathbb{Z}_{6,I}$ & $m_0,m_2,n_0,n_2$ & $-e^{i\pi/3}m_0 - i\sqrt{3}m_2$ & $-\sqrt{3}e^{i\pi/6}n_0 + n_2$ & $2$ & $1$\\
        \hline
    \end{tabular}
    \caption{Superpotential and $N_\mathrm{flux}$ data for orbifolds with $h^{2,1} = 1$, with the parameterisation \eqref{eq:12}. To avoid a wide table, we introduced the notation $A_{6,IIb} = -3e^{i\pi/3}m_0 + 3i\sqrt{3}m_1 + 3e^{2i\pi/3}m_2 - 3i\sqrt{3}m_3$ and $B_{6, IIb} = 3m_0 + 3i\sqrt{3}m_1 + 3e^{2i\pi/3}m_2$. The integer $\ell$ defined in \eqref{firstdefl} such that $N_\mathrm{flux} \in \ell \mathbb{N}$, comes solely from the orbifold action, not from the quantisation of the flux integers, see \cref{sec:tadpoleintegersmultiple}.
    }\label{tab:5}
\end{table}

\subsection{Supergravity effective theory, moduli stabilisation and vacuum solutions}\label{sec:modstab}
The $\mathcal{N}=1$  effective supergravity theory is described by the K\"ahler potential $\mathcal{K}$ and superpotential $W$. In this work, we only consider the flux induced superpotential \eqref{eq:10}. In our conventions, the tree-level K\"ahler potential $\mathcal{K}$ reads:
\begin{equation}\label{Kahlerpot}
\mathcal{K}=-2 \log (\mathcal{V}) - \log \Big(-i (\mathcal{S}-\bar{\mathcal{S}})\Big) -\log\Big( i \int \Omega \wedge \bar \Omega \Big),
\end{equation}
where the three terms correspond respectively to the K\"ahler moduli, axio-dilaton and complex structure moduli. The part depending on the internal volume $\mathcal{V}$, parameterised by the K\"ahler moduli, satisfies the famous no-scale structure~\cite{Cremmer:1983bf, Ellis:1983ei}. The complex structure moduli part can be written \cite{Candelas:1990pi,Ceresole:1995ca,Taylor:1999ii,Blumenhagen:2003vr} in terms of the projective coordinates \eqref{omegaXaGa} as:
\begin{equation}\label{eq:prepotential}
e^{-\mathcal{K}_{\rm c.s.}}= i \int \Omega \wedge \bar \Omega = -i \Big(X^a\bar{\mathcal{G}_a}-\bar X^a \mathcal{G}_a\Big).
\end{equation}
The last equality including the derivative $\mathcal{G}_a$ of the prepotential $\mathcal{G}$ comes from the symplectic structure of the special geometry of $\mathcal{N}=2$ moduli space \cite{Candelas:1990pi,Ceresole:1995ca,Taylor:1999ii,Blumenhagen:2003vr}. The supergravity scalar potential is eventually obtained by:
\begin{equation}\label{eq:21}
    V = e^\mathcal{K}\Big(\mathcal{K}^{i\bar j}D_i W\bar{D}_{\bar j} \overline{W} -3W\overline{W}\Big),
\end{equation}
where the indices run over all moduli fields. The K\"ahler covariant derivatives are defined as $D_i=\partial_i+\mathcal{K}_i$. Due to the tree-level no-scale structure of the K\"ahler sector, the sum over the K\"ahler moduli cancels the negative contribution, leading to the remaining scalar potential:
\begin{equation}
  V = e^\mathcal{K}\mathcal{K}^{a\bar b}D_aW\bar{D}_{\bar b}\overline{W}, \label{scalarpot2}
  \end{equation}
where now the  $a, \bar b$ indices only run over the complex structure moduli and the axio-dilaton. This scalar potential is positive and is minimised at points where $D_aW=0$, for all $a$. These are exactly the supersymmetry conditions for the complex structure moduli. They are sufficient conditions to find a vacuum. Minimising this potential thus stabilises the complex structure moduli and the axio-dilaton totally or partially, depending on the flux background.

\subsection{Orientifolding and tadpole condition} \label{sec:tadpoleintegersmultiple}

\paragraph{Tadpole constraint} In addition to quantisation conditions, the $m_i, n_i, p_i, q_i$ flux parameters should also satisfy the tadpole condition. The latter translates the fact that the total D3-brane charge $Q_{D3}$ supported by the compact manifold must vanish. Using the conventions of \cite{Ibanez:2012zz}, this condition reads
\begin{equation}\label{eq:13}
    N^{\scriptscriptstyle Y}_\mathrm{flux} + N_{D3} = \frac{1}{4} N_{O3},
\end{equation}
where $N^{\scriptscriptstyle Y}_\mathrm{flux}$ is the $Y$ orientifold flux number obtained from $N_{\rm flux}$ of \eqref{eq:9} as described below, around \cref{defNy}. Similarly, $N_{D3}, N_{O3}$ denote the D3-brane and O3-plane charges in the quotient space, obtained from the orbifold charges without counting orientifold images. In the absence of anti-D3-brane charge, $N_{D3} \geq 0$. In that case, as $N^{\scriptscriptstyle Y}_\mathrm{flux}$ is positive at the vacuum solution, it is bounded by the number of O3-planes.

 The number and the loci of O3-planes are fixed by the choice of orientifolding. This is a further quotient of the orbifold by a geometric involution combined with a reversal of worldsheet orientation. 
The number of O3-planes $N_{O3}$ is obtained by counting the number of involution fixed points, their localisations are then simply the fixed points coordinates. 
We consider the simplest reflection involution $x^i \rightarrow - x^i$. It is the involution maximising the number of O3-planes $N_{O3}$, thus giving weakest bound on $N_{\rm flux}$. 

On the torus $T^6$, there are $2^6 = 64$ fixed points, with real coordinates $(\iota_1, \dots, \iota_6)$ on the torus lattice, where $\iota_i = 0$ or $1/2$. Some of these  points are however identified by the orbifold action, acting  through the matrix $M$ introduced in \cref{orbifoldActionreal}. Such points should count only once. We obtain the number of O3-planes reported in table \ref{tab:6} for each orientifold of table \ref{tab:2}. We have some mismatches with the results of \cite{Reffert:2006du}, for the orientifolds $T^6/\mathbb{Z}_{6,Ia}$ and $T^6/\mathbb{Z}_3\times\mathbb{Z}_3$. In this work, we will not consider  quantised NS-NS $B_2$ field requiring exotic O$p$-planes  and lowering the total $Op$-planes charges \cite{Witten:1997bs,Kakushadze:1998bw,Angelantonj:2002ct}. All O$p$-planes RR charges thus have opposite signs with respect to those of D$p$-branes.

In general Calabi-Yau compactifications, O$7$-planes and D$7$-branes wrapped around $4$-dimensional submanifolds of the internal space also induce geometric contribution to the D3-brane charge \cite{Blumenhagen:2008zz}. This geometric contribution is proportional to the submanifold Euler characteristic. In the case of toroidal orbifolds, the wrapped submanifolds have vanishing Euler characteristic and the geometric contribution vanishes. In presence of world-volume magnetic fluxes, D7-branes can also induce D3-brane charge.  Depending on the choice of fluxes in toroidal orientifolds, magnetised D7-branes thus also contribute to the D3 tadpole of toroidal orientifolds \cite{Bachas:1995ik,Marino:1999af,Angelantonj:2000hi,Blumenhagen:2006ci}. We come back to this point in \cref{subsec:magnetizedbranes}.

\begin{table}%[ht!]
    \centering
    \begin{tabular}{|c|c||c|c||c|c|}
    \hline
        orbifold & $N_{O3}$ & orbifold & $N_{O3}$ & orbifold & $N_{O3}$ 
        \\
        \hline\hline
        $\mathbb{Z}_3$ & $22$ & 
        $\mathbb{Z}_7$ & $10$ & 
        $\mathbb{Z}_2\times \mathbb{Z}_2$ & $64$ \\
        $\mathbb{Z}_{4,a}$ & $22$ & 
        $\mathbb{Z}_{8,Ia}$ & $12$ & 
        $\mathbb{Z}_2\times \mathbb{Z}_4$ & $40$ \\
        $\mathbb{Z}_{4,b}$ & $28$ & 
        $\mathbb{Z}_{8,Ib}$ & $22$ & 
        $\mathbb{Z}_2\times \mathbb{Z}_{6,I}$ & $24$\\
        $\mathbb{Z}_{4,c}$ & $40$ & 
        $\mathbb{Z}_{8,IIa}$ & $16$ & 
        $\mathbb{Z}_2\times \mathbb{Z}_{6,II}$ & $22$ \\
        $\mathbb{Z}_{6,Ia}$ & $14$ & 
        $\mathbb{Z}_{8,IIb}$ & $24$ & 
        $\mathbb{Z}_3\times \mathbb{Z}_3$ & $10$\\
        $\mathbb{Z}_{6,Ib}$ & $22$ & 
        $\mathbb{Z}_{12,Ia}$ & $8$ & 
        $\mathbb{Z}_3\times \mathbb{Z}_6$ & $17$\\
        $\mathbb{Z}_{6,IIa}$ & $16$ & 
        $\mathbb{Z}_{12,Ib}$ & $14$ & 
        $\mathbb{Z}_4\times \mathbb{Z}_4$ & $28$\\
        $\mathbb{Z}_{6,IIb}$ & $24$ & 
        $\mathbb{Z}_{12,II}$ & $16$ & 
        $\mathbb{Z}_6\times \mathbb{Z}_6$ & $17$\\
        \cline{3-6}
        $\mathbb{Z}_{6,IIc}$ & $16$ \\
        $\mathbb{Z}_{6,IId}$ & $24$ \\
        \cline{1-2}
    \end{tabular}
    \caption{Numbers of O3-planes in the orientifolds with involution $x^i\rightarrow -x^i$ constructed from the orbifolds of \cref{tab:2}.}
    \label{tab:6}
\end{table}

\paragraph{A bit more on flux quantisation in orbifolds: quantised quanta}  At this stage, we shall introduce an additional fact about the quantisation of the flux integers \cite{Frey:2002hf,Cascales:2003zp,Blumenhagen:2003vr,Blumenhagen:2005tn}. Namely, to avoid subtleties associated with additional $3$-cycles that are not present in the covering $T^6$, we take the flux quanta to be multiples of $2|G|$, where $|G|=N,NM$ is the cardinal of the orbifold group for  $T^6/\mathbb{Z}_N$ or $T^6/\mathbb{Z}_N\times\mathbb{Z}_M$ orbifolds. The factor $|G|$ comes from the orbifold action and the factor of $2$ comes from the $\mathbb{Z}_2$ involution in the orientifold action. 

Such quantisation can be understood from the fact that in \eqref{eq:8} we defined the flux integers on the cohomology of the \emph{covering torus} $T^6$. They can indeed be expressed as:
\begin{equation}
    m_i^H = \int_{A^i} H_3,
\end{equation}
where $A^i$ is the $3$-cycle that is Poincar\'e dual to $\alpha_i$. However, under the orbifold quotient of the torus by $G$, this cycle is mapped to a cycle $\tilde A^i$, which  is $|G|$ times smaller. More precisely, the $3$-cycles have $|G|$ homologically equivalent images under the orbifold action from the torus point of view. All of them are identified to a single $3$-cycle in the orbifold. The flux integrals over this cycle can be used to define the ``orbifold flux integers" which are thus smaller  than the torus ones. For instance, we get:
\begin{equation}\label{eq:16}
    \tilde m_i^H = \int_{\tilde A^i} H_3 = \frac{1}{|G|}\int_{A^i} H_3
    \quad\rightarrow\quad
    m_i^H = |G| \tilde m^H_i.
\end{equation}
The Dirac quantisation in the orbifold, hence for $\tilde m^H_i$, implies that $m^H_i$ is multiple of $|G|$. Moreover, taking the orientifold quotient, in absence of discrete fluxes on exotic O$p$-planes, imposes an additional factor of $2$ \cite{Frey:2002hf,Coudarchet:2023mmm}. The same quantisation conditions hold for all the flux quanta so that when using the ``torus integers''  of \eqref{eq:8} in the orientifold of $T^6/G$, they should always be multiples of $2|G|$.

In a similar manner,  the flux number appearing in the tadpole constraint should be computed on the orientifold $Y$ rather than on the torus $T^6$. This is the motivation of the definition of $N^{\scriptscriptstyle Y}_{\rm flux}$ used in the tadpole constraint \eqref{eq:13}:
\begin{equation}
N^{\scriptscriptstyle Y}_\mathrm{flux}\equiv \int_Y H_3\wedge F_3 = \frac{1}{2|G|} \int_{T^6} H_3\wedge F_3, \label{defNy}
\end{equation}
with now the fluxes computed using the properly quantised ``torus integers". The ratio between the volumes of the torus and of the orientifold $Y$ gives the factor $2|G|$. The cardinal  $|G|=N, NM$  accounts for the volume of the orbifold fundamental cell, while the additional factor of $2$ accounts for the orientifolding.
From the quantisation of the ``torus integers", we  thus see that the torus $N^{T^6}_\mathrm{flux}$ and the orientifold $N^{\scriptscriptstyle_ Y}_\mathrm{flux}$ are multiple of a minimal integer value, respectively $4|G|^2$ and $2|G|$. Indeed, we have:
\begin{equation}
N^{T^6}_{\rm flux}= 4 |G|^2 N_{\rm flux}, \qquad N^{\scriptscriptstyle Y}_{\rm flux}= 2 |G| N_{\rm flux},  \quad |G|=N, NM, \label{relNy}
\end{equation}
where $N_{\rm flux}$ is computed ignoring the further orbifold quantisation of the ``torus integers", hence with the $\tilde{m}_i, \tilde{n}_i$ (taking arbitrary integer values), instead of the $m_i,n_i$ of \cref{eq:16}. We recall that $N_{\rm flux}$ is however multiple of an integer $k$ or $\ell$ for other reasons, see \cref{Nfluxh21_0,firstdefl}.
 
 In \cite{Blumenhagen:2003vr,Cascales:2003zp}, authors considered the orientifold $T^6/\mathbb{Z}_2\times \mathbb{Z}_2$, for which $|G|=4$ leads to $N^{T^6}_\mathrm{flux}$ being multiple of $8^2=64$. We see that $N^{T^6}_\mathrm{flux}$ overshoots the orientifold charge, which is $N_{O3}/2=32$ for this orientifold, unless one turns-on fluxes on twisted $3$-cycles, which carry smaller quanta. However, the previous discussion shows that we should rather use $N^{\scriptscriptstyle Y}_\mathrm{flux}$ in the tadpole constraint, which is multiple of $8$ and thus seems to invalidate their conclusion\footnote{we thank Ralph Blumenhagen and Tomasz R. Taylor for discussion on this topic.}.

\subsection{Twisted moduli and discrete torsion}\label{sec:6}

We come back to the twisted K\"ahler and complex structure moduli. They are additional degrees of freedom corresponding to strings closing up to the action of the orbifold group, on the covering space of the orbifold. They are necessary for the theory to be well defined on the singular geometry of the orbifold,  in particular to ensure modular invariance of the partition function. They correspond geometrically to deformation parameters allowing to resolve singularities lying at fixed loci of the orbifold action. In the effective theory, they correspond to additional scalars, counted by the twisted Hodge numbers $(\tilde h^{1,1},\tilde{h}^{2,1})$, see \cref{orbiifoldconstruction}. The cohomology basis can be extended by addition of twisted forms. The latter are dual to twisted cycles made from cycles blowing up the orbifold singularities. They are thus located at the orbifold fixed points and their sizes are parameterised by the twisted moduli. 

For the $\mathbb{Z}_2\times \mathbb{Z}_2$ orbifold with discrete torsion (see just below) the twisted cohomology basis can be constructed taking the dual forms of the twisted 3-cycles \cite{Blumenhagen:2005tn}:
\begin{equation}
[E^{g_i}_{\alpha\beta}]\otimes [A^i] \rightarrow \alpha_{i,\alpha\beta}, \qquad [E^{g_i}_{\alpha\beta}]\otimes [B^i] \rightarrow \beta^{i,\alpha\beta}. 
\end{equation}
 The twisted sectors $g_i=\theta_2, \theta_2', \theta_2 \theta_2'$, with $\theta_2$ and $\theta_2'$ the generators of two  $\mathbb{Z}_2$ of \cref{tab:2}, keep the torus $T^2_i$ fixed. The 1-cycles $[A^i]$ and $[B^i]$ are the generators of this torus. The indices $\alpha,\beta=1,\ldots,4$ label the $16$ fixed points of $g_i$ in the two other tori.  In this orbifold, there are thus $2\times3\times 16=96$ elements $\alpha_{i,\alpha\beta},  \beta^{i,\alpha\beta}$ in the twisted cohomology basis. The $[E^{g_i}_{\alpha\beta}]$ 2-cycles are parameterised by $48$ twisted moduli $\mathcal{D}^i_{\alpha\beta}$.

In $T^6/\mathbb{Z}_N\times\mathbb{Z}_M$ orbifolds, there is an arbitrary choice of discrete torsion \cite{Vafa:1986wx}. It corresponds to the possibility of a discrete phase between different twisted sectors of the partition function, keeping modular invariance. The choice of discrete torsion has a non-trivial effect on the geometric interpretation of the twisted moduli \cite{Vafa:1994rv}. As shown in  \cite{Font:1988mk}, it affects the numbers of twisted K\"ahler and complex structure moduli $(\tilde h^{1,1}, \tilde h^{2,1})$. The Hodge numbers of these orbifolds in presence of discrete torsion are explicitly shown in  \cref{tabHodgeDT}.

\begin{table}[ht!]
    \centering
    \begin{tabular}{|c||c|ccc|}
    \hline
     \multirow{2}*{orbifold}   & \multicolumn{2}{c}{$\vphantom{\frac{1^{1^{1}}}{2}}(\tilde h^{1, 1}, \tilde h^{2, 1})$}&& \\
            & without & \multicolumn{3}{c|}{  with discrete torsion} \\
        \hline\hline
        $\mathbb{Z}_2\times\mathbb{Z}_2$ & $(48, 0)$  & $(0, 48)$ && \\
        $\mathbb{Z}_2\times\mathbb{Z}_4$ & $(58, 0)$ & $(18, 8)$ &&\\
        $\mathbb{Z}_2\times\mathbb{Z}_{6,I}$ & $(48, 2)$ & $(16, 18)$ &&\\
        $\mathbb{Z}_2\times\mathbb{Z}_{6, II}$ & $(33, 0)$ & $(13, 15)$ &&\\
        $\mathbb{Z}_3\times\mathbb{Z}_3$ & $(81, 0)$ & $(0, 27)$ &&\\
        $\mathbb{Z}_3\times\mathbb{Z}_6$ & $(70, 1)$ & $(10, 13)$&& \\
        $\mathbb{Z}_4\times\mathbb{Z}_4$ & $(87, 0)$ & $(39, 0)$ & $(3, 12)$ & $(3, 12)$ \\
        $\mathbb{Z}_6\times\mathbb{Z}_6$ & $(81, 0)$ & $(48, 3)$ & $(24, 3)$ & $(6,9)$\\
        \hline
    \end{tabular}
    \caption{Twisted Hodge numbers $(\tilde h^{1, 1}, \tilde h^{2, 1})$ in $\mathbb{Z}_N\times \mathbb{Z}_M$ orbifolds with discrete torsion \cite{Font:1988mk}.} \label{tabHodgeDT}
\end{table}

In particular, in the case of the orbifold $T^6/\mathbb{Z}_2\times\mathbb{Z}_2$, we see that discrete torsion exchanges $\tilde h^{1, 1}$ with $\tilde h^{2, 1}$. This is a particular case where the orbifolds with and without discrete torsion are related by mirror symmetry. This  does not happen for the other orbifolds. 

 If we do not turn on fluxes on the twisted $3$-cycles parameterised by the twisted moduli, the latter are generically stabilised locally at the orbifold point, i.e. with vanishing vacuum expectation value. 
This is due to the discrete symmetry of the moduli space at the orbifold point. By choosing fluxes that are invariant under this symmetry, the vacuum equations of the non-invariant moduli are solved automatically \cite{Cascales:2003zp}. Similar discrete symmetries have been used in the past in more general Calabi-Yau compactifications to reduce the number of complex structure moduli to be stabilised by fluxes \cite{Giryavets:2003vd,Denef:2004dm,Louis:2012nb,Cicoli:2013cha,Lust:2022mhk,Candelas:2023yrg}. To summarise, as long as we do not turn on fluxes on the twisted $3$-cycles the corresponding twisted moduli are automatically stabilised at the orbifold point. We show it explicitly in the case of the $T^6/\mathbb{Z}_2\times\mathbb{Z}_2 $ orbifold in \cref{subsec:twistedmoduliZ2Z2}.

\section{Orbifolds with $h^{2,1}_{\rm untw.}=0,1$: vacuum solutions and string coupling}  \label{sec:gsNflux}

In this section, we study the stabilisation of the untwisted complex structure moduli and axio-dilaton for the orbifolds listed in table \ref{tab:2}, except the $T^6/\mathbb{Z}_2\times\mathbb{Z}_2$ treated in section \ref{T6Z2xZ2}. As reminded in \cref{sec:4}, we search for sets of background fluxes that stabilise the moduli. A vacuum solution is thus a combination of the flux quanta together with a point in moduli space, depending on the flux quanta, which minimises the scalar potential.  The scalar potential depends on the superpotential $W$, which for the orbifolds of \cref{tab:2} was presented and parameterised in \cref{sec:4}. The expression  \eqref{scalarpot2} of the scalar potential shows that its minimisation is ensured for solutions satisfying the supersymmetry conditions $D_aW = 0$. We thus look for such solutions. We remind that these are not a necessary conditions.

We find vacua stabilising all the complex structure moduli and the dilaton for these orbifolds and we exhibit evidence for an exact relation between the minimal value of the string coupling $g_s$ and $N_\mathrm{flux}$. In orbifolds with $h^{2,1}=0,1$, i.e. with zero or one untwisted complex structure modulus, this relation goes in the large flux number limit $N_\mathrm{flux}\gg1$ as:
\begin{equation}\label{eq:rel_1}
    g_{s, \mathrm{min}} \sim \frac{1}{N_\mathrm{flux}},
\end{equation}
The exact relation is given in the next subsections. We comment that this relation does not seem to match in these cases with the one that could be estimated from the early works \cite{Ashok:2003gk,Denef:2004ze} on flux vacua statistics. See discussion around \cref{estimationAD} for such estimate in the case of $\mathbb{Z}_2\times\mathbb{Z}_2$ where it gives the correct result.

When supplemented with the tadpole condition  \eqref{eq:13} satisfied by $N_\mathrm{flux}$, the above relation places a constraint on the value of the string coupling. In absence of negative D3 charge, this constraint is a lower bound, which depends on the particular orbifold.
We stress that all these conclusions only hold for supersymmetric vacua, satisfying $D_aW=0$.

\subsection{Orbifolds with no complex structure moduli}\label{nocs}

\paragraph{Vacuum relation between $g_s$ and $N_{\rm flux}$} We derive relation \eqref{eq:rel_1} for orbifolds with no untwisted complex structure moduli, thus with $h^{2,1}=0$. We parameterised the flux superpotential of such orbifolds in \cref{eq:11}. It reads:
\begin{equation}\label{eq:15}
    W = K(a + \gamma b) \quad\text{where}\quad a = \mathcal{S}a^H + a^F 
    \quad\text{and}\quad b = \mathcal{S}b^H + b^F,
\end{equation}
It only depends on the flux integers and the axio-dilaton $\mathcal{S}$, as long as $a^H$ or $b^H$ is non-vanishing. The flux charge reads $N_\mathrm{flux} = k(a^Hb^F - a^Fb^H)$, with $k\in \mathbb{Z}$, see \cref{Nfluxh21_0}. See \cref{tab:4} for the values of $K, \gamma$ and $k$ for each $h^{2,1}=0$ orbifold.

 Vacua are obtained solving $D_\mathcal{S}W = 0$, which yields:
\begin{equation}\label{eq:rel_6}
    \mathcal{\bar S} = -\frac{a^F + \gamma b^F}{a^H + \gamma b^H}, \qquad \qquad a^H\neq0 \quad {\rm or} \quad b^H\neq0.
\end{equation}
The inverse of the string coupling, defined below \cref{eq:10}, is given by the imaginary part of $\mathcal{S}$. It thus reads:
\begin{align}\label{eq:rel_3}
    \frac{1}{g_s} = \frac{\mathrm{Im}(\gamma)(a^Hb^F - a^Fb^H)}{(a^H + b^H\mathrm{Re}(\gamma))^2 + (b^H\mathrm{Im}(\gamma))^2} &= \frac{N_\mathrm{flux}}{k} \frac{\mathrm{Im}(\gamma)}{(a^H + b^H\mathrm{Re}(\gamma))^2 + (b^H\mathrm{Im}(\gamma))^2} \nonumber\\
    &\leq \frac{N_\mathrm{flux}}{k} \frac{\mathrm{Im}(\gamma)}{\min(\mathrm{Re}^2(\gamma),\mathrm{Im}^2(\gamma))}. 
\end{align}
The last inequality comes from the fact that, when  $(a^H, b^H)$ are integers different from $(0,0)$, the denominator is bounded from below. \Cref{eq:rel_3} leads to a relation between the minimal value for the string coupling and the flux charge $N_{\rm flux}$, as advertised in \eqref{eq:rel_1}. This relation depends on $\gamma$ and $k$, with values for each $h^{2,1}=0$ orbifold shown in table \ref{tab:4}, and reads:
\begin{equation}\label{eq:rel_2}
    g_{s, \mathrm{min}} = \frac{C}{N_\mathrm{flux}}.
\end{equation}
The values of $C$  are shown in \cref{tab:rel_1}. 

Note that the sign of $C$ is the one of $\mathrm{Im}(\gamma)/k$. For  $T^6/\mathbb{Z}_3$, $C$ is negative. Since $N_{\rm flux}$ is positive we deduce that there are no physical vacua with $g_s>0$ at the points $D_\mathcal{S}W = 0$ for the orbifold $T^6/\mathbb{Z}_3$.

\begin{table}%[ht!]
    \centering
    \begin{tabular}{|c|c||c|c|}
    \hline
         orbifold & $C$ & orbifold & $C$\\
         \hline\hline 
         $\mathbb{Z}_{6,Ia}$ & $2\sqrt{3}$
         &$\mathbb{Z}_2\times\mathbb{Z}_{6,II}$ & $2\sqrt{3}$\\
         $\mathbb{Z}_{6,Ib}$ & $2\sqrt{3}$
         &$\mathbb{Z}_3\times\mathbb{Z}_3$ & $2\sqrt{3}$\\
         $\mathbb{Z}_7$ & $2\sqrt{7}$
         &$\mathbb{Z}_3\times\mathbb{Z}_6$ & $2\sqrt{3}$\\
         $\mathbb{Z}_{8,Ia}$ & $8$
         &$\mathbb{Z}_4\times\mathbb{Z}_4$ & $4$\\
         $\mathbb{Z}_{8,Ib}$ & $4$
         &$\mathbb{Z}_6\times\mathbb{Z}_6$ & $2\sqrt{3}$\\
         \cline{3-4}
         $\mathbb{Z}_{12,Ia}$ & $2\sqrt{3}$ \\
         $\mathbb{Z}_{12,Ib}$ & $2\sqrt{3}$ \\
         \cline{1-2}
    \end{tabular}
    \caption{Values of $C$ in the relation \eqref{eq:rel_2} for the toroidal orbifolds with no complex structure moduli.}
    \label{tab:rel_1}
\end{table}

\paragraph{Duality, fundamental domain and equivalent vacua} In this paragraph we comment on the use of $S$-duality to relate seemingly different vacuum solutions. The ${S}$-duality enjoyed by type IIB string theory \cite{Schwarz:1995dk} is implemented by the transformation acting on the axio-dilaton and the fluxes as:
\begin{equation}\label{eq:rel_8}
    \mathcal{S} \quad\rightarrow\quad \frac{a\mathcal{S} + b}{c\mathcal{S}+d}
    \qquad\text{and}\qquad
    \begin{pmatrix}
    H_3\\F_3
    \end{pmatrix}
    \quad\rightarrow\quad
    \begin{pmatrix}
    d & -c \\ -b & a
    \end{pmatrix}
    \begin{pmatrix}
    H_3\\F_3
    \end{pmatrix},
\end{equation}
with $a,b,c$ and $d$ integers satisfying $ad - bc = 1$. This transformation defines the $\mathrm{SL}(2,\mathbb{Z})$ group,  generated by  $\mathcal{S}\rightarrow -1/\mathcal{S}$ and $\mathcal{S} \rightarrow \mathcal{S} + 1$.  

Through their action on the fluxes $H_3$ and $F_3$, these transformations also act on the flux integers. They however leave $N_\mathrm{flux}$ invariant
\begin{equation}\label{eq:rel_9}
    N_\mathrm{flux} = m^H n^F - m^Fn^H \quad\rightarrow\quad (a d - b c)  (m^H n^F - m^F n^H) = N_\mathrm{flux}.
\end{equation}
Such  $\mathrm{SL}(2,\mathbb{Z})$ transformations can be used to bring the dilaton $\mathcal{S}$ in the fundamental domain:
\begin{equation}
    \mathcal{F} = \{-1/2 < \mathrm{Re}(\mathcal{S}) \leq 1/2 \quad\text{and}\quad |\mathcal{S}| \geq 1 \}.
\end{equation}
We can infer that {\it a priori} different vacuum solutions, obtained through \cref{eq:rel_6} from different choices of integers $a^H,b^H,a^F,b^F$, can be mapped by means of ${S}$-duality transformations. The easiest way to compare vacuum solutions is thus to bring  $\mathcal{S}$ to its fundamental domain. 

As an example, we take the $\gamma$ parameter of the superpotential \eqref{eq:15}  with value $\gamma = 1+i$. It does not correspond to any orbifold of table \ref{tab:4} but serves as a simple illustration of the duality. For $N_\mathrm{flux} = 1$ and integers with $|a^H|, |b^H|, \ldots \leq 1$, we obtain $20$ choices of fluxes  stabilising  $\mathcal{S}$ with $\mathrm{Im}(\mathcal{S}) \neq 0$. Not all of these $20$ combinations have $\mathcal{S}$ in the fundamental domain. For instance, the following vacuum solution: 
\begin{equation}
    (a^H,b^H,a^F,b^F) = (1, 1, 0, 1),
    \quad \qquad
    \langle \mathcal{S} \rangle= -\frac{3}{5} + \frac{i}{5},
\end{equation}
can be mapped through $S$-duality to $\langle \mathcal{S}  \rangle \rightarrow i$, its representative in $\mathcal{F}$. The parameters \eqref{eq:rel_8} of this duality transformation are $(a,b,c,d)=(2,1,1,1)$. It brings the flux integers to $(a^H,b^H,a^F,b^F) = (1, 0, -1, 1)$.

We find that all of the $20$ combinations with $N_\mathrm{flux} = 1$ and integers in the range $|a^H|, |a^F|, |b^H|, |b^F| \leq 1$ have the same representative $\mathcal{S} = i$ in the fundamental domain. Even with integers above this range, we checked that all integer combinations with $N_\mathrm{flux} = 1$ lead to $\mathcal{S} = i$ once brought in the fundamental domain. We conclude that the axio-dilaton solution $\mathcal{S} = i$ can be obtained with different choices of flux parameters satisfying $N_\mathrm{flux} = 1$. Although all these choices of parameters give the same $\mathcal{S} $, the specific values of parameters may give different masses for the stabilised moduli. 

\begin{figure}%[ht!]
\vspace{0.4cm}
\begin{center}
\includegraphics[scale=0.5]{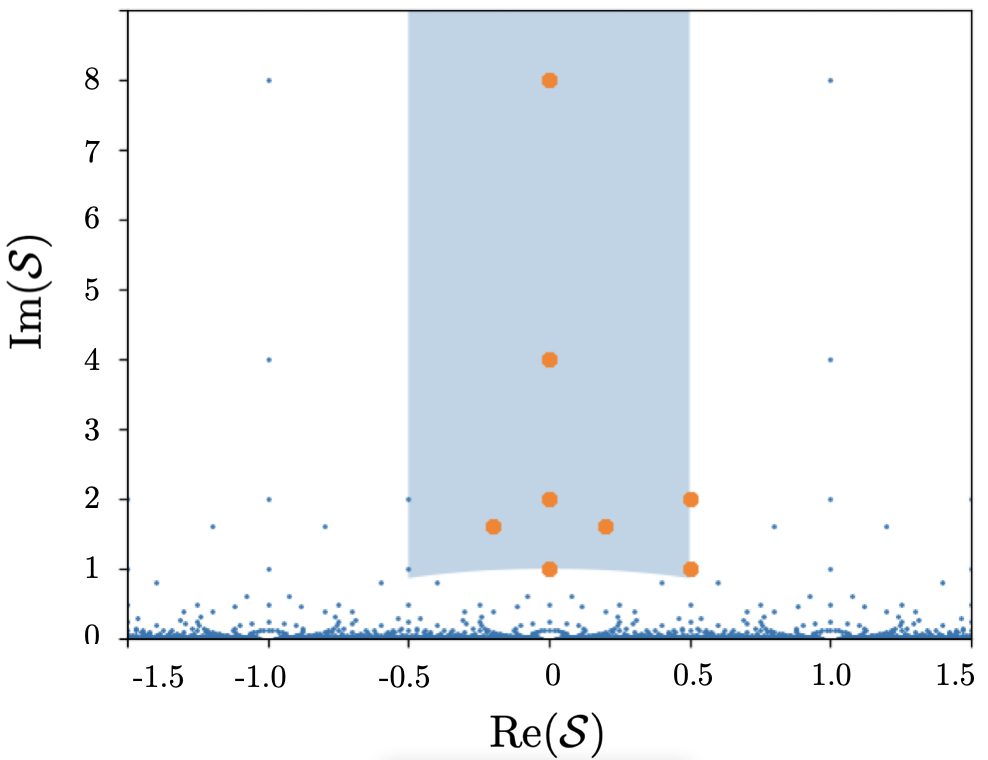}
\end{center}
\vspace{-0.5cm}
\caption{Values of $\mathcal{S}$ for vacuum solutions with $N_\mathrm{flux} = 8$ in the complex plane. Each blue dot corresponds to a different combination of integers. The corresponding values of $\mathcal{S}$ are all mapped to one of the eight orange dots in the fundamental domain (shaded blue region) under $\mathcal{S}$-duality.}
\label{figSsolutions}
\end{figure}

For $N_\mathrm{flux} \geq 1$, we obtain other vacuum solutions for $\mathcal{S}$. They always come in finite numbers for each value of $N_\mathrm{flux}$: when spanning for integers below a certain range $k$, we find a certain range $ k^{ N_{\rm flux}}_{{\rm max} \#}$ above which no new vacua exist. In \cref{tableNfluxh21_0} we give the number of different values of $\mathcal{S}$ found for the first values of $N_\mathrm{flux}$ in the toy orbifold under consideration, with $\gamma = 1+i$ in \eqref{eq:15}. In figure \ref{figSsolutions}, we also plot the locations of the values of $\mathcal{S}$ for $N_\mathrm{flux} = 8$, before and after duality transformation.  All of the eight inequivalent vacua are found within the range $|m|, |n| \leq k^{8}_{{\rm max} \#}=4$. Greater ranges $k\geq k^{N_{\rm flux}}_{\rm \rm max \#}$ lead to the same number of solutions.
\begin{table}[ht!]
    \centering
    \begin{tabular}{|c|c|c|c|c|c|c|c|c|c|c|}
    \hline
         $N_\mathrm{flux}$ & $1$ & $2$ & $3$ & $4$ & $5$ & $6$ & $7$ & $8$ & $9$ & $10$ \\
         \hline
         \# of values for $\mathcal{S}$ & $1$ & $2$ & $2$ & $4$ & $3$ & $6$ & $4$ & $8$ & $7$ & $8$\\
         \hline
    \end{tabular}
    \caption{Numbers of different vacuum solutions for $\mathcal{S}$, at fixed $N_\mathrm{flux}$, obtained solving $D_{\cal S}W=0$ for the axio-dilaton dependent superpotential \eqref{eq:15} with $\gamma = 1+i$.} \label{tableNfluxh21_0}
\end{table}

\paragraph{The tadpole constraint} The previous relation between  the minimal value of the string coupling and the flux charge $N_{\rm flux}$ can be combined with the tadpole constraint \eqref{eq:13}. In absence of negative D3 charge, the latter gives a bound $N_{\rm flux}$ given by the number of O3-planes. For the involution $x^i\rightarrow -x^i$ this number was given in table \ref{tab:6} for each orbifold.  For instance, in $T^6/\mathbb{Z}_{6, Ia}$ in absence of negative D3-charge, the tadpole bound reads:
\begin{equation} \label{tadpoleconsh21_0}
   N^Y_\mathrm{flux}=2 |G| N_{\rm flux}=12 N_{\rm flux} \leq \frac{N_{O3}}{4} = \frac{7}{2},  \qquad   N_\mathrm{flux} =  k n =  3n, \quad n\in \mathbb{N}^*.
\end{equation}
We remind that according to the discussion around equation \eqref{eq:16}, $N^Y_\mathrm{flux}=2|G| N_{\rm flux}=12 N_{\rm flux}$, and that $N_{\rm flux}$ is multiple of $k=3$ in this orbifold, see \cref{tab:4}. Hence, the tadpole condition cannot be satisfied in absence of negative {D}3-charge. The same conclusion is reached for all the orientifolds listed in \cref{tab:4}.

%
%but for this orientifold, $N_\mathrm{flux}$ is a multiple of $3$ as a consequence of the orbifold projection (i.e. it has $k = 3$ in \cref{tab:4}). So $N_\mathrm{flux} \geq 3$ and the tadpole constraint cannot be satisfied. The same conclusion is reached for all the orientifolds listed in \cref{tab:7}.

\subsection{Orbifolds with one complex structure modulus} \label{sec:1}

\paragraph{Vacuum solutions} In toroidal orbifolds with $h^{2,1}=1$, i.e. with one untwisted complex structure modulus, the relation \eqref{eq:rel_1} is analytically harder to derive. We recall that we search for vacua satisfying $D_aW=0$ for the superpotential parameterised in \eqref{eq:12}  by:
\begin{equation}\label{eq:17}
  W = \mathcal{U}A + B,  \qquad\text{where} \quad A \equiv  A^F +  \mathcal{S}A^H , B\equiv B^F +  \mathcal{S}B^H .
  \end{equation}
The expressions of $A, B$ in terms of the flux integers can be found in table \ref{tab:5} for each orbifold. We also parameterise $N_\mathrm{flux} = k(\mathrm{Re}(A^H \bar B^F) - \mathrm{Re}(A^F\bar B^H))$, see \cref{Nfluxh21_1}. Solving $D_\mathcal{U}W = 0$ gives:
\begin{equation}\label{eq:rel_10}
    \mathcal{\bar U} = -\frac{\mathcal{S}B^H + B^F}{\mathcal{S}A^H + A^F}.
\end{equation}
Once plugged in the second equation $D_\mathcal{S}W = 0$, it yields a second order equation for $\mathcal{S}$:
\begin{equation}
    \mathrm{Im}(A^H \bar B^H) \mathcal{S}^2 + \Big(\mathrm{Im}(A^F \bar B^H) + \mathrm{Im}(A^H \bar B^F) \Big)\mathcal{S} + \mathrm{Im}(A^F \bar B^F) = 0.
\end{equation}
The imaginary part of $\mathcal{S}$ thus reads:
\begin{equation}\label{eq:rel_4}
\frac{1}{g_s} = \frac{\sqrt{4\mathrm{Im}(A^H\bar B^H)\mathrm{Im}(A^F\bar B^F)-(\mathrm{Im}(A^H\bar B^F) + \mathrm{Im}(A^F\bar B^H))^2}}{2\mathrm{Im}(A^H\bar B^H)},
\end{equation}
when the argument of the square root is positive. Otherwise, the imaginary part of $\mathcal{S}$ vanishes. Eventual solutions with  $A^H=B^H=0$ do not satisfy this solution either, since the superpotential does not depend on the axio-dilaton $\mathcal{S}$. The latter is thus not stabilised by such fluxes.

Contrary to the solutions of $h^{2,1}=0$ orbifolds, with $g_s$ given in \cref{eq:rel_3}, it seems difficult to identify $N_\mathrm{flux}$ in the current solution \eqref{eq:rel_4}. To extract our relation, we first investigate inequivalent vacua, as in the previous subsection. 

\paragraph{Dualities, fundamental domains and equivalent vacua} We verify again that some different choices of flux integers, leading to vacua with different  $(\mathcal{S}, \mathcal{U})$, can be related by duality transformations. As before, we can bring $\mathcal{S}$ to its fundamental domain using $S$-duality transformations \eqref{eq:rel_8}. Under this transformation $N_\mathrm{flux}$ is unmodified, see \eqref{eq:rel_9}, and so is $ \mathcal{U} $, as can be checked from its vacuum solution \eqref{eq:rel_10}.

In addition to ${S}$-duality, the theory is invariant under the following ${U}$-duality:
\begin{equation}
    \mathcal{U} \quad\rightarrow\quad \frac{a\mathcal{U} + b}{c\mathcal{U}+d}, \qquad \quad \begin{pmatrix} B^{H,F} \\ A^{H,F} \end{pmatrix} =  \begin{pmatrix} d & -c\\ -b & a  \end{pmatrix} \begin{pmatrix} B^{H,F} \\ A^{H,F} \end{pmatrix}. \label{Tduality1}
\end{equation}
It can be understood from the invariance of the superpotential \eqref{eq:17} under $\mathcal{S} \leftrightarrow \mathcal{U}$ and $A^F \leftrightarrow B^H$. In the exchange, $S$-duality is traded with the above $U$-duality. Just like $N_\mathrm{flux}$ and $\mathcal{U}$ are invariant under $S$-duality, one can check using the previous expressions that $N_\mathrm{flux}$ and $\mathcal{S}$ are invariant under this $U$-duality. 

We can thus bring both $\mathcal{S}$ and $\mathcal{U}$ into their fundamental domains by using both duality transformations. For a given value of $N_\mathrm{flux}$, we find a finite number of inequivalent vacua. We proceed by choosing a range for the flux integers, finding combinations of integers in this range which give the correct $N_{\rm flux}$ and finally scanning over all these combinations to find solutions satisfying \cref{eq:rel_4}. We increase this range and observe that after some value there are no new vacua. We show this procedure more explicitly for the case of $T^6/\mathbb{Z}_2\times\mathbb{Z}_2$ in \cref{sec:5}. The boundedness of $g_s$, advertised in \eqref{eq:rel_1}, appears thus as a byproduct of this finiteness of the number of inequivalent vacua.

In \cref{tab:8_1}, we report the number of vacua for the first values of $N_\mathrm{flux}$ for the $T^6/\mathbb{Z}_{4,a}$ orbifold.  We also give the obtained minimal vacuum values of $1/g^{2}_{s, \mathrm{min}}$. The multiplicative factor in $N_\mathrm{flux}$ is the parameter $\ell=4$ of table \ref{tab:5}, coming from the geometry of the orbifold. It does not come from the quantisation of the flux integers explained around equation \eqref{eq:16}.

\begin{table}[ht!]
    \centering
    \vspace{0.3cm}
    \begin{tabular}{|c||c|c|c|c|c|c|c|c|c|c|}
    \hline
         $N_\mathrm{flux}/4$ & $1$ & $2$ & $3$ & $4$ & $5$ & $6$ & $7$ & $8$ & $9$ & $10$ \\
         \hline
         \# of vacua & $0$ & $1$ & $3$ & $9$ & $20$ & $37$ & $66$ & $104$ & $159$ & $226$ \\
         \hline
         $1/g^{2}_{s, \mathrm{min}}$ & $-$ & $3/4$ & $2$ & $15/4$ & $6$ & $35/4$ & $12$ & $63/4$ & $20$ & $99/4$\\
         \hline
    \end{tabular}
    \caption{Number of inequivalent vacua and values of $1/g_{s, \mathrm{min}}^2$, at fixed  $N_\mathrm{flux}/4$, obtained solving $D_{\cal U,\cal S} W=0$ for the $T^6/\mathbb{Z}_{4,a}$ orbifold.}
    \label{tab:8_1}
\end{table}

In the $T^6/\mathbb{Z}_{4,a}$ orbifold of \cref{tab:8_1},  we can guess the explicit relation between $g_{s, \rm min}$ and $N_{\rm flux}$. It reads:
\begin{equation}\label{eq:5}
    g_{s, \mathrm{min}} = \sqrt{\frac{64}{N_\mathrm{flux}^2 - 16}},
\end{equation}
and realises the relation \eqref{eq:rel_1} for large $N_\mathrm{flux}$. Similar results and relations can be obtained for the other orbifolds. We can always parameterise $g_{s, \mathrm{min}}$ as:
\begin{equation}\label{eq:18}
    g_{s, \mathrm{min}} = \biggl(\frac{N_\mathrm{flux}^2}{c_2} - \frac{1}{c_0} \biggr)^{-1/2}.
\end{equation}
The values of the parameters $c_2$ and $c_0$ for the orbifolds under consideration are given in table \ref{tab:7}. They all have $c_2 \neq 0$ and therefore satisfy the advertised relation \eqref{eq:rel_1},  i.e. $g_{s, \mathrm{min}} \sim 1/N_\mathrm{flux}$ in the limit $N_{\rm flux}\gg 1$. For some orbifolds, the parameter  $c_0$  is a function of $r$, defined through $N_{\rm flux}/\ell= r \,\, {\rm mod} \,\,  p$, where the integer $p$ depends on the orbifold, see  \cref{tab:7}. 

Note the absence of the orbifolds $T^6/\mathbb{Z}_{6,IIb}$ and $T^6/\mathbb{Z}_{8,IIa}$ and $T^6/\mathbb{Z}_2\times\mathbb{Z}_4$ from this table. Table \ref{tab:5} shows that they have $k < 0$, see \cref{Nfluxh21_1} for the definition of $k$.  They thus suffer from the same issue as the orbifold $T^6/\mathbb{Z}_3$ explained after equation \eqref{eq:rel_2}: they do not admit solutions of $D_{\cal U, \cal S}W = 0$ with $g_s>0$.

\begin{table}%[ht!]
    \centering
\begin{tabular}{|c||c||c|clll|}
\hline
        orbifold & $\ell$ & $c_2$ & $c_0$&&& \\
        \hline\hline
        $\mathbb{Z}_{4,a}$ & $4$ & $64$ & $4$ &&&\\
        \hline
        \multirow{2}*{$\mathbb{Z}_{4,b}$} & \multirow{2}*{$2$} & \multirow{2}*{$16$} & $1$ & \multirow{2}*{for $N_\mathrm{flux}/\ell \cong \left\{ \begin{array}{ll}  \\ \end{array} \right. $\hspace{-15pt}} &  $ 0 $&$\hspace{-10pt} \ \mathrm{mod}\ 2$  \\
        & & & $2$ &&  $ 1 $&$\hspace{-10pt} \ \mathrm{mod}\ 2$ \\
        \hline
        $\mathbb{Z}_{4,c}$ & $2$ & $16$ & $4$ &&&\\
        \hline
        \multirow{2}*{$\mathbb{Z}_{6,IIa}$} & \multirow{2}*{$6$} & \multirow{2}*{$48$} & $4$ &  \multirow{2}*{for $N_\mathrm{flux}/\ell \cong \left\{ \begin{array}{ll}  \\ \end{array} \right. $\hspace{-15pt}} &  $ 0 $&$\hspace{-10pt} \ \mathrm{mod}\ 2$  \\
        & & & $4/3$ & & $ 1 $&$\hspace{-10pt} \ \mathrm{mod}\ 2$  \\
        \hline
        \multirow{4}*{$\mathbb{Z}_{6,IIc}$} & \multirow{4}*{$2$} & \multirow{4}*{$48$}
        & $4$ & \multirow{4}*{for $N_\mathrm{flux}/\ell  \cong \left\{ \begin{array}{ll}  \\  \\  \\ \end{array} \right.$ \hspace{-2cm}} & $ 0$ & $\hspace{-10pt} \ \mathrm{mod}\ 6$ \\
        & & & $12$ & & $ 1, 5 $&$ \hspace{-10pt} \ \mathrm{mod}\ 6$ \\
        & & & $3$ && $ 2, 4 $&$\hspace{-10pt} \ \mathrm{mod}\ 6$ \\
        & & & $4/3$ && $  3 $&$\hspace{-10pt} \ \mathrm{mod}\ 6$ \\
        \hline
        \multirow{2}*{$\mathbb{Z}_{6,IId}$} & \multirow{2}*{$1$} & \multirow{2}*{$12$} & $1$ &  \multirow{2}*{for $N_\mathrm{flux}/\ell \cong \left\{ \begin{array}{ll}  \\ \end{array} \right. $ \hspace{-15pt}}& $ 0$&$\hspace{-10pt} \ \mathrm{mod}\ 3$ \\
        & & & $3$ & &$ 1, 2 $&$\hspace{-10pt} \ \mathrm{mod}\ 3$ \\
        \hline
        \multirow{3}*{$\mathbb{Z}_{8,IIb}$} & \multirow{3}*{$2$} & \multirow{3}*{$32$} & $4$ & \multirow{3}*{for $N_\mathrm{flux}/\ell \cong \left\{ \begin{array}{ll} \\  \\ \end{array} \right. $\hspace{-15pt}}& $ 0 $&$ \hspace{-10pt} \ \mathrm{mod}\ 4$ \\
        & & & $8$ & & $  1, 3 $&$\hspace{-10pt} \ \mathrm{mod}\ 4$ \\
        & & & $2$ & & $  2 $&$\hspace{-10pt} \ \mathrm{mod}\ 4$ \\
        \hline
        $\mathbb{Z}_{12,II}$ & $2$ & $16$ & $4$ &&&\\
        \hline
        \multirow{2}*{$\mathbb{Z}_{2}\times\mathbb{Z}_{6,I}$} & \multirow{2}*{$1$} & \multirow{2}*{$12$} & $1$ &  \multirow{2}*{for $N_\mathrm{flux}/\ell \cong \left\{ \begin{array}{ll}\\  \end{array} \right. $\hspace{-15pt}}& $ 0 $&$\hspace{-10pt} \ \mathrm{mod}\ 3$ \\
        & & & $3$ & & $ 1, 2 $&$\hspace{-10pt} \ \mathrm{mod}\ 3$ \\
        \hline
    \end{tabular}
    \caption{Values of the parameters in the relation \eqref{eq:18} between $g_{s,\mathrm{min}}$ and $N_\mathrm{flux}$ for the toroidal orbifolds with one complex structure modulus of table \ref{tab:5}. See \eqref{firstdefl} for the definition of $\ell$.}
    \label{tab:7}
\end{table}

\paragraph{Tadpole constraint} As we did for the orientifolds with no complex structure modulus around equation \eqref{tadpoleconsh21_0}, we now discuss the tadpole condition \eqref{eq:13}. For the $x^i \rightarrow -x^i$ involution, the number of O3-planes are given in table \ref{tab:6}. In absence of negative D3-brane charge, the tadpole condition puts a constraint on the flux charge.
For instance, in the orientifold of $T^6/\mathbb{Z}_{4, a}$, containing 22 O3-planes, the constraint reads:
\begin{equation}\label{eq:19}
 N^Y_\mathrm{flux}=2 |G| N_{\rm flux} = 8 N_{\rm flux} \leq \frac{N_{O3}}{4} = \frac{11}{2}, \qquad  N_{\rm flux} = \ell n = 4 n,  \quad n\in \mathbb{N}^*.
\end{equation}
We remind that $N^Y_\mathrm{flux}=2|G|N_{\rm flux}=8N_{\rm flux}$, due to the quantisation of flux integers discussed around equation \eqref{eq:16}. On top of this,  we recalled the quantisation of $N_{\rm flux}$ by $\ell=4$ consequence of the orbifold projection. We deduce that the tadpole constraint cannot be satisfied in absence of negative D3-charge. 

The same goes for all the orientifolds of \cref{tab:7}.

\newpage

\section{The $T^6/\mathbb{Z}_2\times\mathbb{Z}_2$ orbifold: vacuum solutions and string coupling} \label{T6Z2xZ2}

The  $T^6/\mathbb{Z}_2\times\mathbb{Z}_2$ orbifold is unique on several respects. First, it is the only orbifold of our lists with $h^{2,1}=3$, hence three untwisted complex structure moduli $\mathcal{U}_i$, $i=1,2,3$. Second, adding discrete torsion transform all the twisted sector from K\"ahler to complex structure moduli. The latter can be stabilised with 3-form fluxes, leaving behind only three untwisted K\"ahler class moduli.  Third, its orientifold has the greatest number of O3-planes of the list, $N_{O3}=64$.

We treat this example in great detail and show explicitly that we can stabilise all complex structure moduli, untwisted and twisted, for certain choices of flux. We showcase the following relation between the minimal value of the string coupling and $N_{\rm flux}$:
\begin{equation}\label{eq:rel_12}
    g_{s, \mathrm{min}} \sim \frac{16}{N_\mathrm{flux}^2}.
\end{equation}
We also show that for this orbifold, this relation agrees with the one derived from the seminal works on flux vacua statistics \cite{Ashok:2003gk,Denef:2004ze}.

We proceed as follows.
We start by presenting explicitly how to stabilise the twisted moduli at the orbifold point, realising the method described in \cref{sec:6}. We then study analytic solutions to the equations $D_\mathcal{S}W = 0$, $D_{\mathcal{U}_i} W = 0$, ensuring scalar potential minimisation and stabilisation of the untwisted moduli. Next, we show evidence that the number of inequivalent vacua is finite for given value of $N_\mathrm{flux}$ and that they realise the relation \eqref{eq:rel_12}. We then point that in absence of negative D3-charge the tadpole condition cannot be satisfied. We eventually describe how to relax this constraint by introducing supersymmetry breaking magnetised D7-branes. 

\subsection{Twisted moduli stabilisation}\label{subsec:twistedmoduliZ2Z2}

We study the $T^6/\mathbb{Z}_2\times\mathbb{Z}_2$ orbifold with discrete torsion. According to \cref{tabHodgeDT}, it has $h^{2,1}=3$ untwisted complex structure moduli  $\mathcal{U}_i$, $i=1,2,3$, and $\tilde{h}^{2,1}=48$ twisted ones. The latter are denoted $\mathcal{D}^i_{\alpha\beta}$ with $i=1,2,3$ and $\alpha,\beta = 1, \dots, 4$ labelling the fixed points of the twist element $g_i=\theta,\theta',\theta\theta'$. They correspond to the elements of the cohomology basis shown in \cref{sec:6}.
 
The complex structure moduli K\"ahler potential can be expanded around the orbifold point as:
\begin{align}\label{eq:20}
    \mathcal{K} =& -\log\Big(-i(\mathcal{U}^1 - \mathcal{\bar U}^1)(\mathcal{U}^2 - \mathcal{\bar U}^2)(\mathcal{U}^3 - \mathcal{\bar U}^3)
    -\frac{i}{2}(\mathcal{U}^1 - \mathcal{\bar U}^1)(\mathcal{D}^1_{\alpha\beta} - \mathcal{\bar D}^1_{\alpha\beta})^2 \nonumber \\
    &-\frac{i}{2}(\mathcal{U}^2 - \mathcal{\bar U}^2)(\mathcal{D}^2_{\alpha\beta} - \mathcal{\bar D}^2_{\alpha\beta})^2
    -\frac{i}{2}(\mathcal{U}^3 - \mathcal{\bar U}^3)(\mathcal{D}^3_{\alpha\beta} - \mathcal{\bar D}^3_{\alpha\beta})^2 + \mathcal{O}(\mathcal{D}^4)\Big).
\end{align}
The untwisted sector part, depending only on the $\mathcal{U}^i$, can be obtained from the last term of \cref{Kahlerpot} once the $\Omega$ $(3,0)$-form is expressed in terms of the complex structure of \cref{tab:3}, following the method of \cref{orbiifoldconstruction}. In the expansion around the orbifold point, twisted moduli should appear in pairs. Indeed, they are acted upon by a discrete $\mathbb{Z}_2$ symmetry of the orbifold group \cite{Ferrara:1987jr}, and they  do not mix among themselves, which is reminiscent of the fact that the exceptional divisors of a $T^4/\mathbb{Z}_2$ do not intersect one another \cite{Blumenhagen:2002wn}. Their coefficients are related to those of the untwisted moduli \cite{Ferrara:1987jr}, ensuring the form of the expansion \eqref{eq:20}. 

From the above K\"ahler potential, we infer the corresponding prepotential through eq. \eqref{eq:prepotential}: 
\begin{equation}
    \mathcal{G}= \frac{X^1 X^2 X^3}{X^0}+\frac{1}{2}\frac{X^i(X^I)^2}{X^0}= \mathcal{U}^1\mathcal{U}^2\mathcal{U}^3 + \frac{1}{2} \mathcal{U}^i (\mathcal{D}^i_{\alpha\beta})^2 + \mathcal{O}(\mathcal{D}^4).
\end{equation}
In the first equality we kept the projective coordinates, used to compute the derivatives $\mathcal{G}_a$, while in the second we replaced them as explained below \cref{omegaXaGa} by $(X^0,X^i,X^I)=(1,\mathcal{U}^i,\mathcal{D}^i_{\alpha\beta})$.
This prepotential allows to compute the flux superpotential as expressed in \cref{superpotfromG}: 
\begin{align}\label{superpotZ2Z2}
    W =& n_0 + n_1\mathcal{U}^1 + n_2\mathcal{U}^2 + n_3\mathcal{U}^3 + m_1\mathcal{U}^2\mathcal{U}^3 + m_2\mathcal{U}^1\mathcal{U}^3 + m_3\mathcal{U}^1\mathcal{U}^2 - m_0\mathcal{U}^1\mathcal{U}^2\mathcal{U}^3 \nonumber\\
    &+ (n_{i, {\alpha\beta}} + m_{i, {\alpha\beta}}\mathcal{U}^i)\mathcal{D}^i_{\alpha\beta} + (m_i - m_0 \mathcal{U}^i)(\mathcal{D}^i_{\alpha\beta})^2.
\end{align}
The $n_{i, {\alpha\beta}}, m_{i, {\alpha\beta}}$ are the flux parameters on the twisted cycles parameterised by the twisted complex structure moduli  $\mathcal{D}^i_{\alpha\beta}$.

From this superpotential and the K\"ahler potential of \cref{eq:20}, we can derive the scalar potential \eqref{scalarpot2}. If terms of the scalar potential linear in the twisted complex structure moduli  $\mathcal{D}^i_{\alpha\beta}$ (and their conjugates $\bar{\mathcal{D}}^i_{\alpha\beta}$) all vanish at the same time, the twisted moduli are stabilised at the orbifold point, where they all vanish. This is achieved when taking vanishing fluxes on twisted cycles, as we show now. 

Take background fluxes such that $m_{i, {\alpha\beta}} = n_{i, {\alpha\beta}} = 0$, i.e.  with no component along the twisted 3-cycles. In that case, the superpotential \eqref{superpotZ2Z2} contains   quadratic but no linear terms in $\mathcal{D}^i_{\alpha\beta}$ (and conjugates), and so does the K\"ahler potential \eqref{eq:20}. This implies that terms in the scalar potential linear in $\mathcal{D}^i_{\alpha\beta}$ can only come from terms containing exactly one of the following terms: $\partial_{\mathcal{D}^i_{\alpha\beta}}W$, $\mathcal{K}_{\mathcal{D}^i_{\alpha\beta}}\equiv \partial_{\mathcal{D}^i_{\alpha\beta}} \mathcal{K}$ or $\mathcal{K}^{\mathcal{U}^i \, \bar{\mathcal{D}}^j_{\alpha\beta}}\equiv \partial^{\mathcal{U}^i \, \bar{\mathcal{D}}^j_{\alpha\beta}} \mathcal{K}$. The first two appear through the covariant derivatives $D_{\mathcal{D}^i_{\alpha\beta}}$ while the last is just a component of the inverse K\"ahler metric. Such terms are only included in the scalar potential \eqref{scalarpot2} as:
\begin{equation}
V \supset \mathcal{K}^{ \mathcal{D}^i_{\alpha\beta} \bar{\mathcal{D}}^j_{\gamma \delta}} (D_{\mathcal{D}^i_{\alpha\beta}} W )(\bar{D}_{\bar{\mathcal{D}}^j_{ \gamma\delta}} \bar{W}) + \mathcal{K}^{ \mathcal{U}^i \,\mathcal{D}^j_{\alpha \beta}} (D_{\mathcal{U}^i} W)( \bar{D}_{\bar{\mathcal{D}}^j_{ \alpha \beta}} \bar{W} ),
\end{equation}
so that they always come in pairs. Hence, in that case, the scalar potential does not contain linear terms in the twisted moduli $\mathcal{D}^i_{\alpha\beta}$ nor in their conjugate. It however contains positive quadratic terms, which shows that all twisted moduli are stabilised at the orbifold point:
\begin{equation}
\mathcal{D}^i_{\alpha\beta} =0, \qquad i=1,2,3, \quad \alpha,\beta=1,\ldots,4. \label{twistedOrbifoldpoint}
\end{equation}

To summarise, as advertised and previously used in the literature \cite{Cascales:2003zp}, taking vanishing fluxes along the twisted cycles $m_{i, {\alpha\beta}} = n_{i, {\alpha\beta}} = 0$ allows to stabilise the twisted moduli at the orbifold point. We have shown it explicitly. The remaining superpotential is just the one for the untwisted complex structure moduli, the stabilisation of which we study below. We conclude this section mentioning that twisted moduli can also be stabilised with non-vanishing twisted fluxes, albeit away from the orbifold point. This was done for instance in the large complex structures limit in \cite{Demirtas:2023als, Coudarchet:2023mmm} to study new flux vacua.

\subsection{Untwisted moduli stabilisation}\label{sec:2}

\paragraph{Solving the vacuum equations $D_{\mathcal{S},\mathcal{U}^i} W=0$ analytically} Once the twisted moduli are stabilised at the orbifold point \eqref{twistedOrbifoldpoint}, the flux-induced superpotential \eqref{superpotZ2Z2} reduces to:
\begin{equation}\label{eq:2}
    W = n_0 + n_1\mathcal{U}^1 + n_2\mathcal{U}^2 + n_3\mathcal{U}^3 + m_1\mathcal{U}^2\mathcal{U}^3 + m_2\mathcal{U}^1\mathcal{U}^3 + m_3\mathcal{U}^1\mathcal{U}^2 - m_0\mathcal{U}^1\mathcal{U}^2\mathcal{U}^3,
\end{equation}
We recall that the dependence in the axio-dilaton is contained in the flux parameters $m_a, n_a$ through $m_a = \mathcal{S}m_a^H + m^F_a$, \ldots, see \cref{eq:8}. The equations $D_\mathcal{S}W = 0, D_{\mathcal{U}^i} W = 0$ read:
\begin{eqnarray}\label{eq:equations_SUSY_2}
    \bar n_0 + \bar n_1\mathcal{U}^1 + \bar n_2\mathcal{\bar U}^2 + \bar n_3\mathcal{\bar U}^3 +
    \bar m_1\mathcal{\bar U}^2\mathcal{\bar U}^3 + \bar m_2\mathcal{U}^1\mathcal{\bar U}^3 
    + \bar m_3\mathcal{U}^1\mathcal{\bar U}^2 - \bar m_0\mathcal{U}^1\mathcal{\bar U}^2\mathcal{\bar U}^3 = 0, \nonumber\\
    n_0 + n_1\mathcal{U}^1 + n_2\mathcal{\bar U}^2 + n_3\mathcal{U}^3 +
    m_1\mathcal{\bar U}^2\mathcal{U}^3 + m_2\mathcal{U}^1\mathcal{U}^3 
    + m_3\mathcal{U}^1\mathcal{\bar U}^2 - m_0\mathcal{U}^1\mathcal{\bar U}^2\mathcal{U}^3 = 0, \nonumber\\
    n_0 + n_1\mathcal{U}^1 + n_2\mathcal{U}^2 + n_3\mathcal{\bar U}^3 +
    m_1\mathcal{U}^2\mathcal{\bar U}^3 + m_2\mathcal{U}^1\mathcal{\bar U}^3 
    + m_3\mathcal{U}^1\mathcal{U}^2 - m_0\mathcal{U}^1\mathcal{U}^2\mathcal{\bar U}^3 = 0, \nonumber\\
    \bar n_0 + \bar n_1\mathcal{U}^1 + \bar n_2\mathcal{U}^2 + \bar n_3\mathcal{U}^3 +
    \bar m_1\mathcal{U}^2\mathcal{U}^3 + \bar m_2\mathcal{U}^1\mathcal{U}^3 
    + \bar m_3\mathcal{U}^1\mathcal{U}^2 - 
    \bar m_0\mathcal{U}^1\mathcal{U}^2\mathcal{U}^3 = 0.
\end{eqnarray}
The symmetry of this system makes it solvable under certain conditions through the steps described below. As will be clear from the procedure, the necessary condition is that the solution has all imaginary parts of the complex structure moduli and axio-dilaton stabilised and non-vanishing. This condition corresponds to non-vanishing tori angles and string coupling constant and is thus a necessary assumption. It is also consistent with the definitions of the axio-dilaton $\mathcal{S}$ and complex structure moduli $\mathcal{U}_i$ K\"ahler potentials shown in \eqref{Kahlerpot} and \eqref{eq:20}. In our conventions we thus look for solutions satisfying:
\begin{equation}\label{priorassumptions}
\text{Im}(\mathcal{U}^i)<0, \quad \text{Im}(\mathcal{S})>0.
\end{equation}

To solve the system \eqref{eq:equations_SUSY_2}, we first notice that all four equations are linear in each moduli, e.g. in $\mathcal{U}^1$. They can thus be written as:
\begin{equation}\label{eq:1}
    L_k\times
    \left(\begin{array}{cc}
         1\\
         \mathcal{U}^1
    \end{array}\right)
    =0,
\end{equation}
where $L_k$ is a $1\times2$ matrix that depends on $(\mathcal{U}^2, \mathcal{U}^3)$. For instance,
\begin{equation}
    L_1 = (\bar n_0 + \bar n_2\mathcal{\bar U}^2 + \bar n_3\mathcal{\bar U}^3 +
    \bar m_1\mathcal{\bar U}^2\mathcal{\bar U}^3, \quad \bar n_1 + \bar m_2\mathcal{\bar U}^3 
    + \bar m_3\mathcal{\bar U}^2 - \bar m_0\mathcal{\bar U}^2\mathcal{\bar U}^3).
\end{equation}
With pairs of such $L_k$, we can form $2\times2$ matrices which,  by virtue of \eqref{eq:1}, have a vanishing eigenvalue with eigenvector $(1, \mathcal{U}^1)$. Their determinants thus vanish and can be combined to rewrite the system. One combination of such determinants turns out to be particularly useful: 
\begin{equation}
    \det\left(\begin{array}{cc}
         \bar L_1\\
         \bar L_4
    \end{array}\right)
    -
    \det\left(\begin{array}{cc}
         L_2\\
         L_3
    \end{array}\right)
    =0.
\end{equation}
The $\mathcal{U}^2$ dependence of this combination completely factorises as $(\mathcal{U}^2-\mathcal{\bar U}^2)$, which according to contributions \eqref{priorassumptions} is nonzero. We thus obtain a second order equation on $\mathcal{U}^3$ only
\begin{equation}\label{eq:U3}
    n_1n_2-m_3n_0 + (m_0n_0+m_1n_1+m_2n_2-m_3n_3)\mathrm{Re}(\mathcal{U}^3)+(m_0n_3+m_1m_2)|\mathcal{U}^3|^2=0.
\end{equation}
Similar equations can be obtained for $\mathcal{U}^1$ and $\mathcal{U}^2$ by the same procedure. We parameterise them as:
\begin{equation}\label{eq:equation_Ui}
    a_i + b_i \mathrm{Re}(\mathcal{U}^i) + c_i |\mathcal{U}^i|^2 = 0, 
\end{equation}
From the explicit equation \eqref{eq:U3}, we see that for $\mathcal{U}^3$ the $a_3,b_3, c_3$ parameters read:
\begin{equation}\label{eq:coefficients_abc}
    a_3 = n_1 n_2 - m_3 n_0, \quad b_3 = m_0 n_0 - m_3 n_3 + m_1 n_1 + m_2 n_2, \quad c_3 = m_0 n_3 + m_1 m_2.
\end{equation}
Similar expressions hold for  $a_1, b_1, c_1$ and $a_2, b_2, c_2$. Equation \eqref{eq:equation_Ui} is solved by:
\begin{equation}\label{eq:solution_Ui}
    x_i \equiv \text{Re}(\mathcal{U}^i)= \rho_i\cos\theta_i = -\frac{\mathrm{Im}(\bar a_i c_i)}{\mathrm{Im}(\bar b_i c_i)}, \qquad
    \rho_i^2 = |\mathcal{U}^i|^2= \frac{\mathrm{Im}(\bar a_i b_i)}{\mathrm{Im}(\bar b_i c_i)}, \qquad \mathcal{U}^i \equiv \rho_ie^{i\theta_i}.
\end{equation}
These solutions hold as long as the denominator $\mathrm{Im}(\bar b_i c_i)$ does not vanish. We come back to this point in the next paragraph. Note that $\mathcal{U}^i$ is uniquely determined, because we look for solutions with $\mathrm{Im}(\mathcal{U}^i) < 0$, see \cref{priorassumptions}, which uniquely determines $\theta_i$. Moreover, for this solution to make sense, we must ensure that $\rho_i > 0$ and $\rho_i^2 \geq x_i^2$.

So far, we have obtained the complex structure moduli $\mathcal{U}^i$ as functions of the axio-dilaton $\mathcal{S}$, through the expressions of $m_a, n_a$. We can thus obtain an equation on $\mathcal{S}$ by inserting the expressions of $\mathcal{U}^i$ in any of the initial equations \eqref{eq:equations_SUSY_2}. There is however a more convenient way to proceed, making use of the symmetry of the system. We rewrite the system \eqref{eq:equations_SUSY_2}  by making explicit the dependence in $\mathcal{S}$ and hiding the dependence in e.g. $\mathcal{U}^1$. Indeed, the superpotential \eqref{eq:2} can be rewritten as
\begin{equation}\label{eq:3}
    W = q_0 + q_1\mathcal{S} + q_2\mathcal{U}^2 + q_3\mathcal{U}^3 +
    p_1\mathcal{U}^2\mathcal{U}^3 + p_2\mathcal{S}\mathcal{U}^3 
    + p_3\mathcal{S}\mathcal{U}^2 - p_0\mathcal{S}\mathcal{U}^2\mathcal{U}^3,
\end{equation}
with
\begin{align}\label{eq:coefficients_pq}
    &p_0 = -m_1^H+m_0^H\mathcal{U}^1, \quad p_1 = m_1^F-m_0^F\mathcal{U}^1, \quad p_2 = n_3^H+m_2^H\mathcal{U}^1, \quad p_3 = n_2^H+m_3^H\mathcal{U}^1\nonumber\\
    &q_0 = n_0^F+n_1^F\mathcal{U}^1, \quad q_1 = n_0^H+n_1^H\mathcal{U}^1, \quad q_2 = n_2^F+m_3^F\mathcal{U}^1, \quad q_3 = n_3^F+m_2^F\mathcal{U}^1.
\end{align}
We obtain a system of the same form as the previous one, with $\mathcal{U}^1 \leftrightarrow \mathcal{S}$ and $(m_a, n_a) \leftrightarrow (p_a, q_a)$. Note that in our conventions, the K\"ahler potential is not invariant under the exchange $\mathcal{U}^1 \leftrightarrow \mathcal{S}$. We can solve this system as before and obtain an expression for $\mathcal{S}$ as a function of $\mathcal{U}^1$. Indeed, by defining:
\begin{equation}\label{eq:coefficients_stu}
    a_s = q_2 q_3 - p_1 q_0, \quad b_s = p_0 q_0 - p_1 q_1 + p_2 q_2 + p_3 q_3, \quad c_s = p_0 q_1 + p_2 p_3,
\end{equation}
we get the solution:
\begin{equation}\label{eq:solution_S}
    x_s \equiv \text{Re}({\cal S}) = \rho_s\cos\theta_s = -\frac{\mathrm{Im}(\bar a_s c_s)}{\mathrm{Im}(\bar b_s c_s)}, \qquad
    \rho_s^2 = |\mathcal{S}|^2= \frac{\mathrm{Im}(\bar a_s b_s)}{\mathrm{Im}(\bar b_s c_s)}, 
    \quad\text{where}\quad \mathcal{S} \equiv \rho_s e^{i\theta_s}.
\end{equation}
Here again, this holds for non-vanishing $\mathrm{Im}(\bar b_s c_s)$, see next paragraph. 

At this point, \cref{eq:coefficients_abc,eq:solution_Ui} provide $\mathcal{U}^1$ as a function of $\mathcal{S}$ while \cref{eq:coefficients_pq,eq:solution_S} provide $\mathcal{S}$ as a function of $\mathcal{U}^1$. Combining the two thus gives an equation for $\mathcal{S}$. But for that, we should make explicit the $\mathcal{S}$-dependence of $\mathcal{U}^1$ and vice versa. Upon inspection, we obtain that the imaginary parts appearing in \cref{eq:solution_Ui,eq:solution_S} all take the following form:
\begin{align}\label{eq:expression_Im}
    &\mathrm{Im}(\bar a_1 b_1) = y_s(N^0_{ab} + N^1_{ab}x_s + N^2_{ab}\rho_s^2), \quad
    &&\mathrm{Im}(\bar a_s b_s) = y_1(N^0_{ab} + 2N^0_{ac}x_1 + N^0_{bc}\rho_1^2), \nonumber\\
    &\mathrm{Im}(\bar a_1 c_1) = y_s(N^0_{ac} + N^1_{ac}x_s + N^2_{ac}\rho_s^2), \quad
    &&\mathrm{Im}(\bar a_s c_s) = \frac{1}{2} y_1( N^1_{ab} + 2 N^1_{ac}x_1 +  N^1_{bc}\rho_1^2), \nonumber\\
   & \mathrm{Im}(\bar b_1 c_1) = y_s(N^0_{bc} + N^1_{bc}x_s + N^2_{bc}\rho_s^2), \quad
   && \mathrm{Im}(\bar b_s c_s) = y_1(N^2_{ab} + 2N^2_{ac}x_1 + N^2_{bc}\rho_1^2).
\end{align}
Similarly to the real parts $x_s, x_i$ introduced before,  $y_s,y_i$ are the imaginary parts of $\mathcal{S}, \mathcal{U}^i$. The $N$'s are combinations of the integers $m_a^H, n_a^H, m_a^F, n_a^F$ defined as:
\begin{eqnarray} 
    N^0_{ab} &\equiv& a_1^{FF}b_1^{HF} - a_1^{HF}b_1^{FF},\qquad
    N^1_{ab} \equiv 2(a_1^{FF}b_1^{HH} - a_1^{HH}b_1^{FF}),\nonumber\\
    N^2_{ab} &\equiv& a_1^{HF}b_1^{HH} - a_1^{HH}b_1^{HF},
\end{eqnarray}
and similarly for the $N_{ac}$ and $N_{bc}$. We used the following notation, derived naturally from the definition \eqref{eq:coefficients_abc} of the parameters $a_i$:
\begin{eqnarray}
    &&a_1^{FF} = n_2^Fn_3^F - m_1^Fn_0^F, \qquad a_1^{HH} = n_2^Hn_3^H - m_1^Hn_0^H, \nonumber\\
    &&a_1^{HF} = n_2^Hn_3^F + n_2^Fn_3^H - m_1^Hn_0^F - m_1^Fn_0^H,
\end{eqnarray}
and similarly for the $b_i$ and $c_i$. Combining \eqref{eq:solution_Ui}, \eqref{eq:solution_S} and \eqref{eq:expression_Im}, we then obtain:
\begin{equation}\label{eq:system_xs-rhos}
    x_s = -\frac{A_0 + A_1 x_s + A_2 \rho_s^2}{B_0 + B_1 x_s + B_2 \rho_s^2}, \qquad
    \rho_s^2 = \frac{C_0 + C_1 x_s + C_2 \rho_s^2}{B_0 + B_1 x_s + B_2 \rho_s^2},
\end{equation}
with
    \begin{alignat}{6}
   & A_0 = \vec{A} \cdot \vec{N}_0 \quad && A_1 = \vec{A} \cdot \vec{N}_1, \quad && A_2 = \vec{A} \cdot \vec{N}_2,  \quad && B_0= \vec{B} \cdot \vec{N}_0 , \quad &&B_1= \vec{B} \cdot \vec{N}_1 ,  \quad && B_2 = \vec{B} \cdot \vec{N}_2, \nonumber\\
   & C_0 = \vec{C} \cdot \vec{N}_0, && C_1 =  \vec{C} \cdot \vec{N}_1, && C_2 = \vec{C} \cdot \vec{N}_2,  && && &&  \end{alignat}
    computed from the vectors
    \begin{alignat}{3}
   & \vec{A}\equiv \frac{1}{2}(N^1_{ab},-2N^1_{ac},N^1_{bc}), \qquad &&  \vec{B}\equiv (N^2_{ab},-2N^2_{ac},N^2_{bc}), \qquad && \vec{C}\equiv (N^0_{ab},-2N^0_{ac},N^0_{bc}), \\
   & \vec{N}_0\equiv(N^0_{bc},N^0_{ac},N^0_{ab}),&&  \vec{N}_1\equiv(N^1_{bc},N^1_{ac},N^1_{ab}), &&  \vec{N}_2\equiv(N^2_{bc},N^2_{ac},N^2_{ab}).
    \end{alignat}
The $A_i$, $B_i$ and $C_i$ are, again, integers expressed as combinations of the $m_a^H, n_a^H,m_a^F, n_a^F$, obtained combining the previous formulae. Their complete expressions are horrendous. 

\paragraph{Solving for $\mathcal{S}$ } To have a well-defined solution for $\mathcal{S}$ , the denominator appearing in both equations of the system \eqref{eq:system_xs-rhos} shall not be vanishing. The system is then equivalent to:
\begin{align}
B_1 x_s^2 + B_2 x_s \rho_s^2 + (A_1 + B_0) x_s + A_2 \rho_s^2 + A_0 =0, \\
B1 x_s \rho_s^2 + B_2 \rho_s^4 -C_1 x_s + (B_0-C_2) \rho_s^2  -C_0=0. \label{system_conics}
\end{align}
This is a polynomial system of degree two in the two variables $X=x_s$ and $Y=\rho_s^2$. Each of them can be seen as the parameterisation in Cartesian coordinates of a conic section: ellipse, hyperbola or parabola. 
Solving the system \eqref{system_conics} thus amounts to find the real intersections of two conic sections. This is a classical algebraic geometry problem \cite{Richter:2011}. There can be at most four such intersections, except in the case where the two conics are degenerate and share a common line.

Conics are degenerate when their defining second degree polynomial factorises in two first degree ones. Such conics are then unions of two lines. These lines can be coincident or distinct, parallel or intersecting. In the most degenerate case, the conic is just a point, the centre of a degenerate ellipse. Two degenerate conics, hence two pairs of lines, can have an infinite number of intersection points if they share a line. These cases correspond physically to vacua with a flat direction, hence an axio-dilaton not fully stabilised. In such cases the flat direction is parametrised by a line equation, which factorises both equations of \cref{system_conics}.

We wrote an algorithm to solve \eqref{system_conics} in the general case, thus computing the intersection points of the two arbitrary conic sections. For any conic coefficients, thus flux quanta, our algorithm finds the intersection points, and rules out cases with physical flat directions. Our algorithm follows the general procedure detailed in \cite{Richter:2011}.

 In the majority of cases, solutions of the system \eqref{system_conics} can also be derived more directly. Indeed, one can use the first equation to express  $\rho_s^2$ as a function of $x_s$. Inserting this expression in the second equation of  \eqref{eq:system_xs-rhos} yields a polynomial in $x_s$ of degree (at most) three. When the polynomial admits real roots, they can be used back in the first equation of \cref{eq:system_xs-rhos} to find  $\rho_s^2$. However, such method does not  consider  degenerate cases, and one should stick to the general procedure to avoid missing  solutions.

Solutions $(x_s,\rho_s^2)$ of the system \eqref{system_conics}, found through our algorithm, constitute vacuum solutions only if they satisfy additional constraints: $\rho_s^2$ should be positive and greater than $x_s^2$. In this case, ${\rm Im}(\mathcal{S})$ is obtained directly from $\rho_s$ and $x_s$, together with the condition ${\rm Im}(\mathcal{S})>0$ of \cref{priorassumptions}. This solution for $\mathcal{S}$ can then be injected in the solution \eqref{eq:solution_Ui} to obtain $\rho_i$ and $x_i$, and then $\mathcal{U}^i$. Here again they should satisfy consistency conditions similar to those of $\mathcal{S}$, explained below \cref{eq:solution_Ui}. When all these conditions are fulfilled, we end up with a complete solution of the system \eqref{eq:equations_SUSY_2}.

\paragraph{Some comments} Let us make two important comments on the solution we obtained.  First, as already mentioned, the denominators in \cref{eq:solution_Ui,eq:solution_S} must be non-vanishing for the solutions to be well defined. In cases where a denominator vanishes, the solution breaks down, and a special treatment is needed. Actually, this only happens when a modulus is unstabilised. Indeed, the vanishing of the denominator is a condition on the flux integers, which removes the dependence of the superpotential in the real or imaginary part of one of the moduli. For instance, if $b_1 = 0$, the denominator of $\eqref{eq:solution_Ui}$ vanishes. However as $\mathrm{Re}(\mathcal{U}^1)$ drops out of the equation \eqref{eq:equation_Ui}, $\mathrm{Re}(\mathcal{U}^1)$ ends up unstabilised.

Having unstabilised real parts of complex structure moduli is not necessarily problematic. For instance, the unstabilised real part of the modulus associated to a D-brane anomalous $U(1)$ can be eaten by the gauge boson. This scenario implies K\"ahler moduli or the axio-dilaton. If such mechanism occurs for the latter, we should  also consider cases where $x_s$ is not stabilised by the fluxes. In the unitary gauge, we could then impose $x_s = 0$ to remove the dependence of the superpotential in this variable, as discussed above. In the results presented in the following sections, this possibility was not considered. As far as we checked it does not significantly affects our result. It introduced a few additional vacua and hence changed slightly some of the values reported in the following tables. However, all the conclusions remained the same.

Second, the solution involves solving a third order polynomial equation on either $x_s$ or $\rho_s^2$. Such equations can have up to three real solutions, so it seems that some choices of fluxes can lead to multiple vacua. However, the system \eqref{eq:system_xs-rhos} has to be supplemented by constraints ensuring that $x_s$ and $\rho_s^2$ are the real part and the square modulus of $\mathcal{S}$. In particular, we must impose $\rho_s^2 \geq x_s^2$ on the solution. The same goes with $x_i$ and $\rho_i^2$ in \eqref{eq:solution_Ui}. After imposing these constraints, we did not find any choice of fluxes leading to multiple solutions.

\paragraph{Dualities} As for the orbifolds with $h^{2,1}=0,1$, we now study the possibility to go from one solution to another using duality transformations. We reintroduce ${S}$-duality transformations shown in \cref{eq:rel_8}:
\begin{equation}\label{eq:Sdualitbis}
    \mathcal{S} \quad\rightarrow\quad \frac{a\mathcal{S} + b}{c\mathcal{S}+d}
    \qquad\text{and}\qquad
    \begin{pmatrix}
    H_3\\F_3
    \end{pmatrix}
    \quad\rightarrow\quad
    \begin{pmatrix}
    d & -c \\ -b & a
    \end{pmatrix}
    \begin{pmatrix}
    H_3\\F_3
    \end{pmatrix},
\end{equation}
and recall that they leave  $N_\mathrm{flux}$ unchanged. We can also explicitly check on the solutions for $\mathcal{U}^i$ given by \eqref{eq:solution_Ui} that the complex structure moduli are left invariant. This is expected since the theory, and thus equations \eqref{eq:equations_SUSY_2} are themselves invariant. It is a however a non-trivial consistency check.

Let us recall that in orbifolds with one complex structure modulus described in section \ref{sec:1}, the $\mathcal{S}\leftrightarrow\mathcal{U}$ symmetry of the superpotential \eqref{eq:17} trades ${S}$-duality with a ${U}$-duality \eqref{Tduality1} acting on $\mathcal{U}$. For our current orbifold $T^6/\mathbb{Z}_2\times\mathbb{Z}_2$, we encountered a similar symmetry when deriving the analytic solution above. The superpotential is indeed symmetric under $\mathcal{U}^1 \leftrightarrow \mathcal{S}$ and $(m_a, n_a) \leftrightarrow (p_a, q_a)$, see \eqref{eq:3}. This later exchange trades ${S}$-duality with a ${U}_1$-duality acting on $\mathcal{U}^1$ just like \eqref{Tduality1} and leaving $(\mathcal{S}, \mathcal{U}^2, \mathcal{U}^3)$ invariant. Similarly, there exist  ${U}_2$ and ${U}_3$-dualities acting only on $\mathcal{U}^2$ and $\mathcal{U}^3$ respectively. Using all these dualities, we can bring the axio-dilaton $\mathcal{S}$ and all of the complex structure moduli $\mathcal{U}^i$ to their fundamental domain independently:
\begin{equation}\label{funamentaldomains}
    \mathcal{F} = \{-1/2 < \mathrm{Re}(\mathcal{A}) \leq 1/2 \quad\text{and}\quad |\mathcal{A}| \geq 1 \} \quad\text{with}\quad \mathcal{A} = \mathcal{S} \text{ or } \mathcal{U}^i.
\end{equation}

\subsection{Exhausting the finite number of vacua}\label{sec:5}

\paragraph{Counting vacua algorithmically} Once the background fluxes are set to zero on the deformations cycles, thus setting the twisted moduli VEVs to zero, see \cref{twistedOrbifoldpoint}, we are left with the choice of $16$ free flux integers $m_i^F,m_i^H,n_i^F, n_i^H$, $i=0,\ldots,3$. Indeed, contrary to the cases with $h^{2,1}=0,1$ studied in \cref{sec:3,sec:gsNflux}, the presence of  three untwisted complex structure moduli does not introduce any relation between these integers. We recall that $N_{\rm flux}$ is computed from unquantised, thus arbitrary, flux integers.  The orientifold  $N^Y_{\rm flux}$ appearing in the tadpole condition is however multiple of $N_{\rm flux}$, thus taking the quantisation of integers into account, see \cref{relNy}.

 We searched for solutions  proceeding as follows. For a given $N_{\rm flux}$, we chose flux integers in a certain range $k$:
\begin{equation}
|m_i^F|, |m_i^H|, |n_i^F|, |n_i^H|\leq k, \quad \textrm{for a certain fixed range } k \in \mathbb{N} \label{rangek}.
\end{equation}
We constructed algorithmically all combinations of integers in the range $k$ giving flux number $N_{\rm flux}$. For each such flux quanta combination, we computed the coefficients of the systems \eqref{eq:system_xs-rhos} and \eqref{eq:solution_Ui} determining the vacuum solutions, as explained in the previous section. We solved the system \cref{eq:system_xs-rhos} following the procedure described around \cref{system_conics}. When such solution gave a physical vacuum, we brought the moduli and axio-dilaton in their fundamental domains using dualities.

Note that the combinations of flux integers in a range $k$ giving $N_{\rm flux}$ are just a fraction of all combinations of integers in the range $k$. This helps decreasing the number of combinations to be plugged in the analytic solutions. Nevertheless, when increasing the allowed range $k$ for a given $N_\mathrm{flux}$, there is still  a large and rapidly growing number of combinations. Even for solutions preserving some symmetry between the different tori of $T^6 = T_1^2\times T_2^2 \times T_3^2$, this number stays large. We illustrate these facts in \cref{tab:rel_2}, showing the number of combinations of integers giving $N_{\rm flux}=4$ for ranges $k\leq7$. 

\begin{table}[ht!]
\vspace{0.3cm}
    \centering
    \begin{tabular}{|c||r|r|r|}
    \hline
         & \multicolumn{3}{c|}{\# of flux integer combinations, with:} \\         
        $k$ & three eq. tori & two eq. tori & no symmetry \\
         \hline\hline
         $1$ & $416$ & $21492$ & $975968$ \\
         $2$ & $18276$ & $14325076$ & $8726828016$ \\
         $3$ & $99428$ & $386921556$ & $1099101964400$ \\
         $4$ & $622732$ & $5566156388$ & $\cdots$  \\
         $5$ & $1999388$ & $37625301028$ &  \\
         $6$ & $4905948$ & $213491079460$ &\\
         $7$ & $11893404$ &  $\cdots$ &  \\
         %&$8$ & $28118812$ &  &  \\
         %&$9$ & $47647484$ &  &  \\
         %&$10$ & $94295084$ &  &  
         \hline
    \end{tabular}
    \caption{Number of combinations giving $N_\mathrm{flux} = 4$ for flux integers in the range $k$, see \eqref{rangek}, depending on the symmetry between the three tori.}
        \label{tab:rel_2}
\end{table}

The actual number of choices can be reduced by taking into account symmetries of the system. For instance, solutions can be related by reversing the sign of all the flux integers at once. The numbers  of  \cref{tab:rel_2} are thus upper bounds, which however make clear that algorithmic explorations are challenging. We will show evidence that the number of inequivalent vacua for fixed $N_\mathrm{flux}$  is nevertheless finite, and that these vacua realise the relation \eqref{eq:rel_12}. 

\paragraph{Three equivalent tori} Let us start with the symmetric case where the three tori are equivalent. In this case, it is still manageable to compute the analytic solutions for every combination of integers, with given $N_\mathrm{flux}$ and flux integers in the range $k$ defined in \cref{rangek}. For each solution, we bring both $\mathcal{S}$ and $\mathcal{U}^i$ in the fundamental domain \eqref{funamentaldomains}. We then simply count the number of distinct $(\mathcal{S}, \mathcal{U}^i)$ obtained in this way. In \cref{tab:1}, we report the number of inequivalent symmetric vacua found in $T^6/\mathbb{Z}_2\times\mathbb{Z}_2$. Note that our results approximatively match the formula (3.28) of \cite{Denef:2004ze} for the number of supersymmetric vacua of $T^6/\mathbb{Z}_2$ with equivalent tori. According to  \cite{Denef:2004ze}, for $T^6$ with factorised identical tori the number of SUSY flux vacua follows the relation:
\begin{equation}
\mathcal{N}_{SUSY}\approx \, \alpha \, N_{\rm flux}^{K} \approx \, \alpha \, N_{\rm flux}^{4}, \qquad \qquad \alpha=\frac{\pi^4}{5184}.  \label{estimDD}
\end{equation} 
The exponent $K=2m=4$ in the above formula is given by 1+ the number of moduli, which in the case of equivalent tori gives indeed $m=1+1=2$. The coefficient $\alpha$ is found in terms of the intersection form on a cohomology basis and of derivatives of the holomorphic form of the compact space, see \cite{Denef:2004ze}. The value of $\alpha$ given above corresponds to the particular $T^6/\mathbb{Z}_2$ example.
\begin{table}[ht!]
    \centering
   \vspace{0.3cm}
\hspace{2cm} \# solutions imposing three equivalent tori\\
\vspace{5pt}
    \begin{tabular}{|l||c|c|c|c|c|c|c|c|c|c|}
    \hline
         \diagbox[width=5em,height=2.3em,innerleftsep=10pt,innerrightsep=0pt]{$k $}{$N_\mathrm{flux}$  }  &  1 & 2 & 3 & 4 & 5 & 6 & 7 & 8 & 9 & 10 \\
         \hline\hline
         $\hspace{5pt}1$ & $\underline{0}$ & $\underline{0}$ & ${0}$ & $5$ & $7$ & $1$ & $5$ & $1$ & $0$ & $0$\\
         $\hspace{5pt}2$ & $0$ & $0$ & $\underline{1}$ & $\underline{6}$ & $\underline{13}$ & $28$ & $56$ & $82$ & $101$ & $114$\\
         $\hspace{5pt}3$ & $0$ & $0$ & $1$ & $6$ & $13$ & $\underline{30}$ & $\underline{65}$ & $\underline{102}$ & $168$ & $234$\\
         $\hspace{5pt}4$ & $0$ & $0$ & $1$ & $6$ & $13$ & $30$ & $65$ & $102$ & $\underline{171}$ & $\underline{252}$\\
         $\hspace{5pt}5$ & $0$ & $0$ & $1$ & $6$ & $13$ & $30$ & $65$ & $102$ & $171$ & $252$\\
         $\hspace{5pt}6$ & $0$ & $0$ & $1$ & $6$ & $13$ & $30$ & $65$ & $102$ & $171$ & $252$\\
         \hline
    \end{tabular}
    \caption{Number of inequivalent flux vacua for $T^6/\mathbb{Z}_2\times\mathbb{Z}_2$, as a function of $N_\mathrm{flux}$ and the flux integers range $k$, see \eqref{rangek}, for solutions found by imposing three equivalent tori. For each $N_\mathrm{flux}$ we underlined the maximal number of vacua obtained with the smallest range $k^{N_{\rm flux}}_{\rm max \#}$ (e.g. $k^8_{\rm max \#}=4$). Greater ranges $k\geq k^{N_{\rm flux}}_{\rm \rm max \#}$ lead to the same number of solutions.}
    \label{tab:1}
\end{table}

In table \ref{tab:1} we also observe the behaviour mentioned in the previous sections: for given $N_\mathrm{flux}$, when increasing the allowed range $k$ for the integers, we reach a value  $k^{N_{\rm flux}}_{\rm \rm max \#}$ above which we find no new vacua. Even if the table stops at $k = 6$, we  checked beyond this value, e.g. up to $k=30$ for $N_\mathrm{flux} = 4$. This allows us to claim that we have found all the vacua for this value of $N_\mathrm{flux}$, unless a great hierarchy is present between the flux parameters, see \cref{sec:parametriccontrol} for discussion on this point. Note that all the vacua are often found for small $k^{N_{\rm flux}}_{\rm \rm max \#}$. We comment on a subtlety here: as explained above, we found vacua by looking at integer combinations in a range $k$ with given $N_{\rm flux}$, and then bringing the moduli in their fundamental domains by dualities. The integers transform under these dualities as shown in \cref{eq:Sdualitbis}, such that they might exit the range $k$. The ranges $k$ in the tables of this section are thus to be understood as the minimal ranges along the duality orbits. 

\begin{figure}[ht!]
\begin{center}
\includegraphics[scale=.7]{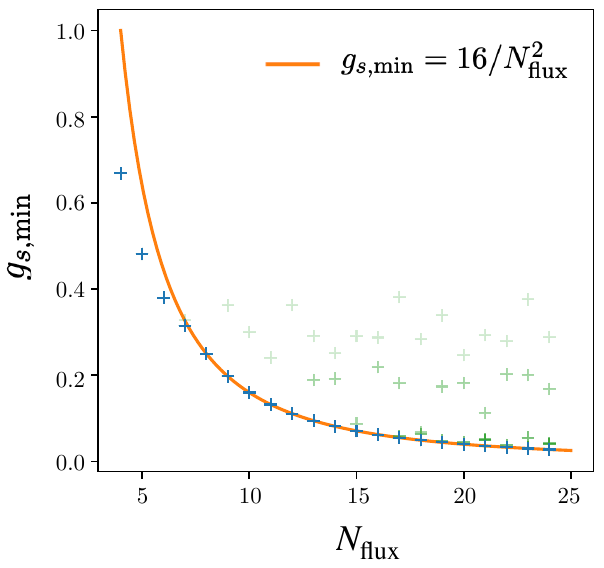}
\end{center}
\vspace{-0.6cm}
\caption{Minimal string coupling $g_{s,\mathrm{min}}$ as a function of $N_\mathrm{flux}$, for our vacuum solutions. The blue points are the exact values found as explained in the text. The orange curve is the fit $g_{s,\mathrm{min}}= 16/N_\mathrm{flux}^2$. The green shaded points are the values $g_{s,\mathrm{min}}$ found with lower ranges of integers. Increasing this range, the points converge towards the orange curve.}
\label{fig:2}
\end{figure}

The fact that $g_s$ is bounded, as advertised in \eqref{eq:rel_12}, is thus a byproduct of the finiteness of the number of vacua. In table \ref{tab:rel_3}, we give the values of $1/g_{s, \mathrm{min}}$ as a function of $N_\mathrm{flux}$ and the range $k$ for the integers.  Here again,  for a given $N_{\rm flux}$ the minimal value is reached at certain range $k^{N_{\rm flux}}_{\rm \,min}$, and above this range no new minimal value is found. When $N_\mathrm{flux} = 4p$ for integer $p\geq 2$, we find the exact formula:
 \begin{equation}
 g_{s, \mathrm{min}} = \frac{16}{N_\mathrm{flux}^2} \qquad {\rm for} \quad N_\mathrm{flux}= 4 p, \quad p\in \mathbb{N}\setminus \{0,1\}. \label{exactgsminNflux4p}
 \end{equation}
It obviously realises  the relation \eqref{eq:rel_12}. For other values of $N_\mathrm{flux}$, this formula does not hold exactly, but still gives a very good fit, exception made of the very first values of $N_\mathrm{flux}$. This is shown in figure \ref{fig:2}. Such conclusions could be infer from past work on flux vacua statistics \cite{Ashok:2003gk}. Imposing bounds on the coupling constant amounts to reduce the integration domain of the axio-dilaton integral appearing in their computation of the number of vacua. They thus obtain:
\begin{equation}
\mathcal{N}_{SUSY\,\, g_s^2 < \epsilon} \sim 3 \epsilon \mathcal{N}_{SUSY}.
\end{equation} 
Using this formula, the minimum value of $g_s$ can be estimated by finding $\epsilon_{\rm min}=g^2_{s, \, \rm min \, AD}$ such that $\mathcal{N}_{SUSY\,\, g_s^2 < \epsilon_{\rm min}}=1$, giving:
\begin{equation}
g^2_{s, \, \rm min \, AD}=\epsilon_{\rm min}=\frac{1}{3  \mathcal{N}_{SUSY}} \sim \frac{1}{3 \alpha N_{\rm flux}^4}, \label{estimationAD}
\end{equation} 
where in the last line we used the estimate \eqref{estimDD} of the number of vacua. We see that this estimate, derived from the flux vacua statistics of \cite{Ashok:2003gk,Denef:2004ze}, is consistent with our result \eqref{exactgsminNflux4p}.
\begin{table}%[ht!]
    \centering 
    \hspace{2cm} $1/g_{s, \rm min}$ for solutions imposing three equivalent tori\\
    \vspace{5pt}
        \begin{tabular}{|l||c|c|c|c|c|c|c|c|}
    \hline
         \diagbox[width=5em,height=2.3em,innerleftsep=10pt,innerrightsep=0pt]{$k $}{$N_\mathrm{flux}$  }  &3 &  4 &5 & 6 & 7 & 8 & 9 & 10 \\
         \hline\hline
         $\hspace{5pt}1$ & none  & $1.148$ & $1.263$ & 1.520 & $1.990$ & $1.000$ & none & none\\
         
         $\hspace{5pt}2$ & $\underline{0.8660} $ & $\underline{1.495}$ & $\underline{2.076}$ & $\underline{2.632}$ & $3.056$ & $\underline{4.000}$ & $2.760$ & $3.331$\\
         
         $\hspace{5pt}3$ & $0.8660 $ & $1.495$ & $2.076$ & $2.632$ & $\underline{3.175}$ & $4.000$ & $\underline{5.061}$ & $\underline{6.249}$\\
         
         $\hspace{5pt}4$ & $0.8660 $ & $1.495$ & $2.076$ & $2.632$ & $3.175$ & $4.000$ & $5.061$ & $6.249$\\
         
         $\hspace{5pt}5$ &  $0.8660 $ &$1.495$ & $2.076$ & $2.632$ & $3.175$ & $4.000$ & $5.061$ & $6.249$ \\
         %&$6$ & $1.495$ & $2.076$ & $2.632$ & $3.175$ & $4.000$ & $5.061$ & $6.249$ \\
                  \hline
    \end{tabular}
    \caption{Values of $1/g_{s, \mathrm{min}}$ for $T^6/\mathbb{Z}_2\times\mathbb{Z}_2$, as a function of $N_\mathrm{\rm flux}$ and the flux integers range $k$,  see \eqref{rangek}, for solutions found by imposing three equivalent tori. For each $N_{\rm flux}$, we underlined the maximal inverse string coupling constant $1/g_{s, \mathrm{min}}$ obtained with the smallest range $k^{N_{\rm flux}}_{\rm \,min}$ (e.g. $k^7_{\rm min}=3$). Greater ranges $k\geq k^{N_{\rm flux}}_{\rm \,min}$ lead to the same value.}
    \label{tab:rel_3}
\end{table}

\newpage

\paragraph{Only two equivalent tori} So far, we discussed solutions preserving three equivalent tori. Relaxing this condition, we still expect to find a finite number of inequivalent vacua at given $N_\mathrm{flux}$. This number is higher, since relaxing the symmetry between the three tori allows for more freedom. In table \ref{tab:8}, we show the number of vacuum solutions preserving only two equivalent tori. 
\begin{table}[ht!]
    \centering
\hspace{2cm} \# solutions imposing only two equivalent tori\\
\vspace{5pt}
    \begin{tabular}{|l||c|c|c|c|c|c|c|c|c|c|}
             \hline
         \diagbox[width=5em,height=2.3em,innerleftsep=10pt,innerrightsep=0pt]{$k $}{$N_\mathrm{flux}$  }  &  1 & 2 & 3 & 4 & 5 & 6 & 7 & 8 & 9 & 10 \\
         \hline\hline
         $\hspace{5pt}1$ & $0$ & $0$ & $\underline{1}$ & $9$ & $66$ & $16$ & $17$ & $1$ & $0$ & $0$\\
         $\hspace{5pt}2$ & $0$ & $0$ & $1$ & $\underline{11}$ & $\underline{94}$ & $408$ & $1182$ & $2730$ & $4891$ & $6899$ \\
          $\hspace{5pt}3$ & $0$ & $0$ & $1$ & $11$ & $94$ & $412$ & $1249$ & $3051$ & $6881$ & $13388$ \\
         %& $4$
         \hline
    \end{tabular}
    \caption{Number of inequivalent flux vacua for $T^6/\mathbb{Z}_2\times\mathbb{Z}_2$, as a function of $N_\mathrm{flux}$ and the allowed flux integers range $k$,  see \eqref{rangek}, for solutions found by imposing two equivalent tori. For each $N_\mathrm{flux}$ we underlined the maximal number of vacua obtained with the smallest range $k^{N_{\rm flux}}_{\rm max \#}$ ($e.g.$ $k^5_{\rm max \#}=2$). Greater ranges $k\geq k^{N_{\rm flux}}_{\rm \rm max \#}$ lead to the same number of solutions.}
    \label{tab:8}
\end{table}
In principle, the values of $g_{s, \mathrm{min}}$ could differ from the case imposing three equivalent tori. As we can see in table \ref{tab:9}, this is the case for some values of $k$, but as far as we explored, the absolute $g_{s, \mathrm{min}}$ for given $N_\mathrm{flux}$ remains the same as in the symmetric case in table \ref{tab:rel_3}. This suggests that  $g_{s,\rm min}$ is obtained from solutions with three equivalent tori, even looking for solutions imposing only two equivalent ones.  We checked  that explicitly in the cases with $N_{\rm flux} \leq 10$. This would mean that the relation between $g_{s, \mathrm{min}}$ and $N_\mathrm{flux}$ remains unchanged even when not imposing symmetry in the solution among the three tori. The number of choices of integers  is however too large to check this explicitly for larger values of $N_\mathrm{flux}$.
\begin{table}[ht!]
    \centering
    \hspace{1.7cm} $1/g_{s,\rm min}$ for solutions imposing only two equivalent tori\\
    \vspace{5pt}
    \begin{tabular}{|l||c|cc|c|c|c|c|c|}
             \hline
         \diagbox[width=5em,height=2.3em,innerleftsep=10pt,innerrightsep=0pt]{$k $}{$N_\mathrm{flux}$  } &3 &  4 &5 & 6 & 7 & 8 & 9 & 10 \\
         \hline\hline
         $\hspace{5pt}1$ & $\underline{0.8660 }$ &$1.148$ & $1.654$ & $1.925$ & $1.990$ & $1.000$ & none & none\\
         $\hspace{5pt}2$ &$0.8660 $ & $\underline{1.495}$ & $\underline{2.076}$ & $\underline{2.632}$ & $3.056$ & $\underline{4.000}$ & $5.032$ & $5.396$\\
          $\hspace{5pt}3$ &$0.8660 $ & $1.495$ & $2.076$ & $2.632$ & $3.175$ & $4.000$ & $5.061$ & $6.249$\\
   %& $4$
            \hline
    \end{tabular}
    \caption{Values of $1/g_{s, \mathrm{min}}$ for $T^6/\mathbb{Z}_2\times\mathbb{Z}_2$ as a function of $N_\mathrm{flux}$ and the range $k$ of the integers, for solutions found imposing two equivalent tori. For each $N_{\rm flux}$, we underlined the maximal inverse string coupling constant $1/g_{s, \mathrm{min}}$ obtained with the smallest range $k^{N_{\rm flux}}_{\rm \,min}$ (e.g. $k^8_{\rm min}=2$). Greater ranges $k\geq k^{N_{\rm flux}}_{\rm \,min}$ lead to the same value.}
    \label{tab:9}
\end{table}

Let us eventually comment that, in the present case with only two equivalent tori, some ``equivalent'' vacua give different values of the invariant $|\tilde W_0|$ defined in \cref{invariantsuperpot}. In other words, different flux quanta gives the same VEVs for the axio-dilaton and the two independent complex structure moduli, with however different $|\tilde W_0|$. With our way of counting vacua, introduced at the beginning of the section, such vacua are called ``equivalent'', whereas they can have different physical properties. This point does not affect our conclusions on the minimal string coupling nor the exhaustion of vacua for given flux charge $N_{\rm flux}$ and range $k$, as it only concerns the way of counting them.

\paragraph{Tadpole constraint and minimal $g_s$ in absence of negative D3-charge} We continue this discussion by studying the tadpole condition \eqref{eq:13}. For the $x^i \rightarrow -x^i$ involution, we read the number of O3-planes from table \ref{tab:6}, $N_{O3}=64$. In absence of negative D3-brane charge, the tadpole condition thus puts the following constraint on the flux charge:
\begin{equation}
N^Y_\mathrm{flux} = 2 |G| N_{\rm flux} = 8 N_{\rm flux} \leq \frac{N_{O3}}{4} =16, \qquad  N_{\rm flux} = n, \quad n \in \mathbb{N}^*. \label{tadpoleconstraintZ2Z2}
\end{equation}
We recall that $N^Y_\mathrm{flux}=2|G|N_{\rm flux}=8N_{\rm flux}$ due to the integer quantisation discussed around equation \eqref{eq:16}. 
In absence of negative D3 charge, the tadpole condition \cref{tadpoleconstraintZ2Z2} constrains the flux charge $N_{\rm flux}\leq 2$. As can be seen from \cref{tab:1,tab:8}, there are no solutions below this bound. 
We insist on the fact that this conclusion is reached in constructions without negative D3-charges. In \cref{subsec:magnetizedbranes} we show how it can be lifted in presence of negative D3-charges induced by magnetised D7-branes.

\begin{figure}[ht!]
\vspace{1.5cm}
\begin{center}
\includegraphics[scale=.87]{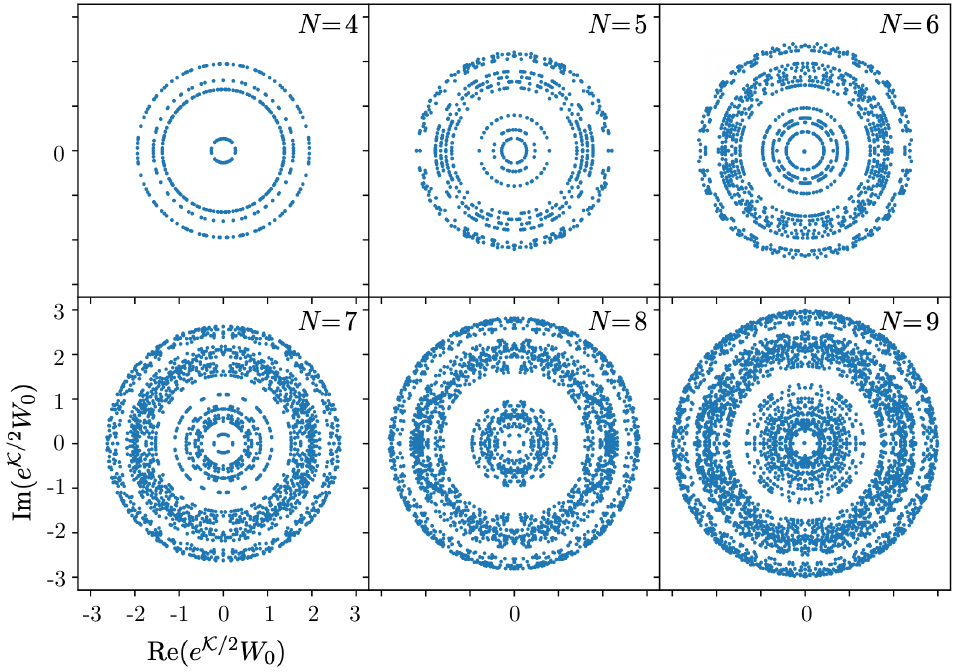}
\end{center}
\vspace{-0.6cm}
\caption{Distributions of $\tilde{W_0}$ in the orientifold $T^6/\mathbb{Z}_2\times\mathbb{Z}_2$ with three equivalent tori, for values  $N_{\rm flux}=4,\ldots,9.$}
\label{fig:3}
\end{figure}

\paragraph{Residual constant superpotential} We eventually introduce the following K\"ahler invariant quantity, related to the residual constant superpotential $W_0$ after complex structure moduli stabilisation:
\begin{equation}\label{invariantsuperpot}
\tilde W_0 \equiv  \mathcal{V} e^{{\mathcal{K}}/{2}}W_0 = \mathcal{V}   m_{3/2},
\end{equation}
where the volume factor $\mathcal{V}$ compensate the K\"ahler moduli part of the K\"ahler potential \eqref{Kahlerpot}. It is related to the complex gravitino mass parameter $m_{3/2}$ of the supergravity effective Lagrangian \cite{Freedman:2012zz}. It is also invariant under the $S$- and $U$-dualities described above. In figure \ref{fig:3}, we show the distributions of $\tilde W_0$ for our vacuum solutions with three equivalent tori and $N_\mathrm{flux}\leq 9$. In this plot we did not bring the moduli to their fundamental domains through the $S$- and $U$-dualities:  the different values of $\tilde W_0$ organise in the complex plane in circles of fixed $|\tilde W_0|$, corresponding to different orbits of these dualities. The fact that we obtain distinct circles shows that $|\tilde W_0|$ is discrete, which is again a consequence of the finiteness of flux vacua. The number of circles approximatively matches the number of vacua found in table \ref{tab:1}. There is however a small discrepancy because few distinct vacua give the same value of $|\tilde W_0|$. In other words,  different solutions with $\mathcal{S}$ and $\mathcal{U}_i$ in their fundamental domains give the same value of  $|\tilde W_0|$, so that some of the circles correspond to superposed duality orbits.

Eventually, note that for three equivalent tori, we obtain solutions with a vanishing superpotential only for $N_{\rm flux}$ multiple of $3$. These are the solutions studied in detail in \cite{Ishiguro:2020tmo}.

\newpage

\subsection{Eventual solutions with flux integers hierarchy and parametric control on $g_s$}\label{sec:parametriccontrol}

We recall that we found our vacua from the analytic solutions to the $D_\mathcal{S}W = 0, D_{\mathcal{U}^i} W = 0$ supersymmetric equations by scanning over a range of flux integers $|m^{H,F}|,|n^{H,F} |\leq k$ and increasing the range $k$. As explained in the previous section, the maximal number of vacua for a given $N_{\rm flux}$ was obtained at a certain range $k^{N_{\rm flux}}_{\max \#}$. Increasing the range further we found no new vacua and stopped the search after a while. For instance, for $N_{\rm flux}=4$ we obtain $k^4_{\rm max \#}=2$ and we searched until  $k= 30$. One drawback of this procedure is that we might be missing solutions with a huge hierarchy between some integers. Indeed, a solution with $n^{H}_0=1$ and $m^F_2=2048$, which can in principle have small $N_{\rm flux}$ if some flux quanta cancel between each others, could only be reached for the range $k=2048$. Reaching this range by scanning over all combinations is way too time consuming. If such solution exists, we expect that it comes together with an entire family of solutions, parameterised by the ratio of some integers or, equivalently, parameterised by one or several of the integers. These families could allow for parametric control on $g_s$.

We thus searched for infinite families of solutions of the system $D_\mathcal{S}W = 0, D_{\mathcal{U}^i} W = 0$, parameterised by the choice of one flux integer. We would say that we have parametric control on $g_s$ if the family is such that we can vary the integers while keeping $N_\mathrm{flux}$ constant, keeping the complex structure moduli in a physical region, and sending $y_s = 1/g_s$ to infinity. If such a family exists, we should be able to solve the system order by order in the quantities (integers and moduli) that become large. We recall that in the $T^6/\mathbb{Z}_2\times\mathbb{Z}_2$ orbifold, the flux charge defined in \cref{eq:9} reads:
\begin{equation}
    N_\mathrm{flux} = m^H n^F - m^F n^H.
\end{equation}
The summation over the indices labelling the cohomology basis is implicit, see below \cref{eq:9}.

We show in few examples different obstructions to the existence of such family of solutions, and hence of parametric control. Expanding the system $D_\mathcal{S}W = 0, D_{\mathcal{U}^i} W = 0$  in real and imaginary parts leads to: 
\begin{align}
   & n^F_1x_1 + n^F_2x_2 + n^F_3x_3 + m^F_1x_2x_3 + m^F_2x_1x_3 + m^F_3x_1x_2 &&\nonumber \\ 
  & \qquad + x_s(n^H_1x_1 + n^H_2x_2 + n^H_3x_3 + m^H_1x_2x_3 + m^H_2x_1x_3 + m^H_3x_1x_2) && \nonumber\\ 
&\hspace{2.1cm} + n^F_0 + n^H_0x_s - m^F_0x_1x_2x_3 - m^H_0x_1x_2x_3x_s + m^H_0y_1y_2y_3y_s &&\hspace{-0.8cm}= 0,  \label{eq:asy_2}  \\
  &   n^H_1x_1y_s + n^H_2x_2y_s + n^H_3x_3y_s + m^H_1x_2x_3y_s + m^H_2x_1x_3y_s&&\nonumber\\ 
    &\qquad \,\, + m^H_3x_1x_2y_s + n^H_0y_s  - m^F_0y_1y_2y_3 - m^H_0x_sy_1y_2y_3 - m^H_0x_1x_2x_3y_s &&\hspace{-0.8cm}= 0,   \label{eq:asy_3} \\
   & m^F_1y_2y_3 + m^H_1x_sy_2y_3 - m^H_2x_3y_1y_s - m^H_3x_2y_1y_s- n^H_1y_1y_s&&\nonumber\\
    &\hspace{4.5cm}  -m^F_0x_1y_2y_3 - m^H_0x_1x_sy_2y_3 + m^H_0x_2x_3y_1y_s && \hspace{-0.8cm} = 0, \label{eq:asy_4} \\
  &  n^F_1y_1 + m^F_2x_3y_1 + m^F_3x_2y_1 + n^H_1x_sy_1 + m^H_1y_2y_3y_s + m^H_2x_3x_sy_1&& \nonumber\\
    &\qquad \qquad \qquad + m^H_3x_2x_sy_1 - m^F_0x_2x_3y_1 - m^H_0x_2x_3x_sy_1 - m^H_0x_1y_2y_3y_s &&\hspace{-0.8cm}= 0.  \label{eq:asy_5} 
\end{align}
and cyclic permutations of these two last equations. We again used the notation $\mathcal{S} = x_s + i y_s$ and $\mathcal{U}^i = x_i + iy_i$.

\paragraph{Case 1: $y_s\rightarrow +\infty$  with one integer going to infinity} We first try simple limits where only one integer goes to infinity, while the imaginary part $y_s$ blows together with this integer, $y_s\rightarrow + \infty$, and other quantities stay of order $x_s, x_i, y_i \sim \mathcal{O}(1)$. 

Take for instance $n^F_0 \rightarrow +\infty$. In this case, \eqref{eq:asy_2} becomes, at leading order, $n^F_0 + m^H_0y_1y_2y_3y_s=0$. We must impose $m^H_0 = 0$ for $N_\mathrm{flux}$ to remain finite, so we end up with $n^F_0 = 0$, which is in contradiction with $n^F_0 \rightarrow \infty$. 

One can also choose $m^F_0\rightarrow+\infty$, with $n^H_0 = 0$. In this case, the system \cref{eq:asy_2,eq:asy_3,eq:asy_4,eq:asy_5} at leading order gives $y_s = (m^F_0 x_1x_2x_3)/(m^H_0 y_1 y_2 y_3)$ and $x_i^2 + y_i^2 = (m^H_i/m^H_0) x_i$, along with the constraints $m^H_0n^H_1 + m^H_2m^H_3 = 0$, with cyclic permutations, and $m^H_1m^H_2m^H_3 = 0$. Hence, one of the $m^H_i$ needs to be zero. However, for the corresponding $i$ we then have $x_i^2 + y_i^2 = 0$. This leads to a vanishing untwisted complex structure modulus. This is not a valid solution, as can be seen from the $T$-dual theory where it corresponds to the decompactification limit.

\paragraph{Case 2: $y_s\rightarrow +\infty$  with two integers going to infinity} A more elaborate example would be with two integers going to infinity, for instance $n^F_0, m^F_0 \rightarrow +\infty$, with $m^H_0, n^H_0 = 0$. We can further simplify by assuming that the three tori are equivalent ($1 = 2 = 3 = i$). In this case, the system of \cref{eq:asy_2} to \cref{eq:asy_5} becomes, at leading order: 
\begin{align}
    n^F_0-m^F_0 x_i^3 &= 0,  \qquad       &-m^F_0 y_i^3+3 m^H_i x_i^2 y_s+3 n^H_i x_i y_s &= 0, \nonumber \\
    m^F_0 x_i y_i^2+2 m^H_i x_i y_i y_s+n^H_i y_i y_s &= 0,  \qquad    &m^F_0 x_i^2 y_i-m^H_i y_i^2 y_s &= 0,     \label{eq:asy_24}
\end{align}
The first, third and last equations give directly:
\begin{equation}
    x_i^3 = \frac{n^F_0}{m^F_0},
    \quad
    y_s^3 = \frac{m^F_0 (n^F_0)^2}{(m^H_i)^3 y_i^3} \quad\text{and}\quad
    y_i^2 = -\frac{(n^F_0)^{1/3} n^H_i}{(m^F_0)^{1/3} m^H_i}-\frac{2 (n^F_0)^{2/3}}{(m^F_0)^{2/3}},
\end{equation}
while the second equation becomes a constraint on the integers that can be written as:
\begin{equation}
    \Big((m^F_0)^{1/3}n^H_i + (n^F_0)^{1/3}m^H_i\Big)^2 - (m^F_0)^{1/3} (n^F_0)^{1/3} m^H_i n^H_i = 0.
\end{equation}
This equation is of the form $(x+y)^2 = xy$, and it has no solutions with both $x$ and $y$ real.

\paragraph{Case 3: $y_s,x_s \rightarrow +\infty$  with two integers going to infinity}

\begin{figure}[h!]
 \vspace{0.1cm}
\begin{center}
\includegraphics[scale=0.5]{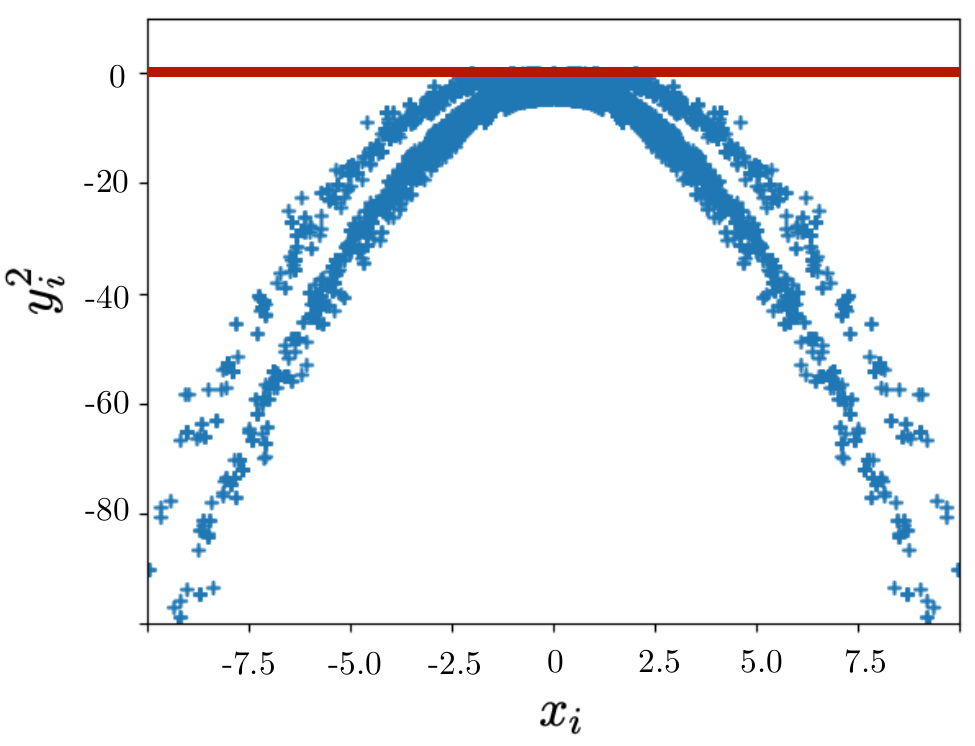}
\vspace{-0.5cm}
\end{center}
\caption{Values of $y_i^2$ obtained from \cref{eq:asy_10}  with $x_i$ numerically evaluated as the a solution of its third order equation \eqref{3rdordereq}, for all flux quanta satisfying $|m_i^{H,F}|, |n_i^{H,F}| \leq 10$. }
\label{figureyi2}
\end{figure}

Finally, we can imagine limits where both $x_s$ and $y_s$ go to infinity with the integers that go to infinity, while $x_i, y_i \sim \mathcal{O}(1)$. Let us still take again $n^F_0, m^F_0 \rightarrow +\infty$ and $m^H_0, n^H_0 = 0$, still with equivalent tori. In this rather specific case, the system \cref{eq:asy_2} to \cref{eq:asy_5} reads at leading order:
\begin{align}
    3n^H_ix_ix_s + 3m^H_ix_i^2x_s
    + n^F_0 - m^F_0x_i^3 &= 0, \label{eq:asy_8}\\ 
    3n^H_ix_iy_s + 3m^H_ix_i^3y_s - m^F_0y_i^3 &= 0, \\
    m^H_ix_sy_i^2 - 2m^H_ix_iy_iy_s - n^H_iy_iy_s
     - m^F_0x_iy_i^2 &= 0, \\
    n^H_ix_sy_i + m^H_iy_i^2y_s + 2m^H_ix_ix_sy_i
     - m^F_0x_i^2y_i &= 0. \label{eq:asy_9}
\end{align}
The first two equations are solved by:
\begin{equation}
x_s = \frac{m^F_0 x_i^3-n^F_0}{3 x_i (m^H_i x_i+n^H_i)} \quad \text{and} \quad 
y_s = \frac{m^F_0 y_i^3}{3 x_i (m^H_i x_i+n^H_i)}, \label{solxsys}
\end{equation}
and the third one by:
\begin{equation}\label{eq:asy_6}
    y_i^2 = -\frac{2 m^F_0 m^H_i x_i^3+3 m^F_0 n^H_i x_i^2+n^F_0 m^H_i}{2 m^F_0 m^H_i x_i+m^F_0 n^H_i}.
\end{equation}
Once plugged in \cref{eq:asy_9} these solutions give a cubic equation in $x_i$, with coefficients depending on the flux quanta only:
\begin{align}
&\Big(4 m^F_0 n^F_0(m^H_i)^3 + 2 (m^F_0)^2 (n^H_i)^3\Big) x_i^3 + 6 m^F_0  n^F_0(m^H_i)^2 n^H_i x_i^2 \nonumber\\
 &\hspace{2cm} +  6 m^F_0 n^F_0m^H_i (n^H_i)^2 x_i -(n^F_0)^2(m^H_i)^3  + m^F_0 n^F_0 (n^H_i)^3 = 0. \label{3rdordereq}
\end{align}
It can be used to eliminate the $x_i^3$ term in the solution for $y_i^2$ of \eqref{eq:asy_6}, yielding the new equation:
\begin{equation}\label{eq:asy_10}
    y_i^2 = -\frac{3 (n^F_0 (m^H_i)^2-m^F_0 (n^H_i)^2 x_i)^2}{m^F_0(m^F_0 (n^H_i)^3+2 n^F_0 (m^H_i)^3)(2 m^H_i x_i+n^H_i)}.
\end{equation}
The numerator being positive, we need a negative denominator to ensure that  $y_i^2$ is positive. It is rather difficult to evaluate its sign using our system of equations. The value of $y_i^2$ obtained from  \cref{eq:asy_10} can however be evaluated by plugging the solutions of the third order equation in $x_i$. We show in figure \ref{figureyi2} the numerical values obtained for all combinations of integers satisfying $|m_i^{H,F}|, |n_i^{H,F}| \leq 10$. We see that the result is always negative, suggesting that the system does not have consistent solution in the limit chosen in the present case.

We considered other limits in addition to the ones displayed here. They all fail to be consistent for similar reasons: they lead to the vanishing of the imaginary part of a complex structure modulus, the impossibility of satisfying certain constraint on the integers, or present other inconsistency as the one just described. There are of course too many possible limits to be exhaustive, and some of them are hard to analyse, but all hints towards the absence of family of solutions and hence parametric control over the string coupling. 

These results thus go in the direction of the finiteness of flux vacua. A family of solutions with parametric control over the string coupling and constant $N_{\rm flux}$ would lead to infinite number of solutions with the dilaton in its fundamental domain and taking different values. The fact that we did not find such families agrees with the fact that we found finite numbers of solutions for each $N_\mathrm{flux}$ through the analysis of previous sections.

\subsection{Magnetised D7-branes} \label{subsec:magnetizedbranes}

We come back to the possibility of evading the tadpole constraint in the $T^6/\mathbb{Z}_2\times\mathbb{Z}_2$ orbifold. As already explained, it requires the presence of negative D3-brane charge, namely $\overline{\mathrm{D3}}$-brane charges. Such charges are directly related to supersymmetry breaking objects. They could be either genuine $\overline{\mathrm{D3}}$-branes or magnetised D7-branes \cite{Cascales:2003zp,Marchesano:2004xz}. As we discuss below, D7-branes being naturally present in most of the toroidal orientifolds, we focus on this latter option. Magnetised D7-branes also play a key role in the fully perturbative K\"ahler moduli stabilisation mechanism using logarithmic loop corrections \cite{Antoniadis:2018hqy,Antoniadis:2021lhi}.

We stress that including magnetised D7-branes does not change the relation \eqref{eq:rel_12} between the minimal string coupling $g_{s, \rm min}$ and the flux number $N_{\rm flux}$. This relation comes directly from complex structure moduli and axio-dilaton stabilisation by quantised background 3-form fluxes and is thus insensitive to the presence of the  D7-branes. Yet, the value of  $N_{\rm flux}$ in the construction is directly related to the brane configuration.

\paragraph{Worldvolume fluxes and RR charges}
In the  $T^6/\mathbb{Z}_2\times\mathbb{Z}_2$ orbifold, a stack $a$ of magnetised D7-branes, with worldvolumes along two tori and localised in the third torus, carry magnetic fields $F_a$ associated to their $U(1)$ gauge group. The latter satisfy the standard Dirac quantisation on fluxes: 
\begin{equation}
m^i_a \int_{T^2_i} F^i_a = 2\pi n^i_a,  \label{magfluxquantification}
\end{equation}
for each of the two wrapped tori $T^2_i$. The wrapping numbers $m^i_a$ and flux quanta $n^i_a$ shall be coprime integers. Moreover, due to the $\mathbb{Z}_2$ orientifold quotient, $n^{i}_a$ can take half-integer values. Through their Chern-Simons couplings, such magnetised D7-branes induce RR D3-charges on top of their D7-charges. Similarly, D7-branes can themselves be seen as magnetised D9-branes with vanishing wrapping number on the torus where they are localised \cite{Cascales:2003zp}. We thus assign vanishing wrapping numbers $m^k_a=0$ and unit fluxes $n^k_a=1$ on the torus where D7-branes are localised, on top of the flux quanta of \cref{magfluxquantification} on the tori wrapped by their worldvolumes. E.g., a stack $a$ of $N_a$ D7-branes localised in the first torus $T^2_1$ has magnetic numbers: 
\begin{alignat}{4}
& &&\,\,\,\, T^2_1 &&\,\,  \quad  T^2_2   &&\,\, \quad T^2_3  \nonumber  \\
N_a \rightarrow \,\, &D7_a \quad &&(1,0)\,\,\,\,  &&(n^2_a,m^2_a) \,\,\,\, &&(n^3_a,m^3_a) \label{conventionstack}
\end{alignat}
In these conventions, the magnetic numbers of a standard D7-brane wrapping the second and third tori without magnetic flux are:
\begin{equation}
D7_{F=0} \quad \,\,\,(1,0)\,\,\,\,  \,\,\,\,(0,1) \,\,\,\, \,\,\,(0,-1) \label{unfluxedbrane}
\end{equation}
 In terms of these magnetic numbers, the RR charges of a stack $a$ of  $N_a$ D7-branes read \cite{Cascales:2003zp}:
\begin{align}
&Q^a_{D3}= N_a n^1_a n^2_a n^3_a, \\
&^{\scriptscriptstyle(i)}Q^a_{D7}= - N_a n^i_a m_a^jm_a^k, \qquad i\neq j\neq k \neq i. \label{D7charge}
\end{align} 
The stack $a$ only has non-vanishing D7-charge on the torus where it is localised, e.g. $^{\scriptscriptstyle(1)}Q^a_{D7}\neq0$ and $^{\scriptscriptstyle(2)}Q^a_{D7}=\,^{\scriptscriptstyle(3)}Q^a_{D7}=0$ for the stack \eqref{conventionstack}.
We see that magnetised D7-branes can easily induce negative D3-charges, namely $\overline{\mathrm{D3}}$-charges, for flux quanta $n^i_a$ of opposite signs on the wrapped tori. For instance, a single D7$_a$ as in \cref{conventionstack} with opposite fluxes $n^2_a=-n^3_a=1$, induces a negative unit charge $Q^a_{D3}=-1$. Note also that in these conventions, the standard D7-brane without magnetic flux \cref{unfluxedbrane} has positive RR charge $Q_{D7}=1$.

\paragraph{Tadpole conditions with magnetised D7-branes}
The induced charge $Q^a_{D3}$ should be included in the D3-brane number $N_{D3}$ appearing in the tadpole condition \eqref{eq:13}. As advertised, magnetised D7-branes inducing negative D3-charges can relax the constraint imposed by the tadpole condition on the $G_3$ flux charge $N_{\rm flux}$. In $T^6/\mathbb{Z}_2\times\mathbb{Z}_2$ the tadpole constraint obtained in absence of negative D3-charge was given in \cref{tadpoleconstraintZ2Z2}. It  can thus {\it a priori} be relaxed. Consistency of the orientifold construction however also requires the cancellation of the RR tadpole related to the D7-charge. Such tadpole condition relates the magnetised brane charges  \eqref{D7charge} to the charges of the O7-planes present in the construction. Hence for a fixed orientifold geometry, one cannot choose arbitrary D7 magnetic fluxes $n^a_i$ and wrappings $m^a_i$.
 
The $T^6/\mathbb{Z}_2\times\mathbb{Z}_2$ orientifold with involution reversing all coordinates $x_j\rightarrow -x_j$, $j=1,\ldots,6$, contains $3\times4$ $\mathrm{O7}_i$-planes, $i=1,2,3$. Each of them wraps two tori and is localised in the third torus $T^2_i$ at one of the four fixed points of the orbifold action, see \cref{sec:tadpoleintegersmultiple}. For instance, the four  $\mathrm{O7}_2$-plane are localised at  $(0,0,\iota_3,\iota_4,0,0)$ with $\iota_3,\iota_4=0,1/2$. 
%There is a important subtlety in presence of discrete torsion \cite{Angelantonj:1999ms,Blumenhagen:2003vr}\footnote{In \cite{Blumenhagen:2003vr} the authors called the orbifold $T^6/\mathbb{Z}_2\times\mathbb{Z}_2$ with $(h^{1,1},h^{2,1})=(3,51)$ ``without discrete torsion" whereas with our conventions it is the case with discrete torsion.}. It requires the presence of ``exotic" orientifold planes, denoted $\overline{Op}$  \cite{Angelantonj:1999ms} or $Op^{(+,+)}$ \cite{Blumenhagen:2003vr}.  They carry reversed RR charges with respect to the standard ones. In the $T^6/\mathbb{Z}_2\times\mathbb{Z}_2$ orientifold, an odd number of the four classes of $Op$-planes must be constituted of such exotic planes. In the original setup of \cite{Angelantonj:1999ms} such classes correspond to the $O9$- and $O5_i$-planes whereas in our T-dual case they are the $O3$- and $O7_i$-planes. When taking the three classes of $O7_i$-planes to be exotic,

The RR tadpole cancellation then requires the total magnetised $\mathrm{D7}_a$ charges \eqref{D7charge} to satisfy \cite{Cascales:2003zp}:
\begin{align}
&\sum_a Q^a_{D3}  + N^Y_{\rm flux} = \sum_a N_a n^1_a n^2_a n^3_a + N^Y_{\rm flux} =  \frac{1}{4} N_{O3}=16, \label{D3tadpolemagD7}\\
&\sum_a  \,^{\scriptscriptstyle(1)}Q^a_{D7}  = - \sum_a N_a n^1_a m^2_a m^3_a = 4 N_{O7_1}=16, \label{D71tadpolemagD7}\\
&\sum_a \,^{\scriptscriptstyle(2)}Q^a_{D7}  = -\sum_a N_a m^1_a n^2_a m^3_a = 4 N_{O7_2}=16, \label{D72tadpolemagD7} \\
&\sum_a \,^{\scriptscriptstyle(3)}Q^a_{D7}  = - \sum_a N_a m^1_a m^2_a n^3_a = 4 N_{O7_2}=16. \label{D73tadpolemagD7}
\end{align}
The first line is a rewriting of the tadpole condition \eqref{eq:13} associated to the D3-charge,  expressing explicitly the charge from magnetised D7-branes.
These four tadpole conditions are the same as in the T-dual model of \cite{Cvetic:2001tj,Cvetic:2001nr} with D6-branes at angles \cite{Aldazabal:2000cn,Aldazabal:2000dg}. 

\paragraph{Solutions relaxing the constraint on $N_{\rm flux}$} The tadpole condition \eqref{D3tadpolemagD7} shows that each stack $a$ with an odd number of negative $n^i_a$ contributes negatively to the D3-charge, thus increasing the allowed $N^Y_{\rm flux}=2|G|N_{\rm flux}$. In the conventions \eqref{conventionstack} the flux number corresponding to the torus not wrapped by the stack is always $n^k_a=1$, so that to have negative contributions we need one positive and one negative flux numbers for the two wrapped tori. Moreover, as explained under \cref{D7charge} a single stack $N_a$ contributes to only one among the tadpole conditions \eqref{D71tadpolemagD7} to \eqref{D73tadpolemagD7}. For this contribution to be of the correct sign, we need one positive and one negative wrapping number, as for the unmagnetised brane \eqref{unfluxedbrane}.

We consider the simple configuration of three D7-branes stacks with magnetic numbers:
\begin{alignat}{3}
&\quad \,\,T^2_1 &&\,\,  \quad  T^2_2   &&\,\, \quad T^2_3  \nonumber  \\
N_1 \rightarrow D7_1 \quad &\,\,\,\,\,(1,0)\,\,\,\,  &&(n_1^2,-1) \,\,\,\, &&(-n_1^3,1) \nonumber\\
N_2 \rightarrow D7_2 \quad &\,\,(-n^1_2,1)\,\,\,\,  &&\,\,\,\,(1,0) \,\, &&(n^3_2,-1) \nonumber\\
N_3 \rightarrow D7_3 \quad &\,\,(n^1_3,-1)\,\,\,\,  &&(-n^2_3,1) \,\,\,\, &&\,\,\,\,\, (1,0) \label{config}
\end{alignat}
with all $n^i_a>0$. The configuration satisfies all the tadpole conditions related to the D7-charges for $N_a=16$, $a=1,2,3$.

 We recall that a negative $Q^a_{D3}$ only requires one positive and one negative flux numbers, so that we made an arbitrary choice for the relative signs between the fluxes of different stacks and tori. We chose to take unit wrapping numbers $m^i_a=\pm1$. A configuration with the same $^{\scriptscriptstyle(i)}Q^a_{D7}$ charge, hence satisfying the tadpole condition, but with greater wrapping numbers $m_a^i$ would require lower $N_a$, thus leading to a smaller absolute value of the $Q^a_{D3}$ charge. The total D3-charge for the configuration \eqref{config} is:
\begin{equation}
Q_{D3}=\sum_{a=1}^3 Q^a_{D3}=\sum_{a=1}^3 N_a n^1_a n^2_a n^3_a = - 16 (n_1^2n_1^3 + n_2^1n_2^3 + n_3^1n_3^2). \label{D3chargeconfig}
\end{equation}
For all the stacks $a$ to preserve supersymmetry, their flux numbers  need to satisfy a constraint, which in the present context reads:
\begin{equation}
\sum_{i=1}^3 \zeta_a^{\scriptscriptstyle (i)} \equiv \sum_i \frac{1}{\pi} {\rm Arctan}(2\pi \alpha' F_a^i)=\sum_i \frac{1}{\pi} {\rm Arctan}\left(\frac{ m_a^i \alpha'}{n_a^i \mathcal{A}_i}\right)=0. \label{susycondition}
\end{equation}
The $\zeta_a^{\scriptscriptstyle (i)}$ correspond to oscillator shifts of open string modes caused by the modification of boundary conditions by magnetic fields \cite{Abouelsaood:1986gd,Bachas:1995ik}.  The second equality used the explicit magnetic flux quantisation condition \eqref{magfluxquantification} for wrapped tori $T^2_i$ of area $4\pi^2\mathcal{A}_i$.

We see that in the configuration \eqref{config}, all D7-brane stacks break SUSY. They cannot satisfy the condition \eqref{susycondition}. Such configuration generically produce tachyons, coming from open strings with endpoints on the same or different stacks, called doubly charged or mixed states. The mass of such states were explicitly written in e.g. \cite{Antoniadis:2021lhi}.  Doubly charged tachyons can be eliminated by introducing separations between branes and their orientifold images, namely by moving the branes away from the orientifold planes \cite{Antoniadis:2021lhi}. This allows to increase the mass of such states to positive values. On the other hand, the following conditions on $\zeta_a^{(i)}$ allow to cancel the mass of all mixed tachyonic states \cite{Antoniadis:2021lhi}:
\begin{alignat}{2}
&(A- && 1) \qquad \zeta_1^{\scriptscriptstyle(3)}=\zeta_2^{\scriptscriptstyle(1)}=\zeta_3^{\scriptscriptstyle(2)}, \qquad \zeta_1^{\scriptscriptstyle(2)}= \zeta_2^{\scriptscriptstyle(3)}=\zeta_3^{\scriptscriptstyle(1)} ;  \nonumber\\
& \qquad && 2) \qquad \zeta_1^{\scriptscriptstyle(3)}=\zeta_2^{\scriptscriptstyle(1)}=-\zeta_3^{\scriptscriptstyle(2)}, \qquad \zeta_1^{\scriptscriptstyle(2)}= \zeta_2^{\scriptscriptstyle(3)}=-\zeta_3^{\scriptscriptstyle(1)} ;  \nonumber\\
& \qquad && 3) \qquad \zeta_1^{\scriptscriptstyle(3)}=-\zeta_2^{\scriptscriptstyle(1)}=\zeta_3^{\scriptscriptstyle(2)}, \qquad \zeta_1^{\scriptscriptstyle(2)}= -\zeta_2^{\scriptscriptstyle(3)}=\zeta_3^{\scriptscriptstyle(1)} ;  \nonumber\\
&\qquad && 4) \qquad \zeta_1^{\scriptscriptstyle(3)}=-\zeta_2^{\scriptscriptstyle(1)}=-\zeta_3^{\scriptscriptstyle(2)}, \qquad \zeta_1^{\scriptscriptstyle(2)}=- \zeta_2^{\scriptscriptstyle(3)}=-\zeta_3^{\scriptscriptstyle(1)} ;  \nonumber\\
&(B- &&1) \qquad \zeta_1^{\scriptscriptstyle(2)}=\zeta_1^{\scriptscriptstyle(3)}, \qquad \zeta_2^{\scriptscriptstyle(1)}=\zeta_2^{\scriptscriptstyle(3)}, \qquad  \zeta_3^{\scriptscriptstyle(1)}=\zeta_3^{\scriptscriptstyle(2)} ; \nonumber \\
& &&2) \qquad \zeta_1^{\scriptscriptstyle(2)}=-\zeta_1^{\scriptscriptstyle(3)}, \qquad \zeta_2^{\scriptscriptstyle(1)}=-\zeta_2^{\scriptscriptstyle(3)}, \qquad  \zeta_3^{\scriptscriptstyle(1)}=-\zeta_3^{\scriptscriptstyle(2)}. \label{conditionsentireworldvolumes} 
\end{alignat}
Solution $(B-2)$ satisfies \eqref{susycondition} and thus preserves supersymmetry, with all lowest-lying states remaining massless.
In the solutions $(A-i)$, all the doubly charged states ${\rm D}7_a$--${\rm D}7_a$ have identical tachyonic masses, equal to $\alpha'm^2 =-2\left|\zeta_1^{\scriptscriptstyle(2)}+\zeta_1^{\scriptscriptstyle(3)}\right|$, while for solution $(B-1)$ they can have different masses. The possible arbitrary choices of relative signs between $n_a^i$, in configurations similar to \eqref{config}, always allow to satisfy $(B-1)$ or only one of the $(A-i)$. The specific choice \eqref{config} allows to satisfy $(A-1)$ or $(B-1)$. For these configurations the $Q_{D3}$ charge \eqref{D3chargeconfig} then reads:
\begin{align}
&(A-1) \qquad Q_{D3}= - 48 \, n_1^2 n_1^3,  \\
&(B-1) \qquad Q_{D3}= - 16 \Big((n_1^2)^2 + (n^1_2)^2 + (n^1_3)^2\Big).
\end{align} 
We see that in any of these two cases, the $Q_{D3}$ charge can be made arbitrary large choosing large values of $n_a^i$. The tadpole condition \eqref{D3tadpolemagD7} for these two configurations free of mixed tachyons leads to:
\begin{alignat}{2}
&(A-1) \qquad \frac{1}{4} N^Y_{\rm flux} = 2 N_{\rm flux} = 2 n= 4 (1+ 3 \, n_1^2 n_1^3), \qquad &&n\in \mathbb{N}^*,  \\
&(B-1) \qquad \frac{1}{4} N^Y_{\rm flux} = 2 N_{\rm flux}= 2 n =4 \Big(1+(n_1^2)^2 + (n^1_2)^2 + (n^1_3)^2\Big),  \qquad &&n\in \mathbb{N}^*. \label{B1Nflux}
\end{alignat}
This shows that the constraint \eqref{tadpoleconstraintZ2Z2} obtained without negative D3-charge is, as expected, totally lifted in such a configuration. The flux charge $N_{\rm flux}=n \in \mathbb{N}^*$ can thus in principle be arbitrarily large, while still satisfying the tadpole condition. Therefore, one can get solutions with small values of $g_{s\, {\rm min}}$. 

For instance, in both cases $(A-1)$ or $(B-1)$ with $n_a^i=1$, the tadpole condition \eqref{B1Nflux} forces $N_{\rm flux}=8$. The string coupling found in our vacuum solutions is then bounded following \cref{exactgsminNflux4p}, giving:
\begin{equation}
g_s\geq g_{s, \rm min} = \frac{16}{N_{\rm flux}^2} = 0.25.
\end{equation}
Configurations with greater $n_a^i$ give yet smaller minimal values of $g_s$.

As explained below \cref{susycondition}, such D7-brane setup totally breaks supersymmetry. Condition \eqref{susycondition} shows that supersymmetry is recovered in the limit of large volumes, with tori areas $\mathcal{A}_i\rightarrow +\infty$, which corresponds to diluted magnetic fluxes. This SUSY breaking configuration induces Fayet-Ilipopoulos (FI) terms in the $4$d effective theory, which depend on the K\"ahler moduli related to the areas $\mathcal{A}_i$ \cite{Jockers:2004yj,Blumenhagen:2006ci,Antoniadis:2021lhi}. When these FI terms are associated with logarithmic loop corrections to the K\"ahler potential of these moduli, the latter can be stabilised at a metastable de Sitter vacuum \cite{Antoniadis:2018hqy,Antoniadis:2020stf}. Hence, with the complex structure moduli stabilisation presented in the present paper, this leads to a toroidal orbifold model with metastable de Sitter vacuum and all moduli stabilised explicitly.  The explicit model with all moduli stabilised in this way is left for future work.

\section{Conclusions}\label{conclusions}

In this work, we studied the stabilisation by fluxes of the axio-dilaton and complex structure moduli of simple ${\cal N}=1$ orientifolds of orbifold compactifications of type IIB string theory. In this simple but calculable setup, we showed how the finiteness of (inequivalent up to ${S}$- and ${U}$-duality) flux vacua manifests itself. Enumerating all flux integer combinations for fixed 3-form flux contribution to the D3-brane tadpole $N_\mathrm{flux}$ and fixed range of the integers $|m|, |n| \leq k$, there is simply a value of $k$ above which no new vacua are found numerically. 

We also found explicit expressions for the minimal string coupling $g_{s, \mathrm{min}}$ of the form $g_{s, \mathrm{min}} \sim 1/N_\mathrm{flux}^\alpha$, with $\alpha = 1$ for orbifolds with zero or one complex structure modulus, and $\alpha = 2$ in the case of $T^6/\mathbb{Z}_2\times\mathbb{Z}_2$. Since $N_\mathrm{flux}$ is bounded by the tadpole constraint, these relations can be used in principle to obtain the lowest value of the string coupling achievable by flux compactification on these orbifolds, in the absence of magnetised D7-branes. In the case of orbifolds with zero or one untwisted complex structure moduli, the tadpole constraint turns out to be too constraining for turning on fluxes in the first place. While for $T^6/\mathbb{Z}_2\times\mathbb{Z}_2$ this is not true, there are still no vacua satisfying the tadpole cancellation condition.
The simplicity of toroidal orbifolds allows to be relatively explicit, but it may lack some important features of a Calabi-Yau compactification. %Therefore, it would be interesting to obtain similar relations between $g_{s, \mathrm{min}}$ and $N_\mathrm{flux}$ in other cases, and investigate whether there is any pattern. This will probably be the subject of upcoming work.

Magnetised D7-branes can give a negative contribution to the D3-brane tadpole and the strict bound on $N_\mathrm{flux}$ is in general relaxed, allowing for much smaller values of the string coupling. The case of $T^6/\mathbb{Z}_2\times\mathbb{Z}_2$
with discrete torsion is particularly interesting since after stabilisation of all complex structure moduli, one is left with only three untwisted K\"ahler moduli that may be stabilised based on perturbative corrections and the use of magnetic fluxes on the three sets of D7-branes. It remains to be seen if a concrete physically interesting example exists with all closed string moduli stabilised in a controllable way.

\section*{Acknowledgements}
IA is supported by the Second Century Fund (C2F), Chulalongkorn University. OL thanks Thibaut Coudarchet and Michele Cicoli for useful discussions.

\appendix

\section{Complex structures of the orbifolds}\label{app:1}

In this Appendix, we give the complex structure of all the orbifolds listed in table \ref{tab:2}. Details of their computation were given in section \ref{sec:3}.

\begin{table}[ht!]
    \centering
    \hspace{2cm} coefficients of the complex structure\\
    \vspace{5pt}
      \begin{tabular}{|c||c||c|c|c|c|c|c|}
    \hline
        \multicolumn{1}{|l}{{\hspace{-3pt}orbifold  }} & & $x_1$ & $x_2$ & $x_3$ & $x_4$ & $x_5$ & $x_6$\\
        \hline\hline
        & $z_1$ & $1$ & $e^{2i\pi/3}$ & $0$ & $0$ & $0$ & $0$ \\
        $\mathbb{Z}_{3}$ & $z_2$ & $0$ & $0$ & $1$ & $e^{2i\pi/3}$ & $0$ & $0$ \\
        & $z_3$ & $0$ & $0$ & $0$ & $0$ & $1$ & $e^{2i\pi/3}$\\
        \hline
        & $z_1$ & $1$ & $i$ & $-1$ & $0$ & $0$ & $0$ \\
        $\mathbb{Z}_{4,a}$ & $z_2$ & $0$ & $0$ & $0$ & $1$ & $i$ & $-1$ \\
        & $z_3$ & $1$ & $-1$ & $1$ & $\mathcal{U}$ & $-\mathcal{U}$ & $\mathcal{U}$\\
        \hline
        & $z_1$ & $1$ & $e^{3i\pi/4}/\sqrt{2}$ & $0$ & $0$ & $0$ & $0$ \\
        $\mathbb{Z}_{4,b}$ & $z_2$ & $0$ & $0$ & $1$ & $i$ & $-1$ & $0$ \\
        & $z_3$ & $0$ & $0$ & $1$ & $-1$ & $1$ & $\mathcal{U}$\\
        \hline
        & $z_1$ & $1$ & $e^{3i\pi/4}/\sqrt{2}$ & $0$ & $0$ & $0$ & $0$ \\
        $\mathbb{Z}_{4,c}$ & $z_2$ & $0$ & $0$ & $1$ & $e^{3i\pi/4}/\sqrt{2}$ & $0$ & $0$ \\
        & $z_3$ & $0$ & $0$ & $0$ & $0$ & $1$ & $\mathcal{U}$\\
        \hline
        & $z_1$ & $1$ & $e^{5i\pi/6}/\sqrt{3}$ & $0$ & $0$ & $0$ & $0$ \\
        $\mathbb{Z}_{6,Ia}$ & $z_2$ & $0$ & $0$ & $1$ & $e^{2i\pi/3}$ & $-1$ & $e^{i\pi/3}$ \\
        & $z_3$ & $0$ & $0$ & $1$ & $e^{2i\pi/3}$ & $1$ & $-e^{i\pi/3}$\\
        \hline
        & $z_1$ & $1$ & $e^{5i\pi/6}/\sqrt{3}$ & $0$ & $0$ & $0$ & $0$ \\
        $\mathbb{Z}_{6,Ib}$ & $z_2$ & $0$ & $0$ & $1$ & $e^{5i\pi/6}/\sqrt{3}$ & $0$ & $0$ \\
        & $z_3$ & $0$ & $0$ & $0$ & $0$ & $1$ & $-e^{i\pi/3}$\\
        \hline
        & $z_1$ & $1$ & $e^{i\pi/3}$ & $e^{2i\pi/3}$ & $-1$ & $-e^{i\pi/3}$ & $0$ \\
        $\mathbb{Z}_{6,IIa}$ & $z_2$ & $1$ & $e^{2i\pi/3}$ & $-e^{i\pi/3}$ & $1$ & $e^{2i\pi/3}$ & $0$ \\
        & $z_3$ & $1$ & $-1$ & $1$ & $-1$ & $1$ & $\mathcal{U}$\\
        \hline
        & $z_1$ & $1$ & $e^{i\pi/3}$ & $-1$ & $-1$ & $0$ & $0$ \\
        $\mathbb{Z}_{6,IIb}$ & $z_2$ & $0$ & $0$ & $0$ & $0$ & $1$ & $e^{2i\pi/3}$ \\
        & $z_3$ & $1$ & $-1$ & $\mathcal{U}$ & $1-\mathcal{U}$ & $0$ & $0$\\
        \hline
        & $z_1$ & $1$ & $-e^{2i\pi/3}$ & $-e^{i\pi/3}$ & $-e^{2i\pi/3}$ & $0$ & $0$ \\
        $\mathbb{Z}_{6,IIc}$ & $z_2$ & $1$ & $e^{i\pi/3}$ & $-e^{2i\pi/3}$ & $-e^{i\pi/3}$ & $0$ & $0$ \\
        & $z_3$ & $0$ & $0$ & $0$ & $0$ & $1$ & $\mathcal{U}$\\
        \hline
        & $z_1$ & $1$ & $e^{5i\pi/6}/\sqrt{3}$ & $0$ & $0$ & $0$ & $0$ \\
        $\mathbb{Z}_{6,IId}$ & $z_2$ & $0$ & $0$ & $1$ & $e^{2i\pi/3}$ & $0$ & $0$ \\
        & $z_3$ & $0$ & $0$ & $0$ & $0$ & $1$ & $\mathcal{U}$\\
        \hline
    \end{tabular}
    \caption{Complex structures of the orbifolds listed in table \ref{tab:2} (part 1 of 3). Here, $\mathcal{U}$ is a complex structure modulus, and the table reads in a straightforward way. For instance, for the orbifold $T ^6/\mathbb{Z}_3$, the complex coordinates are given by $z^1 = x^1 + e^{2i\pi/3}x^2, z^2 = x^3 + e^{2i\pi/3}x^4, z^3 = x^5 + e^{2i\pi/3}x^6$, etc.}
\end{table}

\begin{table}[ht!]
\centering
    \hspace{2cm} coefficients of the complex structure\\
    \vspace{5pt}
    \begin{tabular}{|c||c||c|c|c|c|c|c|}
       \hline
         \multicolumn{1}{|l}{{\hspace{-3pt}orbifold  }} & & $x_1$ & $x_2$ & $x_3$ & $x_4$ & $x_5$ & $x_6$\\
        \hline\hline
        & $z_1$ & $1$ & $e^{2i\pi/7}$ & $e^{4i\pi/7}$ & $e^{6i\pi/7}$ & $-e^{i\pi/7}$ & $-e^{3i\pi/7}$ \\
        $\mathbb{Z}_{7}$ & $z_2$ & $1$ & $e^{4i\pi/7}$ & $-e^{i\pi/7}$ & $-e^{5i\pi/7}$ & $e^{2i\pi/7}$ & $e^{6i\pi/7}$ \\
        & $z_3$ & $1$ & $-e^{i\pi/7}$ & $e^{2i\pi/7}$ & $-e^{3i\pi/7}$ & $e^{4i\pi/7}$ & $-e^{5i\pi/7}$\\
        \hline
        & $z_1$ & $1$ & $i$ & $-1$ & $-e^{i\pi/4}$ & $e^{3i\pi/4}$ & $e^{i\pi/4}$ \\
        $\mathbb{Z}_{8,Ia}$ & $z_2$ & $1$ & $-1$ & $1$ & $i$ & $-i$ & $i$ \\
        & $z_3$ & $1$ & $i$ & $-1$ & $e^{i\pi/4}$ & $-e^{3i\pi/4}$ & $-e^{i\pi/4}$\\
        \hline
        & $z_1$ & $1$ & $e^{i\pi/4}$ & $i$ & $-(1+\sqrt{2}+i)/2$ & $0$ & $0$ \\
        $\mathbb{Z}_{8,Ib}$ & $z_2$ & $0$ & $0$ & $0$ & $0$ & $i$ & $e^{3i\pi/4}/\sqrt{2}$ \\
        & $z_3$ & $1$ & $-e^{i\pi/4}$ & $i$ & $-(1-\sqrt{2} + i)/2$ & $0$ & $0$\\
        \hline
        & $z_1$ & $1$ & $e^{i\pi/4}$ & $i$ & $-(1+\sqrt{2}+i)/2$ & $-(1+\sqrt{2}+i)/2$ & $0$ \\
        $\mathbb{Z}_{8,IIa}$ & $z_2$ & $1$ & $e^{3i\pi/4}$ & $-i$ & $-(1-\sqrt{2}-i)/2$ & $-(1-\sqrt{2}-i)/2$ & $0$ \\
        & $z_3$ & $0$ & $0$ & $0$ & $1$ & $-1$ & $\mathcal{U}$\\
        \hline
        & $z_1$ & $1$ & $e^{i\pi/4}$ & $i$ & $-(1+\sqrt{2}+i)/2$ & $0$ & $0$ \\
        $\mathbb{Z}_{8,IIb}$ & $z_2$ & $1$ & $e^{3i\pi/4}$ & $-i$ & $-(1-\sqrt{2}-i)/2$ & $0$ & $0$ \\
        & $z_3$ & $0$ & $0$ & $0$ & $0$ & $1$ & $\mathcal{U}$\\
        \hline
        & $z_1$ & $1$ & $e^{i\pi/6}$ & $e^{i\pi/3}$ & $-1$ & $-e^{i\pi/6}$ & $-\sqrt{2}e^{i\pi/12}$ \\
        $\mathbb{Z}_{12,Ia}$ & $z_2$ & $1$ & $e^{2i\pi/3}$ & $-e^{i\pi/3}$ & $1$ & $e^{2i\pi/3}$ & $0$ \\
        & $z_3$ & $1$ & $-e^{i\pi/6}$ & $e^{i\pi/3}$ & $-1$ & $e^{i\pi/6}$ & $-\sqrt{2}e^{7i\pi/12}$\\
        \hline
        & $z_1$ & $1$ & $e^{i\pi/6}$ & $e^{11i\pi/12}/\sqrt{2}$ & $-e^{i\pi/12}/\sqrt{2}$ & $0$ & $0$ \\
        $\mathbb{Z}_{12,Ib}$ & $z_2$ & $0$ & $0$ & $0$ & $0$ & $1$ & $e^{2i\pi/3}$ \\
        & $z_3$ & $1$ & $-e^{i\pi/6}$ & $e^{5i\pi/12}/\sqrt{2}$ & $-e^{7i\pi/12}/\sqrt{2}$ & $0$ & $0$\\
        \hline
        & $z_1$ & $1$ & $e^{i\pi/6}$ & $e^{11i\pi/12}/\sqrt{2}$ & $-e^{i\pi/12}/\sqrt{2}$ & $0$ & $0$ \\
        $\mathbb{Z}_{12,II}$ & $z_2$ & $1$ & $e^{5i\pi/6}$ & $-e^{7i\pi/12}/\sqrt{2}$ & $e^{5i\pi/12}/\sqrt{2}$ & $0$ & $0$ \\
        & $z_3$ & $0$ & $0$ & $0$ & $0$ & $1$ & $\mathcal{U}$\\
        \hline
    \end{tabular}
    \caption{Complex structures of the orbifolds listed in table \ref{tab:2} (part 2 of 3). Here, $\mathcal{U}$ is a complex structure modulus, and the table reads in a straightforward way. For instance, for the orbifold $T ^6/\mathbb{Z}_7$, the complex coordinates are given by $z^1 = x^1 + e^{2i\pi/7}x^2 + e^{4i\pi/7}x^3 + e^{6i\pi/7}x^4 - e^{i\pi/7}x^5 - e^{3i\pi/7}x^6$, etc.}
\end{table}

\begin{table}[ht!]
\centering
    \hspace{2cm} coefficients of the complex structure\\
    \vspace{5pt}
    \begin{tabular}{|c||c||c|c|c|c|c|c|}
       \hline
         \multicolumn{1}{|l}{{\hspace{-3pt}orbifold  }} & & $x_1$ & $x_2$ & $x_3$ & $x_4$ & $x_5$ & $x_6$\\
        \hline\hline
        & $z_1$ & $1$ & $\mathcal{U}^1$ & $0$ & $0$ & $0$ & $0$ \\
        $\mathbb{Z}_{2}\times\mathbb{Z}_2$ & $z_2$ & $0$ & $0$ & $1$ & $\mathcal{U}^2$ & $0$ & $0$ \\
        & $z_3$ & $0$ & $0$ & $0$ & $0$ & $1$ & $\mathcal{U}^3$\\
        \hline
        & $z_1$ & $1$ & $\mathcal{U}$ & $0$ & $0$ & $0$ & $0$ \\
        $\mathbb{Z}_{2}\times\mathbb{Z}_4$ & $z_2$ & $0$ & $0$ & $1$ & $e^{3i\pi/4}/\sqrt{2}$ & $0$ & $0$ \\
        & $z_3$ & $0$ & $0$ & $0$ & $0$ & $1$ & $-e^{i\pi/4}/\sqrt{2}$\\
        \hline
        & $z_1$ & $1$ & $\mathcal{U}$ & $0$ & $0$ & $0$ & $0$ \\
        $\mathbb{Z}_{2}\times\mathbb{Z}_{6,I}$ & $z_2$ & $0$ & $0$ & $1$ & $e^{5i\pi/6}/\sqrt{3}$ & $0$ & $0$ \\
        & $z_3$ & $0$ & $0$ & $0$ & $0$ & $1$ & $e^{2i\pi/3}$\\
        \hline
        & $z_1$ & $1$ & $e^{5i\pi/6}/\sqrt{3}$ & $0$ & $0$ & $0$ & $0$ \\
        $\mathbb{Z}_{2}\times\mathbb{Z}_{6,II}$ & $z_2$ & $0$ & $0$ & $1$ & $-e^{i\pi/3}$ & $0$ & $0$ \\
        & $z_3$ & $0$ & $0$ & $0$ & $0$ & $1$ & $e^{5i\pi/6}/\sqrt{3}$\\
        \hline
        & $z_1$ & $1$ & $e^{2i\pi/3}$ & $0$ & $0$ & $0$ & $0$ \\
        $\mathbb{Z}_{3}\times\mathbb{Z}_3$ & $z_2$ & $0$ & $0$ & $1$ & $e^{2i\pi/3}$ & $0$ & $0$ \\
        & $z_3$ & $0$ & $0$ & $0$ & $0$ & $1$ & $-e^{i\pi/3}$\\
        \hline
        & $z_1$ & $1$ & $e^{2i\pi/3}$ & $0$ & $0$ & $0$ & $0$ \\
        $\mathbb{Z}_{3}\times\mathbb{Z}_6$ & $z_2$ & $0$ & $0$ & $1$ & $e^{5i\pi/6}/\sqrt{3}$ & $0$ & $0$ \\
        & $z_3$ & $0$ & $0$ & $0$ & $0$ & $1$ & $-e^{i\pi/6}/\sqrt{3}$\\
        \hline
        & $z_1$ & $1$ & $e^{3i\pi/4}/\sqrt{2}$ & $0$ & $0$ & $0$ & $0$ \\
        $\mathbb{Z}_{4}\times\mathbb{Z}_4$ & $z_2$ & $0$ & $0$ & $1$ & $e^{3i\pi/4}/\sqrt{2}$ & $0$ & $0$ \\
        & $z_3$ & $0$ & $0$ & $0$ & $0$ & $1$ & $-e^{i\pi/4}/\sqrt{2}$\\
        \hline
        & $z_1$ & $1$ & $e^{5i\pi/6}/\sqrt{3}$ & $0$ & $0$ & $0$ & $0$ \\
        $\mathbb{Z}_{6}\times\mathbb{Z}_6$ & $z_2$ & $0$ & $0$ & $1$ & $e^{5i\pi/6}/\sqrt{3}$ & $0$ & $0$ \\
        & $z_3$ & $0$ & $0$ & $0$ & $0$ & $1$ & $-e^{i\pi/6}/\sqrt{3}$\\
        \hline
    \end{tabular}
    \caption{complex structure of the orbifolds listed in table \ref{tab:2} (part 3 of 3). Here, the $\mathcal{U}$ are a complex structure moduli, and the table reads in a straightforward way. For instance, for the orbifold $T ^6/\mathbb{Z}_2\times\mathbb{Z}_2$, the complex coordinates are given by $z^1 = x^1 + \mathcal{U}^1x^2, z^2 = x^3 + \mathcal{U}^2x^4, z^3 = x^5 + \mathcal{U}^3x^6$.}
\end{table}

\clearpage

%\bibliography{bibliographie}

\end{document}